\documentclass[%
 reprint, 
superscriptaddress,
 amsmath,amssymb,
 aps, prd,
floatfix,
]{revtex4-2}

\usepackage{graphicx}
\usepackage{dcolumn}
\usepackage{bm}

\usepackage[utf8]{inputenc} 
\usepackage[T1]{fontenc}    
\usepackage{xcolor}
\definecolor{darkblue}{RGB}{46,48,147}
\usepackage[colorlinks=true,
            linkcolor=darkblue,
            urlcolor=darkblue,
            citecolor=darkblue]{hyperref}     
\usepackage{url}            
\usepackage{booktabs}       
\usepackage{amsfonts}       
\usepackage{nicefrac}       
\usepackage{microtype}      
\usepackage{lipsum}
\usepackage{graphicx}       
\usepackage{amsmath}
\usepackage{xspace}
\usepackage{array}
\usepackage{adjustbox}
\usepackage{xcolor}
\usepackage[capitalise]{cleveref}
\usepackage[acronym]{glossaries}

\usepackage{array}
\usepackage{scalerel}
\usepackage{tikz}
\usetikzlibrary{svg.path}

\definecolor{orcidlogocol}{HTML}{A6CE39}
\tikzset{
  orcidlogo/.pic={
    \fill[orcidlogocol] svg{M256,128c0,70.7-57.3,128-128,128C57.3,256,0,198.7,0,128C0,57.3,57.3,0,128,0C198.7,0,256,57.3,256,128z};
    \fill[white] svg{M86.3,186.2H70.9V79.1h15.4v48.4V186.2z}
                 svg{M108.9,79.1h41.6c39.6,0,57,28.3,57,53.6c0,27.5-21.5,53.6-56.8,53.6h-41.8V79.1z M124.3,172.4h24.5c34.9,0,42.9-26.5,42.9-39.7c0-21.5-13.7-39.7-43.7-39.7h-23.7V172.4z}
                 svg{M88.7,56.8c0,5.5-4.5,10.1-10.1,10.1c-5.6,0-10.1-4.6-10.1-10.1c0-5.6,4.5-10.1,10.1-10.1C84.2,46.7,88.7,51.3,88.7,56.8z};
  }
}

\newcommand\orcidicon[1]{\href{https://orcid.org/#1}{\mbox{\scalerel*{
\begin{tikzpicture}[yscale=-1,transform shape]
\pic{orcidlogo};
\end{tikzpicture}
}{0}}}}

\newacronym{hpo}{HPO}{hyperparameter optimization}
\newacronym{hpc}{HPC}{high performance computing}
\newacronym{hep}{HEP}{high energy physics}
\newacronym{hp}{HP}{Hyperparameter}
\newacronym{ml}{ML}{machine learning}
\newacronym{nn}{NN}{neural network}
\newacronym{pf}{PF}{particle-flow}
\newacronym{oc}{OC}{object condensation}
\newacronym{gnn}{GNN}{graph neural network}
\newacronym{llm}{LLM}{large language model}
\newacronym{mlpf}{MLPF}{machine-learned particle flow}
\newacronym{iqr}{IQR}{interquartile range}
\newacronym{lhc}{LHC}{Large Hadron Collider}
\newacronym{hllhc}{HL-LHC}{High Luminosity LHC}
\newacronym{fcc}{FCC}{Future Circular Collider}
\newacronym{cms}{CMS}{Compact Muon Solenoid}
\newacronym{lsh}{LSH}{locality sensitive hashing}
\newacronym{cld}{CLD}{CLIC-like detector}
\newacronym{hgcal}{HGCAL}{high-granularity calorimeter}
\newacronym{knn}{kNN}{k-nearest neighbors}

\newcommand{\kt}{\ensuremath{k_{\mathrm{T}}}\xspace}
\newcommand{\ptmomentum}{\ensuremath{p_{\mathrm{T}}}\xspace}
\newcommand{\ptvecmiss}{\ensuremath{{\vec p}_{\mathrm{T}}^{\kern1pt\text{miss}}}\xspace}
\newcommand{\ptmiss}{\ensuremath{\ptmomentum^\text{miss}}\xspace}

\newcommand{\GEANTfour}{{\textsc{Geant4}}\xspace}

\newcommand{\pythia} {{\textsc{Pythia}}\xspace}

\newcommand{\akfourchs}[1]{AK4-CHS\xspace}
\newcommand{\akfourpuppi}[1]{AK4-PUPPI\xspace}
\newcommand{\GeV}{\ensuremath{\,\text{Ge\hspace{-.08em}V}}\xspace}

\newcommand{\PQb}{\ensuremath{\mathrm{b}}\xspace}

\newcommand{\Pe}{\ensuremath{\mathrm{e}}\xspace}

\newcommand{\ttbar}{\ensuremath{\mathrm{t\overline{t}}}\xspace}
\newcommand{\qq}{\ensuremath{\mathrm{q\overline{q}}}\xspace}

\newcommand{\ww}{\ensuremath{\mathrm{WW}}\xspace}

\newcommand{\finetuned} {fine-tuned\xspace}
\newcommand{\fromscratch} {from scratch\xspace}
\newcommand{\Finetuned} {Fine-tuned\xspace}
\newcommand{\Fromscratch} {From scratch\xspace}

\begin{document}

\preprint{APS/123-QED}

\title{\textbf{Fine-tuning machine-learned particle-flow reconstruction for new detector geometries in future colliders} 
}%

\received{25 March 2025}
\accepted{1 May 2025}
\published{29 May 2025}

\author{Farouk Mokhtar\orcidicon{0000-0003-2533-3402}}
 \email{Contact author: fmokhtar@ucsd.edu}
\affiliation{%
    \href{https://ror.org/0168r3w48}{University of California San Diego}, La Jolla, California 90293, USA
}%
\author{Joosep Pata\orcidicon{0000-0002-5191-5759}}
 \email{Contact author: joosep.pata@cern.ch}
\affiliation{
    \href{https://ror.org/03eqd4a41}{National Institute of Chemical Physics and Biophysics}, Tallinn 12618, Estonia
}%
\author{Dolores Garcia\orcidicon{0000-0002-0120-8757}}
\affiliation{%
    \href{https://ror.org/01ggx4157}{European Center for Nuclear Research (CERN)}, Geneva 1211, Switzerland 
}%
\author{Eric Wulff\orcidicon{0000-0002-4681-8516}}
\affiliation{%
    \href{https://ror.org/01ggx4157}{European Center for Nuclear Research (CERN)}, Geneva 1211, Switzerland 
}%
\author{Mengke Zhang\orcidicon{0009-0004-7492-4895}}%
\affiliation{%
    \href{https://ror.org/0168r3w48}{University of California San Diego}, La Jolla, California 92093, USA
}%
\author{Michael Kagan\orcidicon{0000-0002-3386-6869}}
\affiliation{%
    \href{https://ror.org/05gzmn429}{SLAC National Accelerator Laboratory}, Menlo Park, California 94025, USA
}%
\author{Javier Duarte\orcidicon{0000-0002-5076-7096}}%
\affiliation{%
    \href{https://ror.org/0168r3w48}{University of California San Diego}, La Jolla, California 92093, USA
}%

\begin{abstract}

We demonstrate transfer learning capabilities in a machine-learned algorithm trained for particle-flow reconstruction in high energy particle colliders.
This paper presents a cross-detector fine-tuning study, where we initially pre-train the model on a large full simulation dataset from one detector design, and subsequently fine-tune the model on a sample with a different collider and detector design. 
Specifically, we use the Compact Linear Collider detector (CLICdet) model for the initial training set, and demonstrate successful knowledge transfer to the CLIC-like detector (CLD) proposed for the Future Circular Collider in electron-positron mode (FCC-ee).
We show that with an order of magnitude less samples from the second dataset, we can achieve the same performance as a costly training from scratch, across particle-level and event-level performance metrics, including jet and missing transverse momentum resolution.
Furthermore, we find that the fine-tuned model achieves comparable performance to the traditional rule-based particle-flow approach on event-level metrics after training on 100,000 CLD events, whereas a model trained from scratch requires at least 1 million CLD events to achieve similar reconstruction performance.
To our knowledge, this represents the first full-simulation cross-detector transfer learning study for particle-flow reconstruction.
These findings offer valuable insights towards building large foundation models that can be fine-tuned across different detector designs and geometries, helping to accelerate the development cycle for new detectors and opening the door to rapid detector design and optimization using machine learning.\\\\
DOI: \href{https://doi.org/10.1103/PhysRevD.111.092015}{10.1103/PhysRevD.111.092015}
\end{abstract}

\maketitle

\clearpage
\section{Introduction}
\label{intro}

Collision events at high energy physics (HEP) experiments such as at the CERN Large Hadron Collider (LHC) occur at extremely high rates and energies, creating complex, dense detector signatures.
Efficient reconstruction of events is essential to perform precision measurements and search for new physics, which could lead to new discoveries such as the Higgs boson observation~\cite{CMS2012,CMS:2013btf,ATLAS2012}.
Reconstruction algorithms at the LHC generally fall under two categories: local and global.
Local reconstruction algorithms rely on individual detector subsystems to measure the energies or trajectories of outgoing particles via energy deposits or ionization signals recorded as a series of hits as they traverse the detector. 
For instance, ATLAS or CMS track reconstruction algorithms rely on the tracker subsystem to reconstruct \textit{tracks}, while calorimeter clustering algorithms rely on the energy deposits in the calorimeters to build \textit{clusters}.
Subsequently, particle-flow (PF) algorithms combine this information to reconstruct individual particles and improve the global description of the event, especially in terms of the resolution of \textit{jets}, sprays of collimated particles, and the missing transverse momentum.
Such global event reconstruction algorithms have been used by many high energy physics experiments~\cite{Buskulic:1994wz,Aaboud:2017aca,Bocci:2001zx,Connolly:2003gb,Abulencia:2007iy,Sirunyan:2017ulk,Behrend:1982gk,Abazov:2008ff,Abreu:1995uz,H1:2020zpd,Breitweg:1997aa,Breitweg:1998gc}. 
In addition, PF reconstruction is a key driver in the detector design for future lepton colliders~\cite{FCC-ee,Behnke2013,CEPCStudyGroup2018}, where the detector technologies and geometries are still under development.

Current PF reconstruction algorithms are generally based on imperative, rule-based, approaches such as proximity-based linking, and have hundreds of tunable parameters that are detector dependent.
Moreover, PF algorithms are detector specific and have a long development cycle.
On the other hand, machine learning (ML) algorithms can be optimized \textit{ab initio} based on simulation samples while making use of low-level features in particle interactions with the detector that may not be immediately obvious from a first-principles approach based on feature engineering.
ML algorithms can be used to augment rule-based PF algorithms in future high-luminosity scenarios at the LHC, where the events will be more complex and detectors more granular.
In addition, unlike the traditional PF algorithm, an ML-based approach may be easily adapted to new detector concepts and detector geometries, reducing the development cycle for reconstruction software, and maximizing the physics potential of future colliders.
At the same time, highly parallel architectures such as graphics processing units (GPUs) and ML architectures such as neural networks have benefited from co-design, and significant global engineering efforts in the public and private sectors have optimized the inference of neural networks on large input sequences.

Typically, ML algorithms used in HEP experiments are trained using supervised learning.
A simulated dataset is generated with associated ground truth labels, or targets, and the model attempts to predict these targets during supervised training.
To enhance the generalization of the model, sampling from different physics processes is often considered during the training.
However, the detector configuration is typically fixed in this process.
This necessitates the design and training of a new model for every detector configuration of interest, which is time- and resource-intensive, making it challenging to iterate new detector designs and evaluate their effectiveness for the next generation of colliders.

Inspired by the large ``foundation models'' pioneered in natural language processing such as BERT~\cite{Devlin2019}, GPT-4~\cite{OpenAI2023}, and others~\cite{Lewis2019,Brown2020}, and recently explored in the HEP context~\cite{Golling2023arXiv,Li:2024htp,leigh2024token,Mikuni2024,Mikuni:2025tar,harris2024resim,Birk:2024knn,Katel:2024ygn,Birk:2025wai,Amram:2024fjg,vigl2024finetuning,Kishimoto2023}, our aim is to design an ML algorithm for event reconstruction that is sufficiently general and flexible to be easily adapted to different detector configurations.
Such a model could accelerate the development cycle for new detector geometries, and open the door to detector design optimization using machine learning.

We present a cross-detector study to show the transfer learning capabilities of MLPF, where we first train the algorithm on a dataset from the Compact Linear Collider detector (CLICdet) model~\cite{CLICdp:2017vju,CLICdp:2018vnx}, subsequently fine-tune it on a second dataset from the CLIC-like detector (CLD)---one of the detector models proposed for the Future Circular Collider in electron-positron mode (FCC-ee)~\cite{Bacchetta2019}.
To our knowledge, this represents the first cross-detector transfer learning study for PF reconstruction.

The paper is structured as follows: in Sec.~\ref{sec:transfer_learning}, we present a possible direction for the future of ML-based reconstruction algorithms as well as introduce common terminology in the context of transfer learning.
Section~\ref{sec:dataset} describes the datasets used for this study.
In Sec.~\ref{sec:mlpfalgo} we describe the ML model and target definition developed to run the ML training.
Finally, results from the cross-detector transfer learning study are presented in Sec.~\ref{sec:results}.

\subsection{Related work}

There has been significant activity on ML-based event reconstruction methods, including for PF reconstruction.
The first approaches for PF reconstruction based on computer vision were investigated in Ref.~\cite{DiBello:2020bas}.
A loss function that is capable of reconstructing a variable number of events without padding was proposed in Ref.~\cite{Kieseler:2020wcq}.
In Ref.~\cite{Pata:2021oez}, the machine-learned particle-flow (MLPF) algorithm, a graph neural network (GNN) based approach, was introduced to reconstruct events with high particle multiplicity in a scalable and accurate manner.
This approach has also been developed in the CMS experiment~\cite{Pata:2022wam,Mokhtar:2023fzl}.
Recently, Refs.~\cite{DiBello:2022iwf,Kakati:2024dun} demonstrated that particle-flow reconstruction can be formulated as a hypergraph learning task, and approached by transformers, improving the jet performance over the baseline.
In parallel, clustering using ML has been demonstrated for high-granularity calorimeter reconstruction~\cite{CMS:2022txh}.
In Ref.~\cite{Wahlen:2024rxt}, transformer-based architectures have been studied for a future collider setup on single-particle gun samples, and in Ref.~\cite{garcia2023towards}, geometric GNN approaches have been applied to the FCC-ee, achieving the same performance as the baseline.
Additionally, several studies have considered transfer learning in HEP through supervised learning in one dataset with subsequent application in a different dataset~\cite{Kuchera:2018djs,Tombs2022,Dillon2022,Dillon2023,Chappell:2022yxd,Dreyer:2022yom,Qu:2022mxj,Beauchesne:2023vie,Mikuni2024,Mikuni:2025tar,Li:2024htp}. 

\section{Cross-detector transfer learning}
\label{sec:transfer_learning}

\begin{figure*}[htbp]
    \centering
    \includegraphics[width=0.49\textwidth]{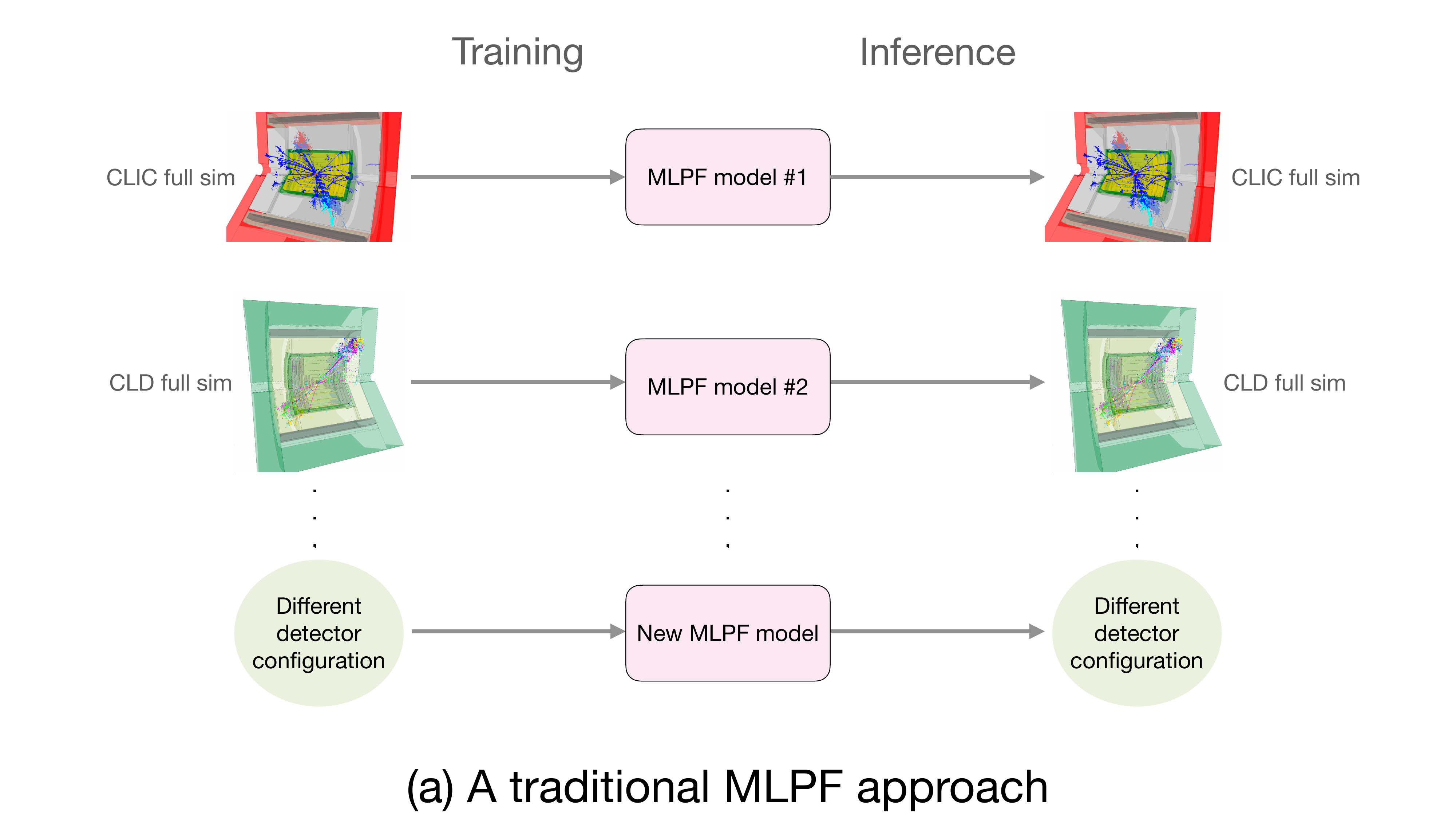}
    \includegraphics[width=0.49\textwidth]{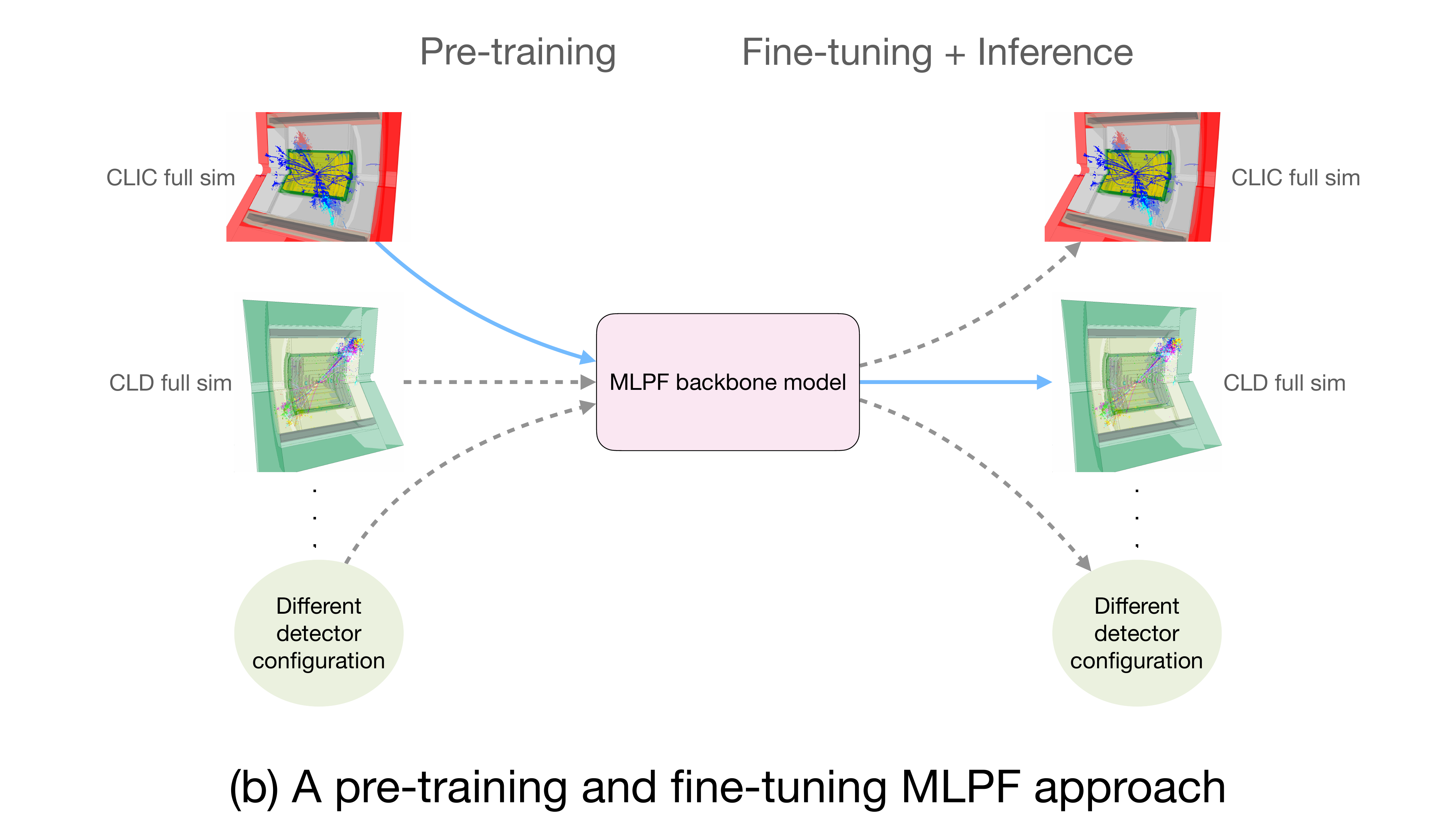}
    \caption{A comparison between traditional ML-based reconstruction algorithms (left), and the foundation model approach to reconstruction (right).
    We highlight with the blue, solid arrows the transfer learning study presented in the paper.}
    \label{fig:finetuning_structure}
\end{figure*}

We build a large ML model for PF reconstruction with the goal of adapting it to different detector configurations.
We refer to this model as the ``MLPF backbone model'' in Fig.~\ref{fig:finetuning_structure} and in the remainder of the paper.
The training procedure of the MLPF backbone model involves a first \textit{pre-training} stage, where the model is trained on a large dataset of one detector configuration to learn representations that are useful for particle-flow reconstruction, and a second \textit{fine-tuning} stage where the model is trained on a smaller dataset from a different detector configuration.
The dataset used during the fine-tuning stage is referred to as the \textit{downstream} dataset.
For the remainder of this work, we use the words transfer learning and fine-tuning interchangeably.
A successful backbone model is able to efficiently learn useful representations during the pre-training stage that will help generalize to different downstream datasets during the fine-tuning stage.
As the representations are already specialized for the pre-training task and not randomly initialized, the model requires significantly less computational resources to learn the downstream task than a model designed specifically for that purpose.
Pre-training methods can be categorized as supervised or self-supervised.
Self-supervised learning (SSL) uses a proxy task to learn useful representations and is needed to make use of large, unlabeled datasets, such as real collision data.
SSL methods adapted to HEP data are an active area of study~\cite{assran2023selfsupervisedlearningimagesjointembedding,Favaro_2025}.
When large labeled datasets are available, such as in HEP simulation, pre-training can also be defined based on a more physics-relevant supervised task, in this case, particle reconstruction.

\section{Dataset}
\label{sec:dataset}

For the purpose of this study, we produce two datasets consisting of $\Pe^+\Pe^-$ collision events with \pythia (v8.306)~\cite{Bierlich:2022pfr}, each followed by a different detector simulation based on \GEANTfour~\cite{GEANT42003,GEANT42006,GEANT42016} (v11.0.2), and the \textsc{Key4HEP} software~\cite{ganis2022key4hep}, along with the \textsc{Marlin} reconstruction code~\cite{Gaede:2006pj}, and the \textsc{Pandora} package~\cite{Marshall:2012hh,Marshall:2012ry,Marshall:2015rfa} for a baseline PF implementation.
We document the exact design configurations used for the detector models in the following subsections.
Due to the specific nature of $\Pe^+\Pe^-$ collision events, the events are expected to provide cleaner environments for reconstruction, having neither pileup nor underlying event contributions.

\subsection{Pre-training dataset}

The initial CLICdet training dataset is generated using the setup introduced in Ref.~\cite{Pata:2023rhh}.
We extended the statistics by approximately four times, and have made it publicly available~\cite{Pata2025}.
Additionally, the definition of target particles has been updated as discussed in Sec.~\ref{sec:targetdef}.
The simulation samples are produced based on the CLICdet model, including a full detector simulation with tracking material, realistic track reconstruction, and calorimeter calibration.
The CLICdet design draws inspiration from the CMS detector at CERN~\cite{CMS:2008xjf}, and the Silicon Detector (SiD) from the International Linear Collider (ILC)~\cite{Aryshev:2815947}, incorporating a superconducting solenoid with a 7\,m internal diameter that generates a 4\,T magnetic field at the detector's core. 
Within this solenoid are the silicon pixel and strip trackers, as well as an electromagnetic (ECAL) and hadron calorimeter (HCAL), each divided into barrel and two endcap sections. 
The ECAL consists of a finely segmented array featuring 40 layers of silicon sensors interspersed with tungsten plates, while the HCAL comprises 60 layers of plastic scintillator tiles with silicon photomultiplier readouts and steel absorber plates. 
The muon detection system envelops the solenoid and includes six active layers in the endcaps and seven in the barrel, interleaved with steel yoke plates. 
Additionally, two compact electromagnetic calorimeters, LumiCal and BeamCal, are situated in the extreme forward areas on either side of the interaction point~\cite{CLICdp:2017vju,CLICdp:2018vnx}.

The pre-training dataset consists of 24 million total events produced at a center-of-mass energy $\sqrt{s}=380\GeV$, including 8 million inclusive \ttbar events, 8 million fully hadronic \ww events, and 8 million \qq events.
The dataset is split into training and validation according to a 90:10 split, resulting in a total of 21.6 million events for training and 2.4 million events for testing and validation.
The dataset includes information about the generator particles, reconstructed tracks, calorimeter clusters, and reconstructed particles from the baseline \texttt{Pandora} algorithm saved in the \texttt{EDM4HEP} format.
For each event, jets are defined by clustering different sets of particles with the generalized \kt algorithm ($R=0.4, p=-1$) for $\Pe^{+}\Pe^{-}$ colliders~\cite{Cacciari:2008gp, Cacciari:2011ma} with a minimum $\ptmomentum \ge 5$\GeV.
The missing transverse momentum vector (\ptvecmiss) is calculated as the negative vector sum of the transverse momenta of all particles in an event, and its magnitude is denoted as \ptmiss.

\subsection{Downstream dataset}

The downstream dataset consists of physics samples based on the CLD detector, which has a similar design to the CLICdet model but with different parameters for the detector subsystems.
The main differences between the two detectors are: the lower solenoidal field of CLD (2\,T instead of 4\,T) which is compensated by a larger outer radius of the silicon tracker, and a shallower hadronic calorimeter due to the lower maximum center of mass energy of FCC-ee~\cite{bacchetta2019cld}.
The detector simulation and reconstruction software tools used to produce the CLD dataset are developed together with the linear collider community. 
The DD4HEP~\cite{Frank2013} detector simulation and geometry framework was developed in the AIDA and AIDA-2020 projects~\cite{AIDASoftware}.
The CLD geometry version used for this study is \texttt{CLD\_o2\_v05}~\cite{CLDConfig2023}, and the key design parameter values are listed in Table~\ref{tab:cld_clicdet_comparison}, along with a comparison with the CLICdet design parameters.

\begin{table}[b]
\caption{\label{tab:cld_clicdet_comparison}%
Comparison of key parameters for the CLICdet (with geometry version \texttt{CLIC\_o3\_v14}) and the CLD (with geometry version \texttt{CLD\_o2\_v05}) detector.
The inner radius of the calorimeters is given by the smallest distance of the calorimeter (dodecagon) to the main detector axis. 
``HCAL ring'' refers to the part of the HCAL endcap surrounding the ECAL endcap.
We highlight the differences in bold.
}
\begin{ruledtabular}
\footnotesize
\begin{tabular}{lcr}
Detector &
CLICdet &
CLD \\
\colrule
\textbf{Vertex inner radius [mm]} & \textbf{30} & \textbf{12.5} \\
\textbf{Vertex outer radius [mm]} & \textbf{116} & \textbf{111} \\
Tracker technology & Silicon & Silicon \\
Tracker half length [m] & 2.3 & 2.3 \\
Tracker inner radius [m] & 0.061 & 0.061 \\
\textbf{Tracker outer radius [m]} & \textbf{1.5} & \textbf{2.1} \\
\textbf{Inner tracker support cylinder radius [m]} & \textbf{0.580} & \textbf{0.696} \\ 
ECAL absorber & W & W \\
ECAL $X_0$ & 22 & 22 \\
\textbf{ECAL barrel $r_{\text{min}}$ [m]} & \textbf{1.5} & \textbf{2.15} \\
ECAL barrel $\Delta r$ [mm] & 202 & 202 \\
ECAL endcap $z_{\text{min}}$ [m] & 2.31 & 2.31 \\
ECAL endcap $\Delta z$ [mm] & 202 & 202 \\
HCAL absorber & Fe & Fe \\
\textbf{HCAL barrel $r_{\text{min}}$ [m]} & \textbf{1.74} & \textbf{2.40} \\
\textbf{HCAL barrel $\Delta r$ [m]} & \textbf{1.59} & \textbf{1.17} \\
HCAL endcap $z_{\text{min}}$ [m] & 2.54 & 2.54 \\
\textbf{HCAL endcap $z_{\text{max}}$ [m]} & \textbf{4.13} & \textbf{3.71} \\
\textbf{HCAL endcap $r_{\text{min}}$ [mm]} & \textbf{250} & \textbf{340} \\
\textbf{HCAL endcap $r_{\text{max}}$ [m]} & \textbf{3.25} & \textbf{3.57} \\
\textbf{HCAL ring $z_{\text{min}}$ [m]} & \textbf{2.36} & \textbf{2.35} \\
HCAL ring $z_{\text{max}}$ [m] & 2.54 & 2.54 \\
\textbf{HCAL ring $r_{\text{min}}$ [m]} & \textbf{1.74} & \textbf{2.48} \\
\textbf{HCAL ring $r_{\text{max}}$ [m]} & \textbf{3.25} & \textbf{3.57} \\
\textbf{Solenoid field [T]} & \textbf{4} & \textbf{2} \\
\textbf{Solenoid bore radius [m]} & \textbf{3.5} & \textbf{3.7} \\
\textbf{Solenoid length [m]} & \textbf{8.3} & \textbf{7.4} \\
\textbf{Overall height [m]} & \textbf{12.9} & \textbf{12.0} \\  
\textbf{Overall length [m]} & \textbf{11.4} & \textbf{10.6} \\  
\end{tabular}
\end{ruledtabular}
\end{table}

The CLD dataset consists of a total of 4.2 million \ttbar events produced at a center-of-mass energy $\sqrt{s}=365\GeV$.
The dataset is split into training and validation according to a 90:10 split, resulting in a total of 3.8 million events for training and 400,000 events for testing and validation.
The \ttbar process is chosen because it provides one of the most complex physics environments for the FCC experiment at the $\Pe^+\Pe^-$ early stage.

The CLD dataset provides access to the same information as the CLICdet dataset.
The size of the downstream dataset is approximately 2\,TB before preprocessing to the ML-specific format using the \textsc{tfds} library~\cite{TFDS}.
The raw datasets in \texttt{EDM4HEP} format, along with the scripts and configurations to generate the data, are publicly available at Ref.~\cite{ZenodoCLD2025raw}, and the datasets in the ML-specific format are available at~\cite{ZenodoCLD2025TFDS}.

\section{MLPF algorithm}
\label{sec:mlpfalgo}

The goal of the MLPF algorithm is to predict a set of target particles from input tracks and calorimeter clusters in an efficient and scalable fashion.
Since the input and output sets have different cardinality, we follow the same procedure as in Ref.~\cite{Pata:2021oez} to assign the target particles to input elements, such that each target particle is associated with a unique input element. 
In the following, we describe the simulation-based target definition (Sec.~\ref{sec:targetdef}), and the backbone model architecture and pre-training (Sec.~\ref{sec:mlpfbackbonepretraining}).

\subsection{Simulation-based target}
\label{sec:targetdef}
What set of particles should a particle-flow algorithm aim to reconstruct and how should it be extracted from the generator and simulator information?
The simplest unambiguous definition is all stable (status 1) \pythia particles, excluding invisible particles such as neutrinos.
We call this set the \textit{truth particles}, as it represents the output of the generator after the hard process and parton shower, but before the detector simulation and any particle decay and production processes therein.

However, not all truth particles may be directly reconstructible, for example, if they do not leave any hits in the detector directly, but only through their descendants in the simulation.
These descendants can be the result of decays or material interactions like electromagnetic showering.
Moreover, we aim to avoid reconstructing truth particles that do not result in significant energy deposits in the detector simulation as that would require the reconstruction algorithm to infer the particles from missing or unmeasured information.
Therefore, we define the set of \textit{target particles} for reconstruction by taking all status 1 \pythia particles that interacted with the detector either directly or through their descendants.
We opt for this choice, rather than choosing a subset of status 1 particles plus their interacting descendants as in~\cite{Kakati:2024dun}, as there is a potential to introduce double counting between the status 1 particles and their descendants.
We note that the current choice is straightforward to implement, but more work is required to define an algorithm that can be applied in all cases, unambiguously and with a minimum of bias between the target and truth particles.

In all cases, the target particle is assigned to a primary track or cluster to allow the classification-based loss to reconstruct the particle.
The primary track or cluster is chosen to be the one into which the target particle (or its descendants) deposited the largest amount of energy, based on the hits in the simulation.

As an illustration, a status 1 photon may not interact with the detector directly, but may instead produce an electromagnetic shower consisting of electrons, positrons and photons, some of which will be captured by the detector.
We include the original status 1 photon in the target particle set if any of the children interacted with the detector, but we do not attempt to reconstruct the children explicitly.
In this case, the target particle (the photon) will be associated with the calorimeter cluster to which the highest amount of energy was deposited by its descendants.

This target definition is an improvement over previous work~\cite{Pata:2023rhh}, where only status 1 particles that interacted directly with the detector, without accounting for possible interactions through the descendants, were used.
With the previous target definition, some status 1 particles were excluded from the target set while leaving a detectable signal in the detector, resulting in the target particles being less aligned with the truth particles, as noted in Ref.~\cite{Kakati:2024dun}.
We also distinguish between the set of target particles used for model training, which may be affected by the fiducial region of the detector and possible algorithmic choices, and the set of detector-independent truth particles, which are used for evaluating the final performance.
An example decay and simulation tree can be found in Fig.~\ref{fig:decay_tree} in the Appendix.

By choosing only stable particles that interact with the detector directly or through their decay products, we ensure that the reconstruction targets are based on observed quantities.
However, a small degree of uncorrectable smearing is introduced between the truth and target particles because of the detector acceptance and energy thresholds, as truth particles that do not leave any simulated or reconstructed detector hits are not reconstructible following this approach.

We cluster the target particles into jets to form the \textit{target jets}, the baseline PF particles to form the PF jets, and the truth particles to form \textit{truth jets}.
We cross-check the target definition algorithm by matching the \textit{target jets} via a $\Delta R<0.1$ criterion to the \textit{truth jets}.
No additional quality cuts are applied on the jets at this stage.
Based on the jet response distributions for the target (blue) and PF particles (orange) in Fig.~\ref{fig:target_matching}, we find that the jets from the target particles match the truth jets significantly better than those from the baseline PF algorithm, thus providing a well-defined and robust particle-level optimization target.

\subsection{MLPF backbone model and pre-training}
\label{sec:mlpfbackbonepretraining}

We develop a transformer-based architecture~\cite{NIPS2017_3f5ee243}, using self-attention without any approximations.
Previously, a GNN-based model was introduced in ~\cite{joosep_pata_2021_4559587}, and compared against an approximate transformer in Ref.~\cite{Pata:2023rhh}.
We now find that using \textsc{FlashAttention2}~\cite{dao2022flashattention,dao2023flashattention2fasterattentionbetter}, it is feasible and practical to switch to an attention-based model without any approximations while retaining scalability on large events.
The transformer architecture is an excellent candidate for correlating information across large input unordered sets, such as detector-level hits, due to the permutation-invariant nature of the attention mechanism.
The input to the model consists of a set of input tracks and clusters.
The full list of input features, and comparison between the CLICdet dataset and CLD dataset is documented in Appendix~\ref{app:input_feat}.
The input tracks and clusters are each embedded with a multilayer perceptron (MLP) into a 1024-dimensional space, allowing the model to attend to both types of inputs simultaneously.
The embedded tracks and clusters are input to both the classification and the regression transformer encoders.

The classification transformer encoder comprises three transformer blocks, while the regression transformer encoder is composed of another three transformer blocks for a total of six transformer blocks, each with a model dimension of 1024.
While the model is currently trained end-to-end to simultaneously solve the classification and regression tasks, separating the classification and regression layers gives the option to freeze either set of layers and continue training the other in case early convergence or over-training is observed in either loss.
After each linear layer in the model we apply a ReLU activation function.
We also apply a layer normalization prior to each transformer block in our architecture, as this has been shown to improve convergence by requiring significantly less training time and hyperparameter tuning in a wide range of applications~\cite{prelayernorm}.

\begin{figure*}[htbp]
    \centering
    \includegraphics[width=0.99\textwidth]{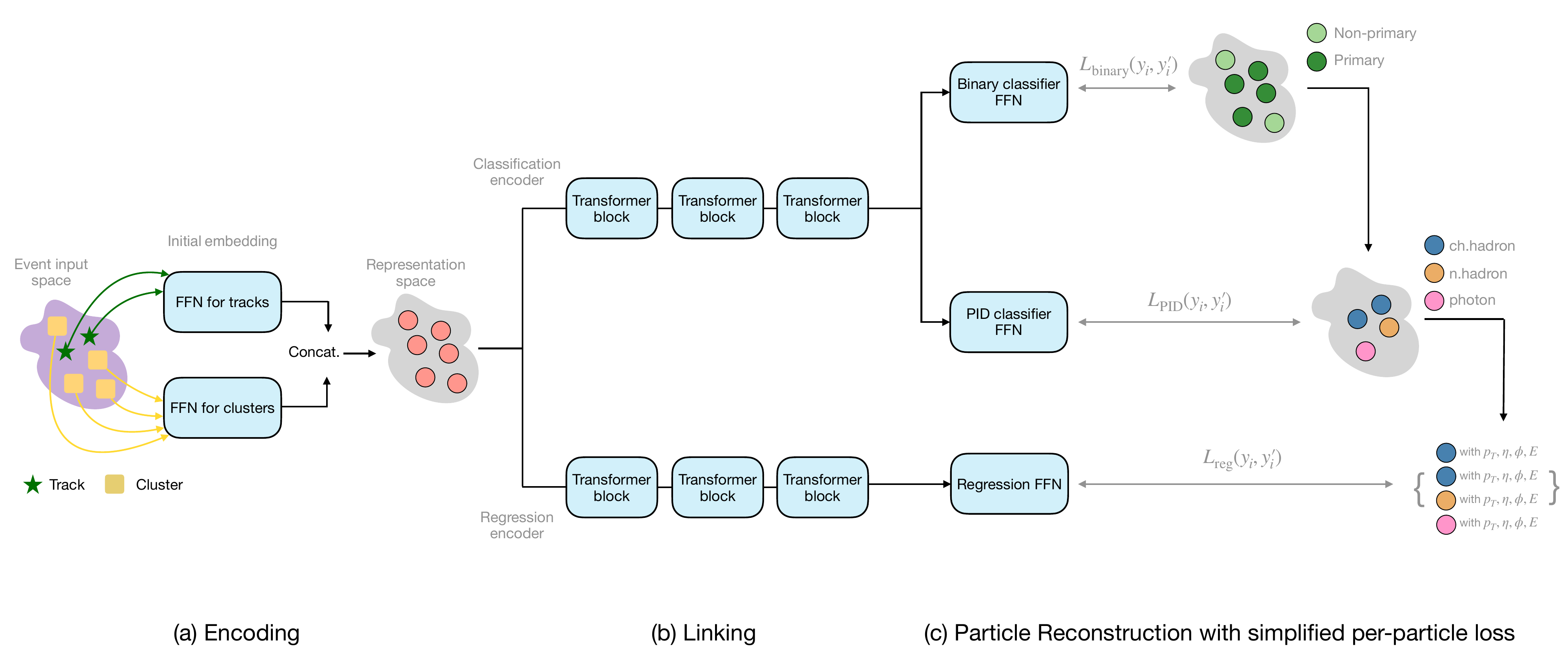}
    \caption{The backbone model architecture. The model is trained in an end-to-end fashion, with the encoder, linker and reconstruction trained simultaneously against the reconstruction loss.}
    \label{fig:model_structure}
\end{figure*}

The transformer blocks are followed by six separate MLP blocks each with two layers, targeting different outputs.
Two MLPs take as input the output from the classification transformer encoder, and are dedicated to the task of particle classification.
The first MLP is tasked with binary classification of primary (target) particles, and the second MLP is tasked with multi-classification of the particle identity (PID).
The remaining four MLPs take as input the output from the regression transformer encoder, and are dedicated to the task of regression.
The output of the four MLPs is the four-momentum of the particle candidates: $[\log{(\ptmomentum/p_{\mathrm{T},\mathrm{orig}})}, \eta, \sin \phi, \cos \phi, \log{(E/E_{\mathrm{orig}})}]$ respectively, where $p_{\mathrm{T},\mathrm{orig}}$ ($E_{\mathrm{orig}}$) refers to the corresponding track or cluster \ptmomentum (energy).
The final output is the set of predicted particle candidates, with a given PID and four-momentum.
The backbone model contains a total of 52 million parameters.
We present an illustration of the backbone model architecture in Fig.~\ref{fig:model_structure}.

The total loss is given by
\begin{equation}
\label{eq_loss}
L(Y,Y') = \sum_{i} L_\mathrm{binary}(y_i, y'_i) + L_\mathrm{pid}(y_i, y'_i) + L_\mathrm{reg}(y_i, y'_i),
\end{equation}
where $L_\mathrm{binary}(y_i, y'_i)$ denotes the binary cross entropy loss between the predicted and target particle, $L_\mathrm{pid}(y_i, y'_i)$ denotes a classification loss between the predicted and target particle type, for which we use the focal loss~\cite{lin2017focal} to address the large class imbalance between the different particle types, and finally, $L_\mathrm{reg}(y_i, y'_i)$ denotes the mean-squared-error loss for the momentum components.

The backbone model is trained for five epochs on the full CLICdet dataset described in Sec.~\ref{sec:dataset} on a single NVIDIA A100 80GB PCIe, with a batch size of 256 events.
In our training configuration, the model is trained with AdamW~\cite{adamw} with an initial learning rate of $10^{-4}$ in conjunction with a cosine decay learning rate schedule. 
This schedule is designed to adjust the learning rate following a cosine curve, gradually decreasing it to $10^{-5}$ by the end of the training.

\begin{figure*}[htbp]
    \centering
    \includegraphics[width=0.3\textwidth]{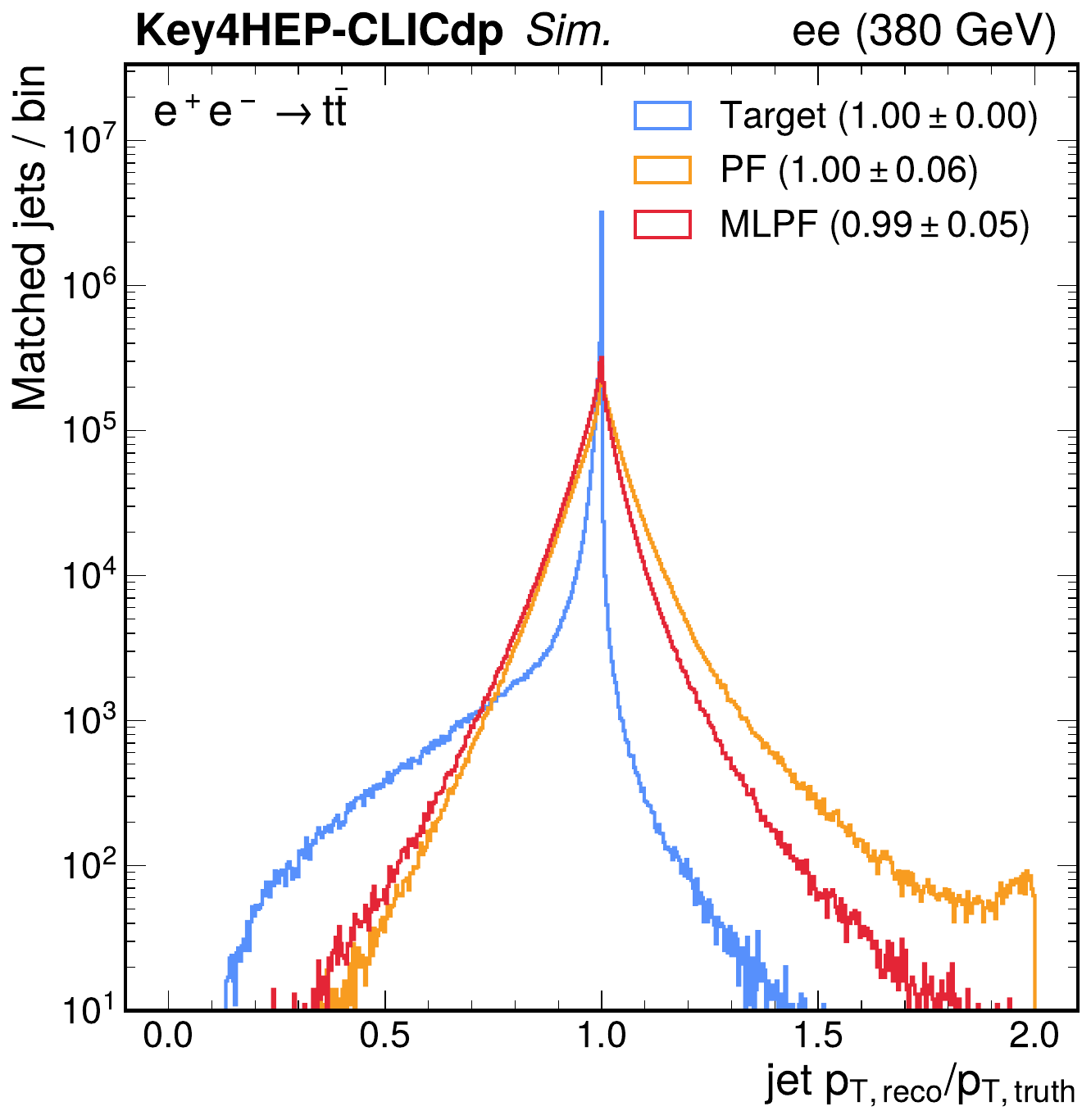}
    \includegraphics[width=0.3\textwidth]{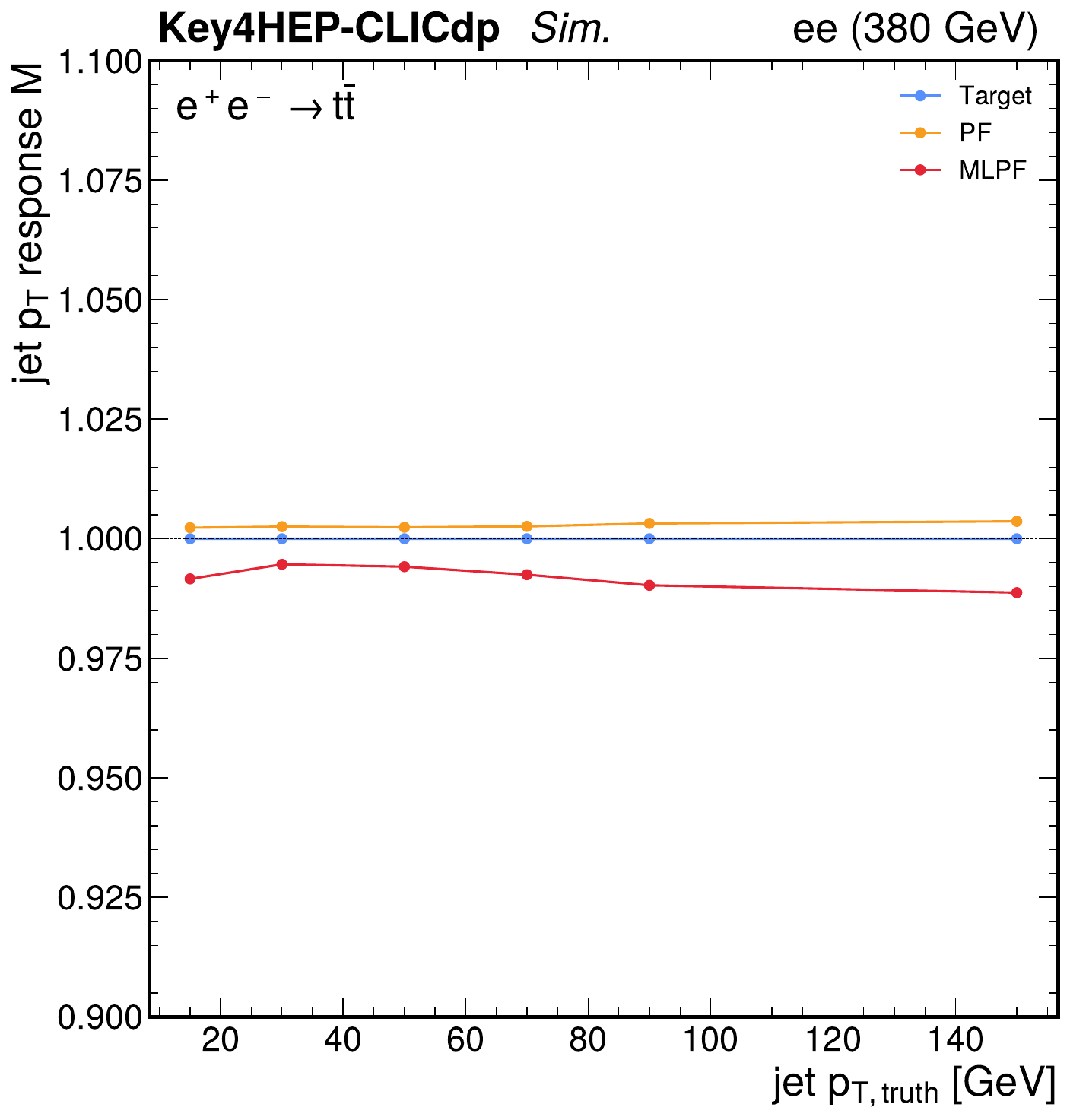}
    \includegraphics[width=0.3\textwidth]{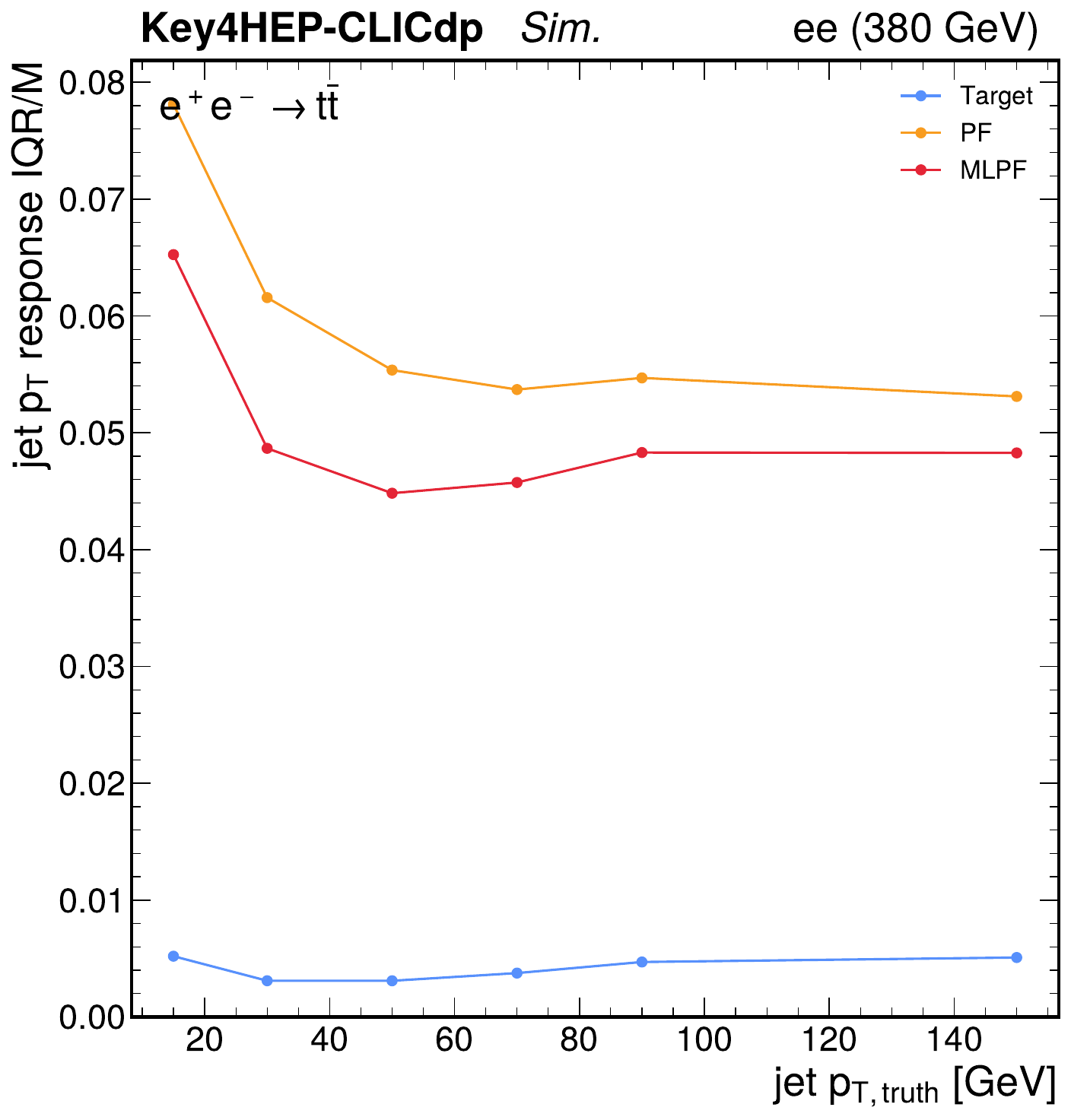}
    \caption{Jet response for target jets (blue), PF (orange), and MLPF  backbone (red); measured with respect to stable \pythia truth particles. The resolution of the target particle distribution, quantified by the interquartile range over the median, is significantly better than the resolution of jets from PF particles, as would be expected. There is some underestimation of the jet \ptmomentum in the target jets compared to the truth, attributable to choosing only particles that resulted in reconstructed hits for the target.}
    \label{fig:target_matching}
\end{figure*}

We investigate the jet performance of the backbone model by comparing MLPF reconstructed jets, PF jets, and \textit{target jets}.
We measure each with respect to \textit{truth jets} (as described in described in Sec.~\ref{sec:targetdef}).
In Fig.~\ref{fig:target_matching}, we present the inclusive jet response, the median of the jet response distribution as a function of \ptmomentum, and the jet resolution as a function of \ptmomentum quantified by the interquartile range over the median ($\mathrm{IQR}/\mathrm{M}$).
We observe that the resolution of the target jets is significantly better than the resolution of jets from PF particles.
We also observe that the resolution of the MLPF reconstructed jets is better than the baseline PF algorithm.

\begin{figure}[htbp]
    \centering
    \includegraphics[width=0.4\textwidth]{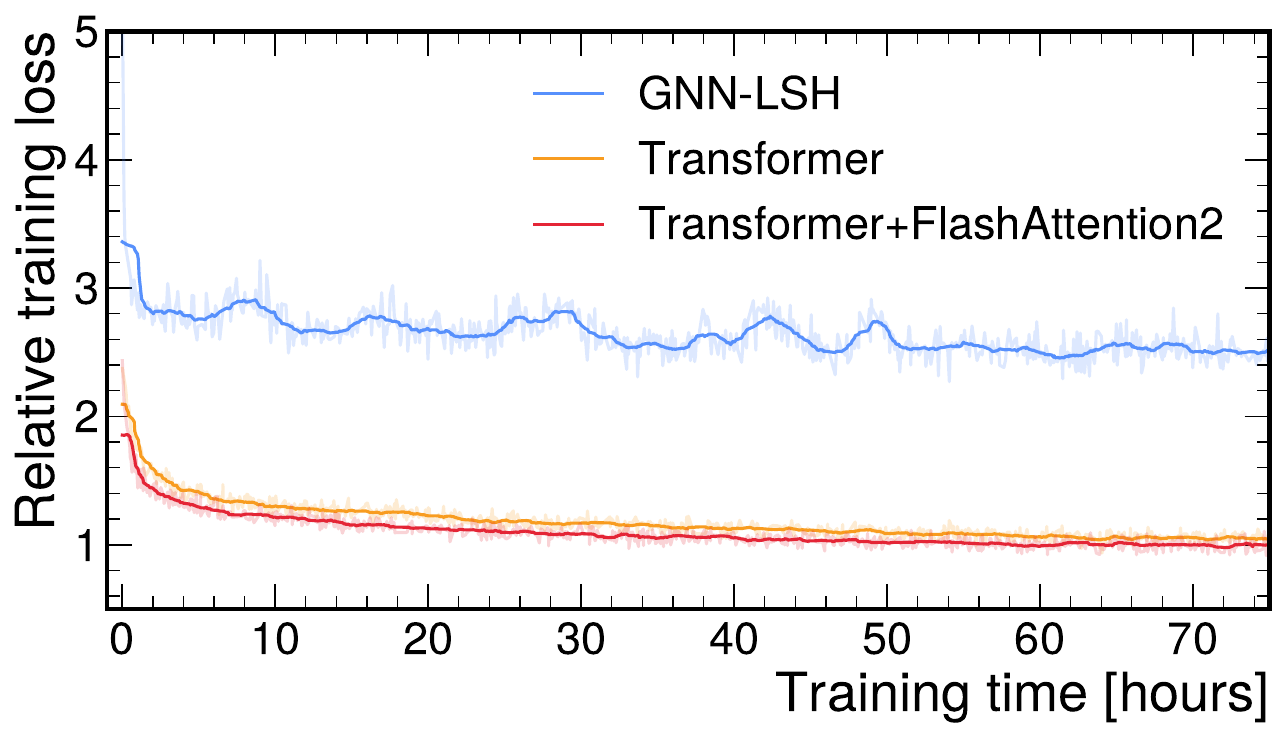}
    \includegraphics[width=0.4\textwidth]{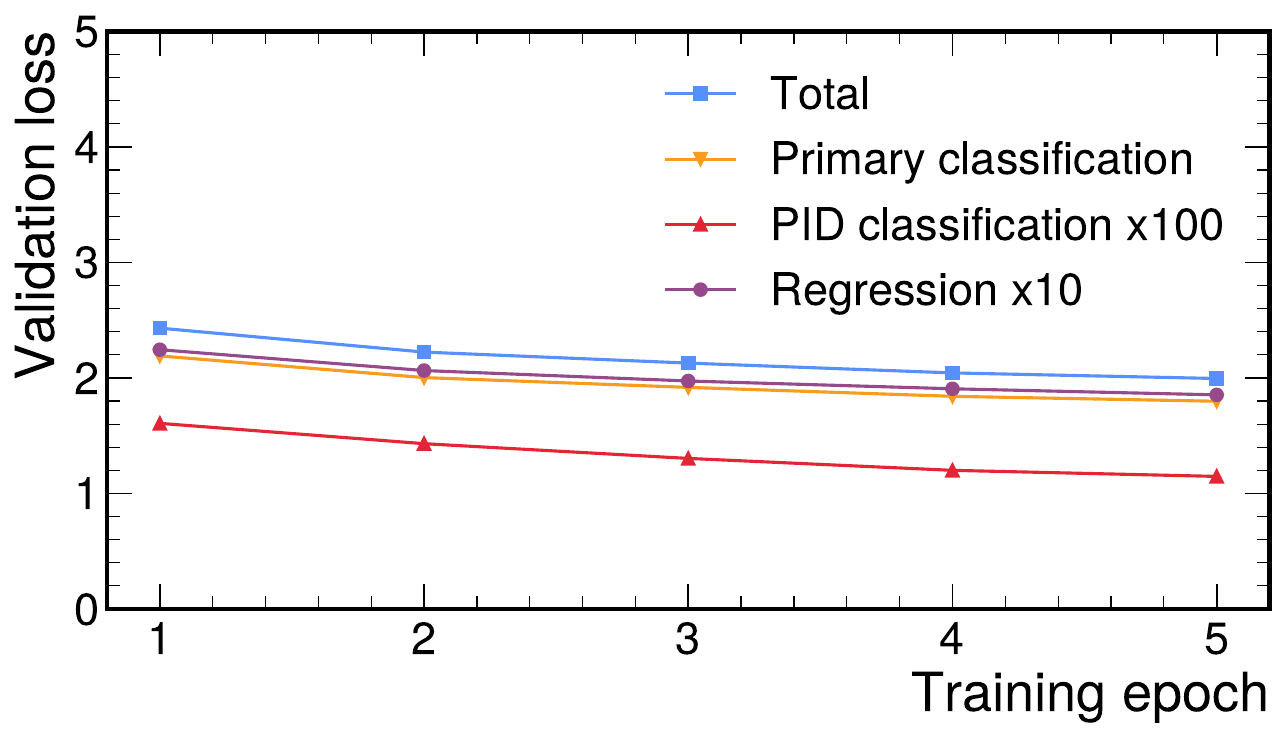}\\
    \includegraphics[width=0.4\textwidth]{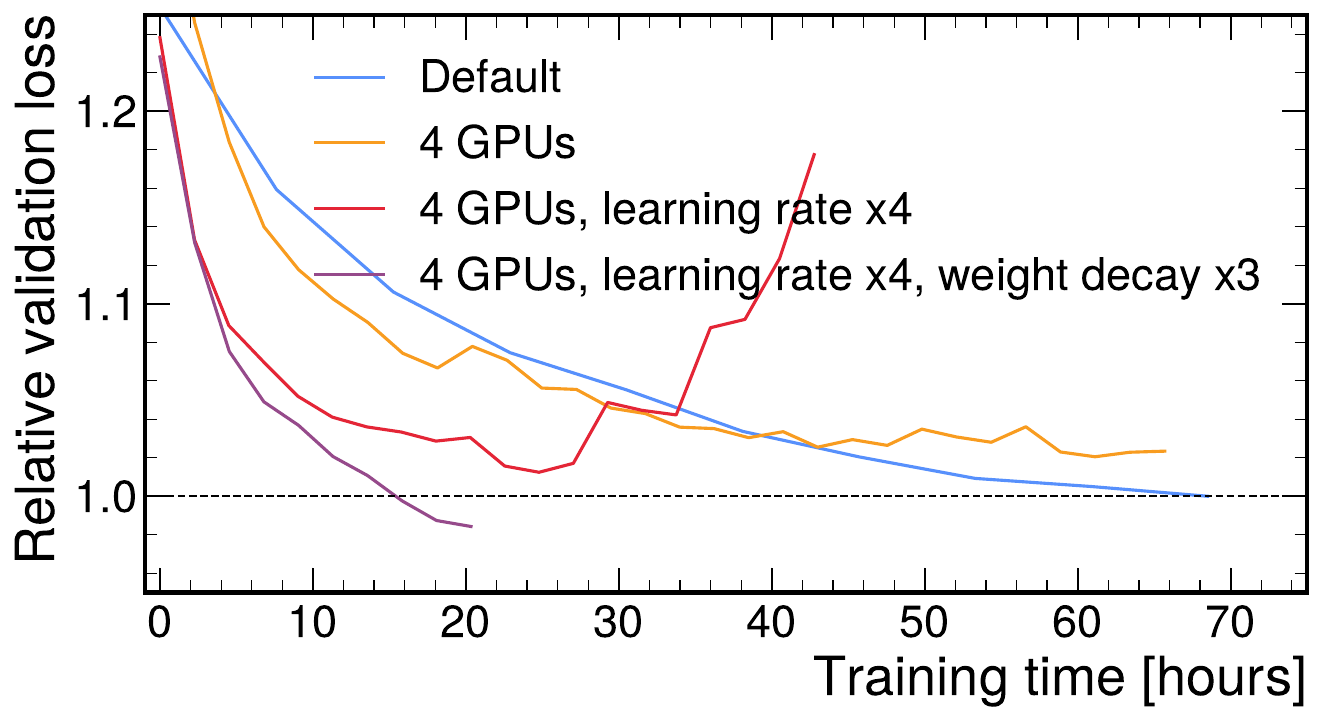}\\
    \caption{Top: the training loss of alternative backbone model candidates, relative to the best achieved training loss.
    With optimizations from \textsc{FlashAttention2}, we are able to achieve significantly lower training losses than with the previously proposed GNN-LSH model within the same training time.
    Middle: the validation loss components for the best model, as a function of the training epoch. 
    In the caption, primary classification refers to the $L_\mathrm{binary}(y_i, y'_i)$ loss term introduced in Eq.~(\ref{eq_loss}).
    The validation loss decreases steadily for all components.
    Bottom: the effect of training the transformer without any approximations with a large batch size with multiple GPUs, comparing to the default training on a single GPU.
    We find that in order to make use of multiple GPUs effectively, the learning rate and weight decay have to be tuned appropriately.}
    \label{fig:loss_backbone}
\end{figure}

We compare the current transformer-based backbone model to the previously-used GNN-LSH implementation, and find that on these $\Pe^+\Pe^-$ events with up to a few hundred particles per event, the transformer architecture significantly outperforms the previously proposed GNN-LSH scheme in terms of training loss convergence.
Moreover, we find that by using \textsc{FlashAttention2}~\cite{dao2023flashattention2fasterattentionbetter}, the convergence speed is further improved by about a factor 2, while also permitting training on significantly larger events on a single GPU.
Figure~\ref{fig:loss_backbone} presents the training and validation loss of the backbone model, comparing alternative models as well as the different loss components.
While in Ref.~\cite{Pata:2023rhh}, we introduced a GNN with locality-sensitive hashing (GNN-LSH) as an effective way to break the quadratic scaling of the training with event size, we find now that with the improvements to the attention computation in \textsc{PyTorch}, including the efficient \textsc{FlashAttention2} kernel, training a transformer without any approximations is feasible both on the relatively low multiplicity events used in this study, as well as larger events consisting of 5,000--10,000 particles that might be encountered in other scenarios.
Moreover, we corroborate the finding from other fields of ML that transformers trained on sufficiently large datasets are often more performant than alternatives.
Therefore, in this work, we use a standard transformer with \textsc{FlashAttention2} for the backbone model, as it requires fewer custom layers, benefits from hardware co-design, and results in significantly lower training and validation losses.

Due to the necessity and utility of training on large datasets, we also investigate strategies to train the model on large batch sizes on multiple GPUs using a data-parallel training strategy.
Figure~\ref{fig:loss_backbone} compares the default training regime using a single GPU with multi-GPU training regimes. 
The batch size per GPU is kept constant, meaning the four-GPU trainings have a global batch size four times larger and therefore four times fewer optimization steps per epoch are performed.
This results in completing one epoch faster, but at a higher validation loss.
To combat this effect, the learning rate was scaled proportionally to the batch size, allowing the optimizer to take larger steps.
We found that this allowed the four-GPU training to reach a validation loss comparable to that of the single-GPU training in a shorter amount of time.
However, we observed that multi-GPU training with a scaled learning rate diverges after approximately 25 hours and is not able to achieve the same validation loss as single-GPU training, as shown in Fig.~\ref{fig:loss_backbone} (bottom).
To address this apparent overfitting, we increased the weight decay parameter by a factor of three, from 0.01 to 0.03.
We hypothesized that this adjustment would introduce a regularizing effect that might have been lost due to the larger global batch size.
Consequently, we witnessed faster convergence and an even lower validation loss compared to the default single-GPU training (see red line in bottom plot of Fig.~\ref{fig:loss_backbone}).

\section{Cross-detector performance}
\label{sec:results}

We study the ability of MLPF to perform event reconstruction on the CLD dataset by comparing two different models,
\begin{itemize}
    \item \textbf{\Finetuned}: a model with the same architecture as the backbone model, initialized with the same weights as the backbone model and trained on a subset of the CLD dataset
    \item \textbf{\Fromscratch}: a model with the same architecture as the backbone model, randomly initialized and trained on a subset of the CLD dataset
\end{itemize}

Both models follow a similar training strategy, except that we train the \finetuned model with a smaller learning rate ($10^{-5}$) compared to the model trained \fromscratch ($10^{-4}$).
This is done to preserve the pre-training performance, increase the stability of the training, and reduce overfitting.
Both models use a batch size of 128 events, and are trained for up to 100 epochs with early stopping on the total validation loss.
We train both models on larger portions of the CLD dataset to evaluate the performance as a function of the downstream dataset size.
Models trained on datasets of size less than $10^5$ are trained with a patience of 20 epochs, models trained with datasets of size $10^5$ are trained with a patience of 10 epochs, and finally, models with datasets of size $10^6$ or larger are trained with a patience of 5 epochs.
All experiments with dataset size less than 1 million events are repeated three times with different random seeds to ensure robustness of the results.
For those experiments, the shaded uncertainty band covers the root-mean-square (RMS) uncertainty, while the dotted line represents the mean performance.
In addition, in order to quantify the variability in the pre-training, we pre-train the backbone model three times with different random initializations, resulting in compatible loss values.

\begin{figure}[htbp]
    \centering
    \includegraphics[width=0.49\textwidth]{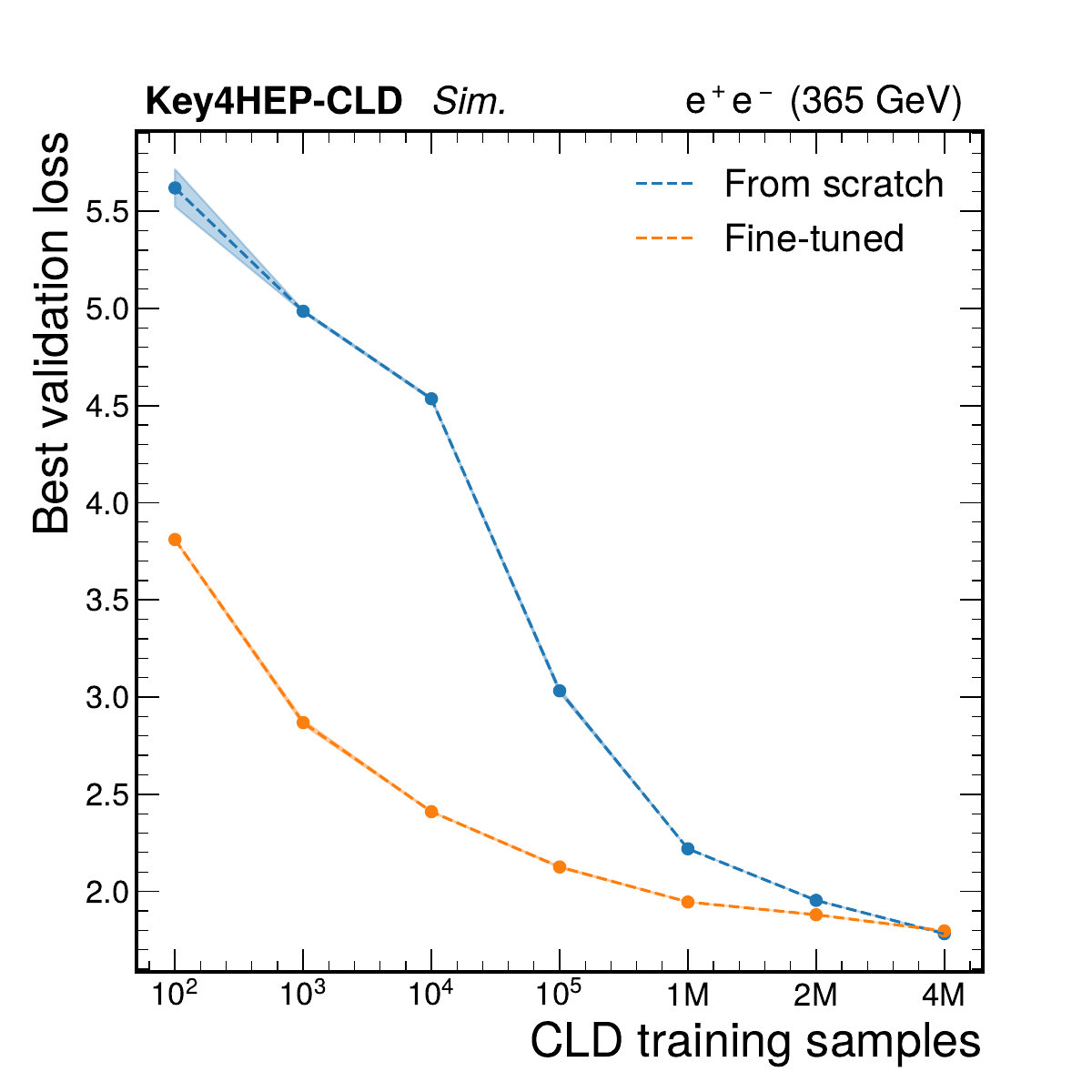}
    \caption{
    The best validation loss achieved on 5,000 validation samples from the CLD dataset as a function of the training dataset size, for the \finetuned model and the \fromscratch model.
    The \finetuned model is able to outperform the \fromscratch model with a factor of 100 less data (1,000 versus 100,000 events).
    We also observe that the best validation loss achieved by the backbone model is only 13.84 indicating significant improvement in the learning after being exposed to as few as 100 CLD events.
    Experiments on $<1 \mathrm{M}$ CLD training samples, are repeated three times with different random seeds, and the shaded uncertainty band covers the RMS uncertainty of the three runs, while the dotted line represents the mean performance.
    }
    \label{fig:fccee_scaling_perf_loss}
\end{figure}

Figure~\ref{fig:fccee_scaling_perf_loss} compares the best validation loss achieved on $5,000$ CLD validation events for the \finetuned and the \fromscratch models.
The \finetuned model achieves the same validation loss as the \fromscratch model with a factor of 100 less downstream data (1,000 versus 100,000 events).
We also observe that the best validation loss achieved by the backbone model is only 13.84, as opposed to 3.81 by the \finetuned model and 5.56 by the \fromscratch model (reported in Fig.~\ref{fig:fccee_scaling_perf_loss}), indicating significant improvement in the learning after being exposed to as few as 100 CLD events.

\begin{figure*}[htbp]
    \centering
    \includegraphics[width=0.3\textwidth]{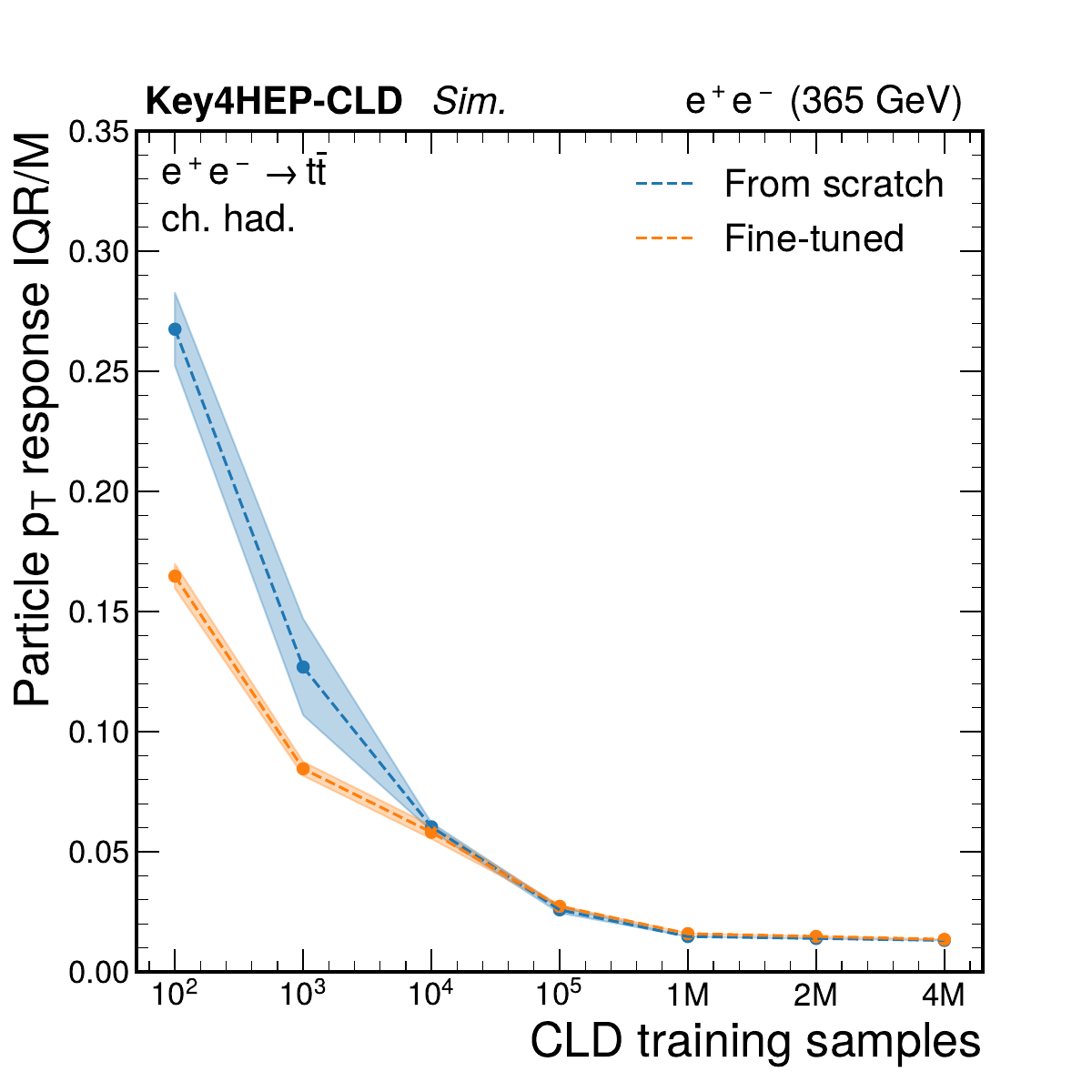}
    \includegraphics[width=0.3\textwidth]{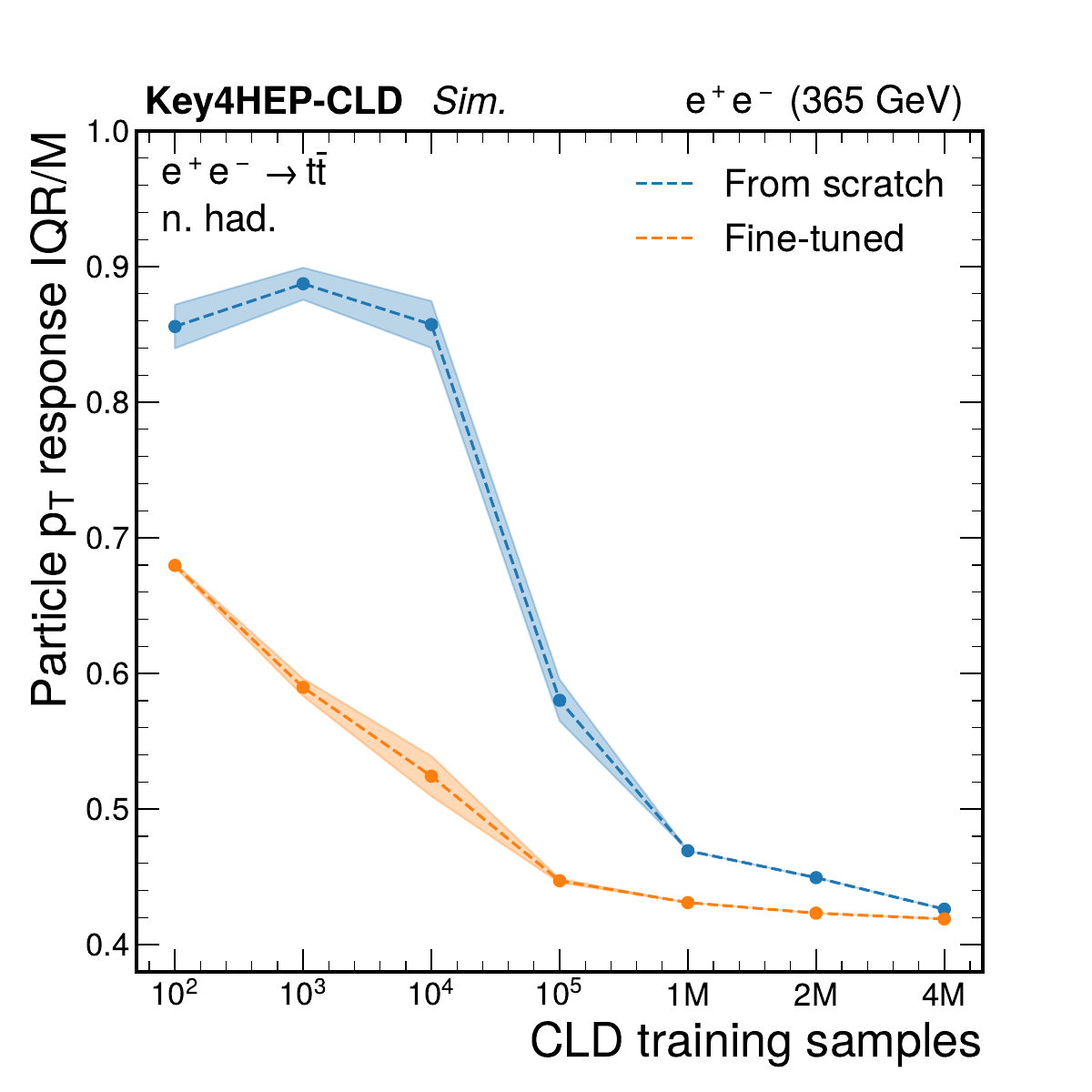}
    \includegraphics[width=0.3\textwidth]{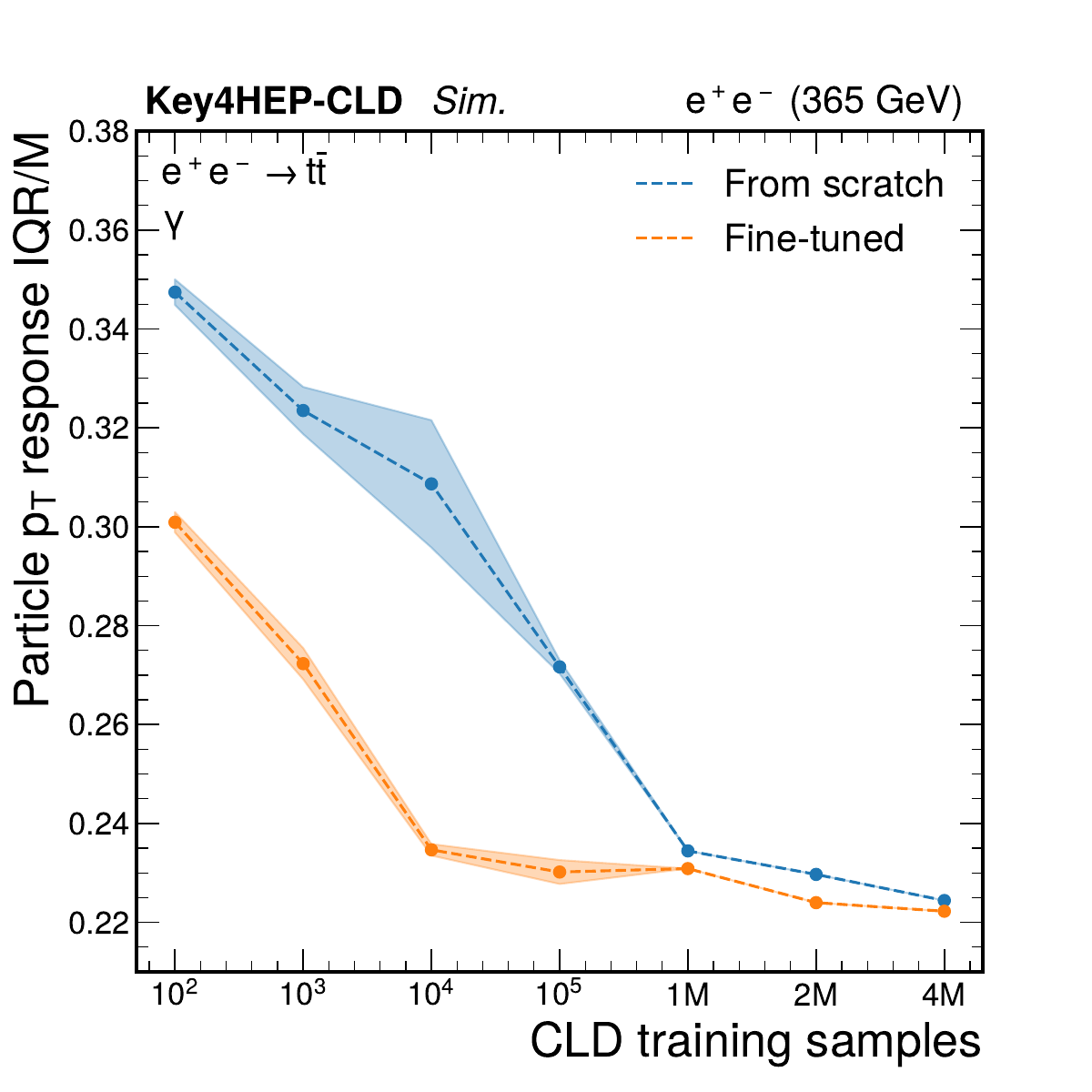} \\
    \includegraphics[width=0.3\textwidth]{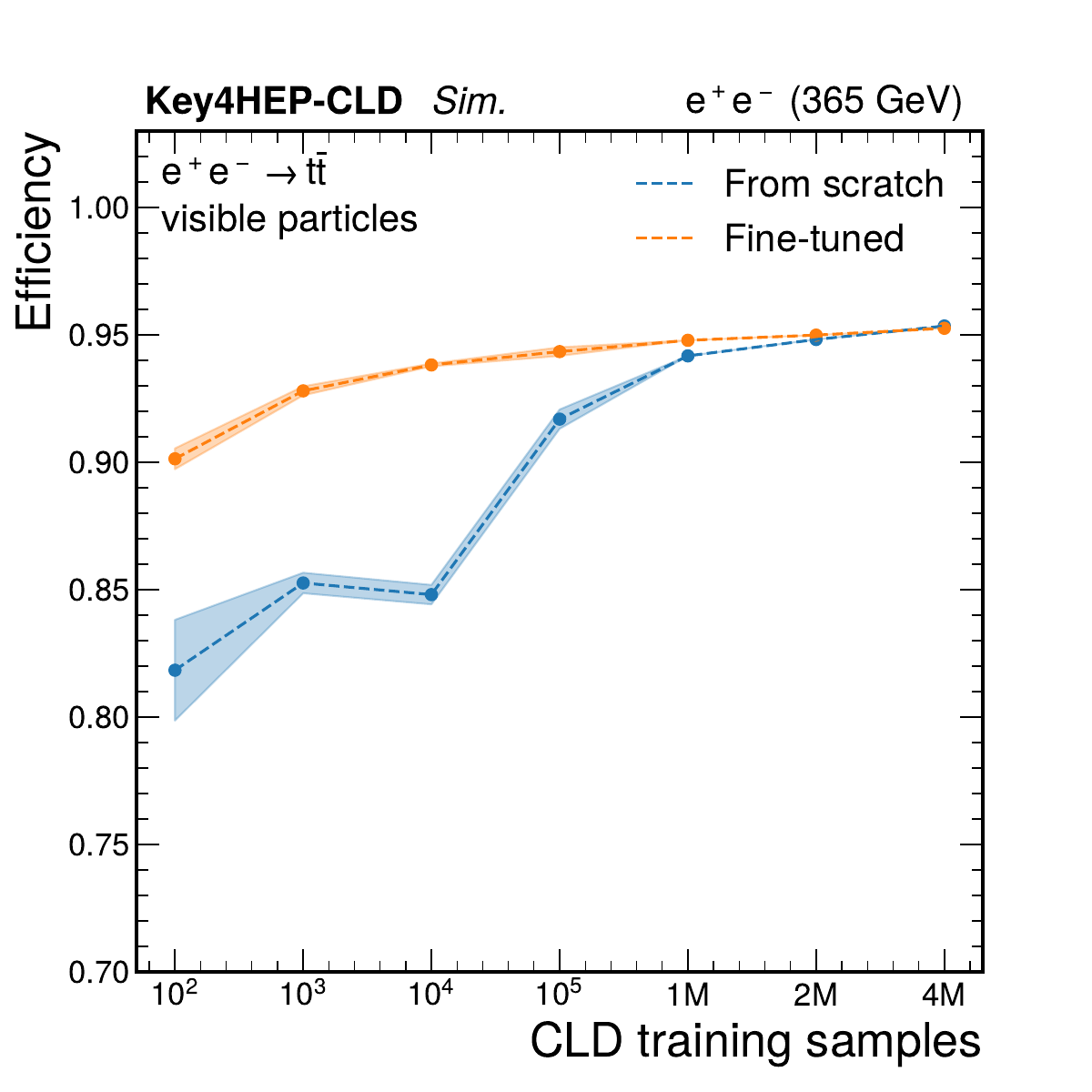}
    \includegraphics[width=0.3\textwidth]{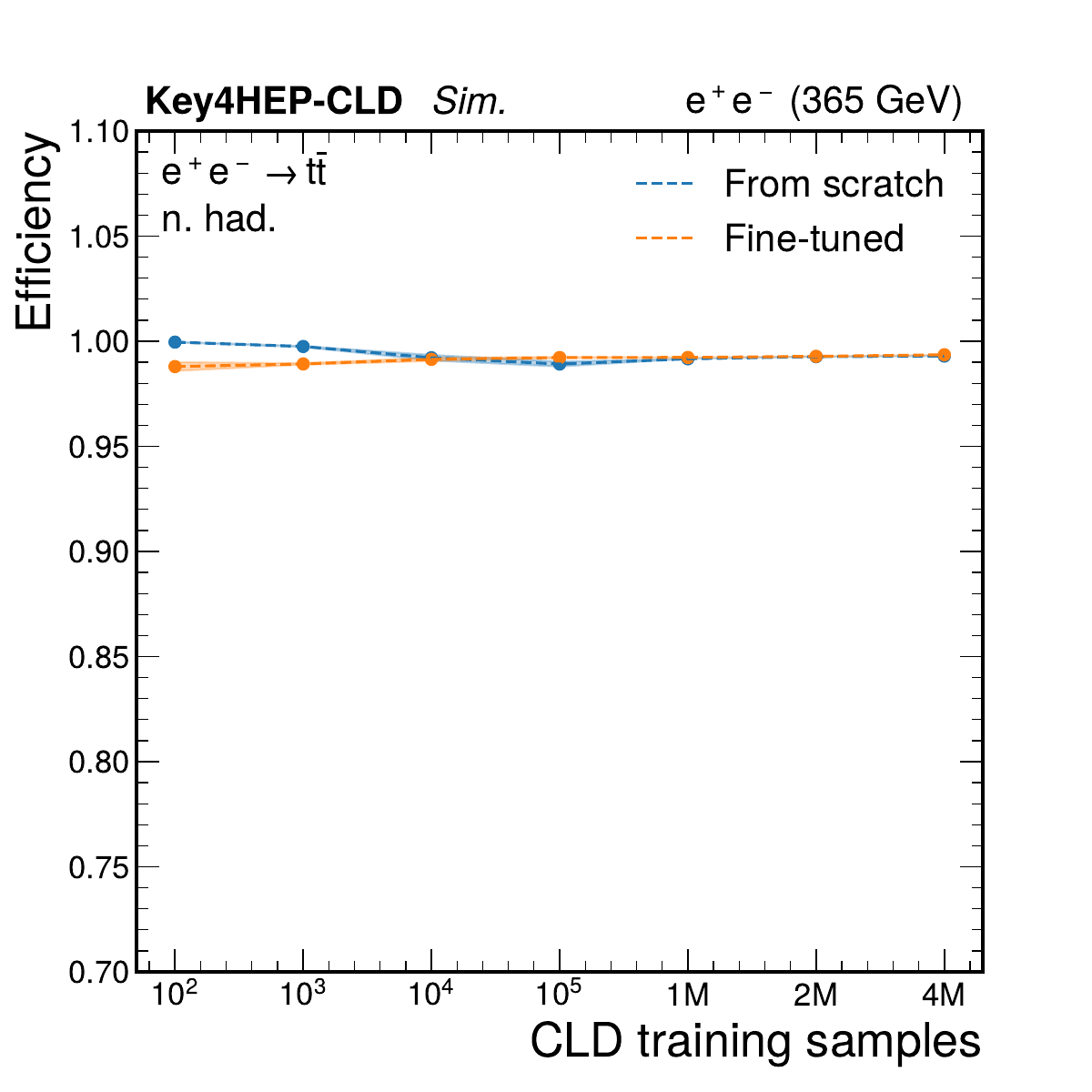}
    \includegraphics[width=0.3\textwidth]{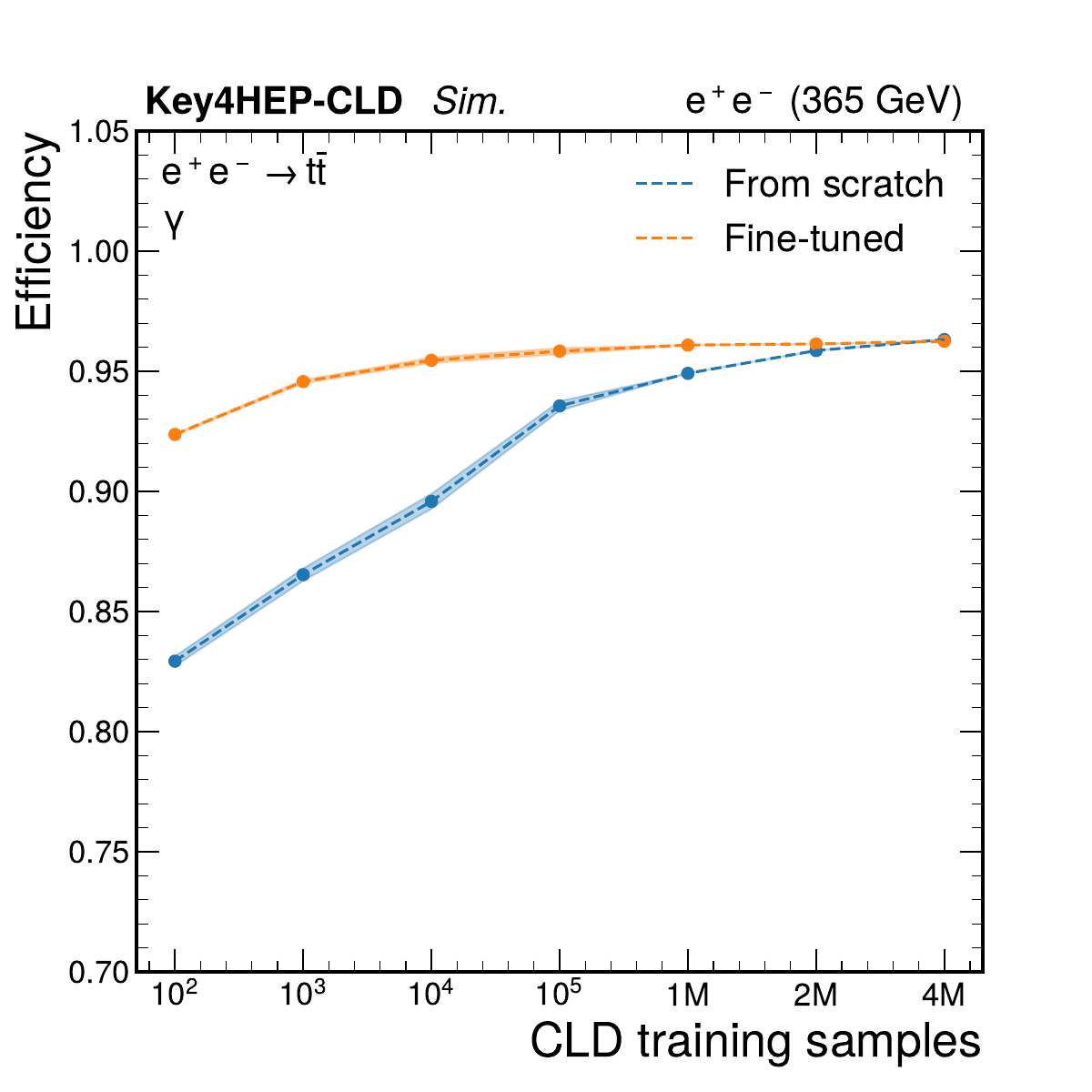} \\
    \includegraphics[width=0.3\textwidth]{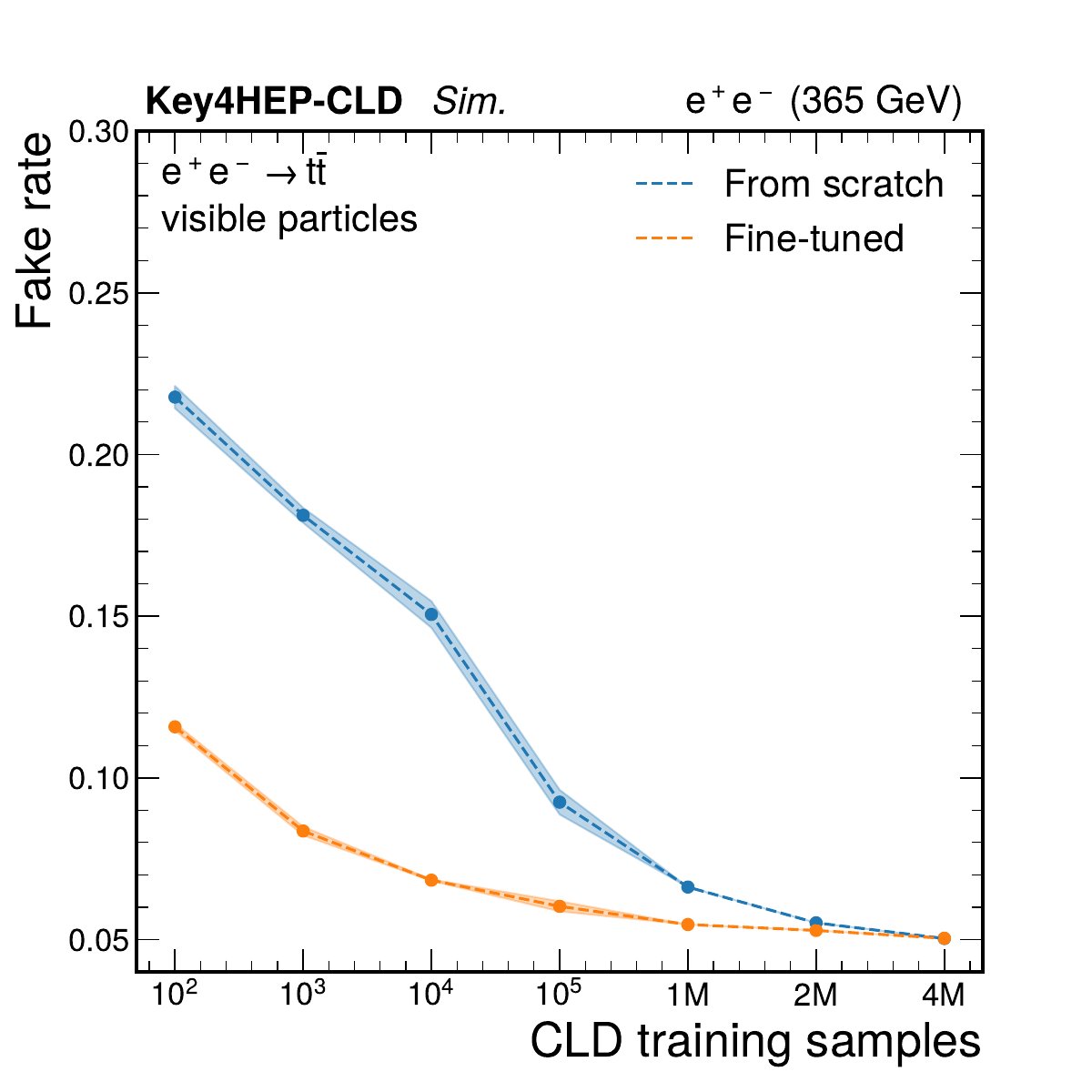}
    \includegraphics[width=0.3\textwidth]{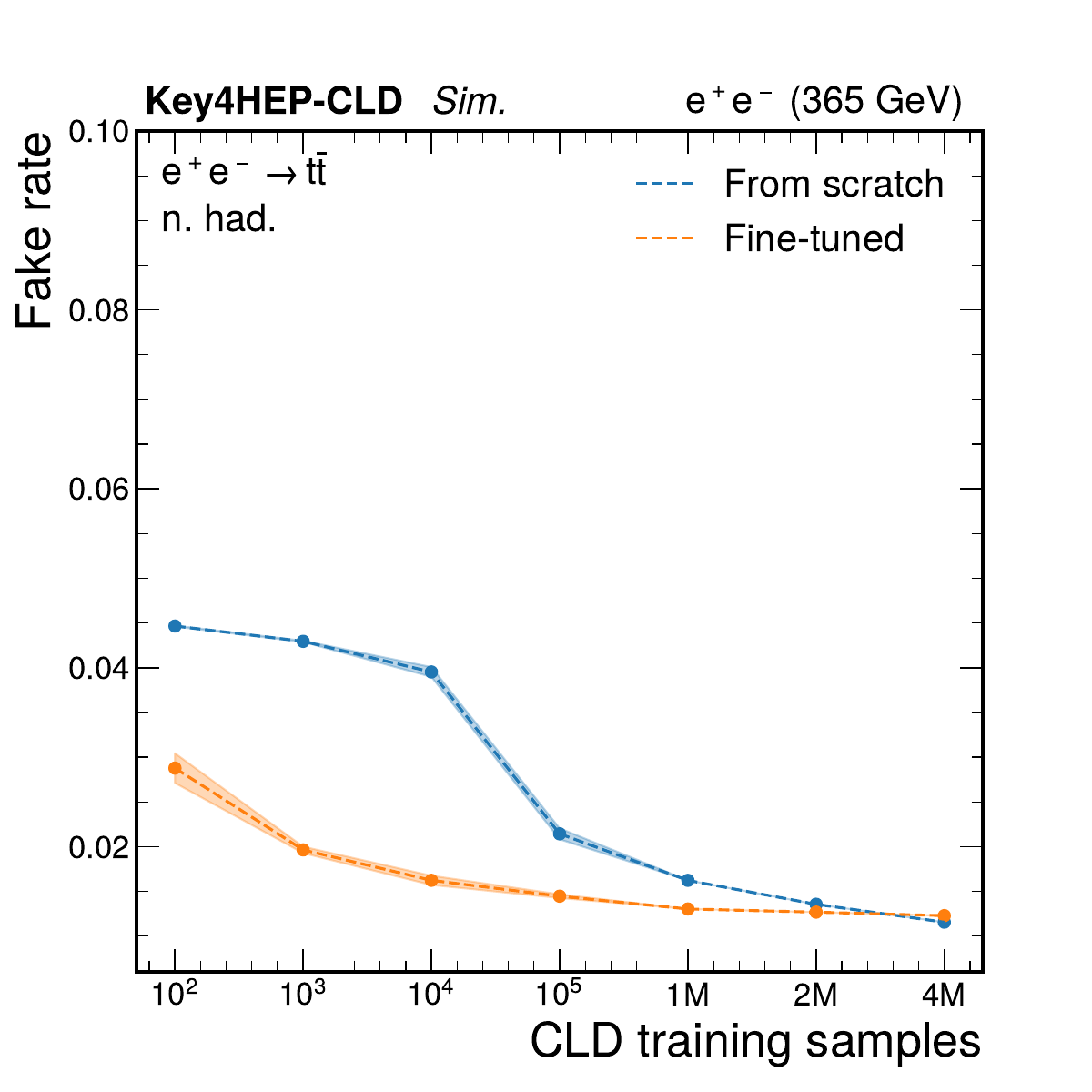}
    \includegraphics[width=0.3\textwidth]{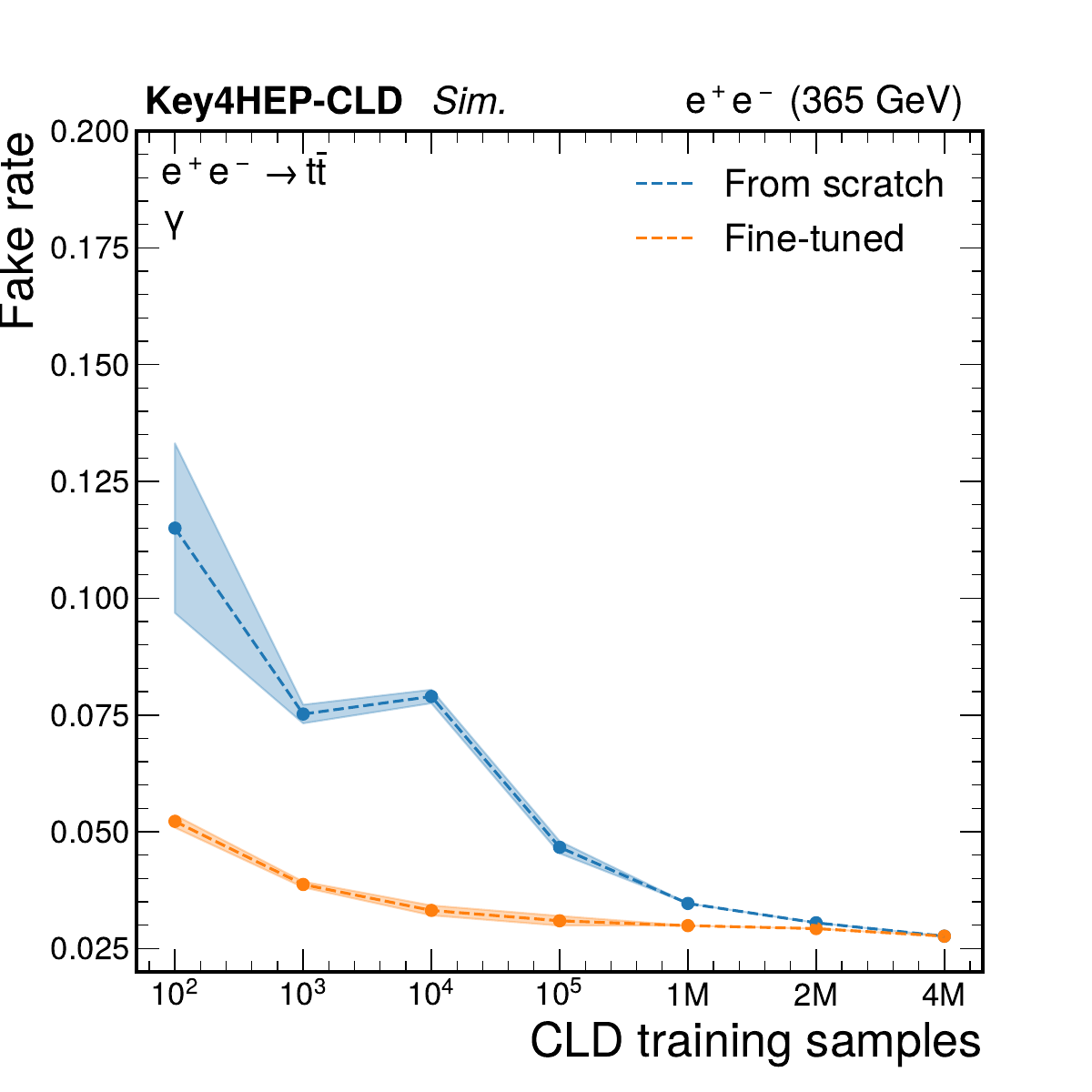}
    \caption{Particle-level performance as a function of the CLD downstream dataset size.
    Particle \ptmomentum resolution presented as response interquartile range over the median (top row) for charged hadrons (left), neutral hadrons (middle), and photons (right).
    Efficiency (middle row) for all visible particles (left), neutral hadrons (middle), and photons (right). 
    Fake rates (bottom row) for all visible particles (left), neutral hadrons (middle), and photons (right).
    The \finetuned model is shown in orange, and the \fromscratch model in blue.
    For nearly all metrics, the \finetuned model trained on 100,000 events matches the performance of the \fromscratch model trained on 4 million events ($40\times$ more data).
    Experiments on ${<}1 \mathrm{M}$ CLD training samples, are repeated three times with different random seeds, and the shaded uncertainty band covers the RMS uncertainty of the three runs, while the dotted line represents the mean performance.    
    }
    \label{fig:fccee_scaling_particleperf}
\end{figure*}

We study particle-level physics performance in terms of several key metrics, including: particle \ptmomentum resolution for charged hadrons, neutral hadrons, and photons; efficiency and fake rate for all visible particles, neutral hadrons, and photons.
The metrics are evaluated against the set of target particles that were introduced in Sec.~\ref{sec:targetdef}.
The \ptmomentum resolution is quantified by the interquartile range over the median of the \ptmomentum response distribution.
In Fig.~\ref{fig:fccee_scaling_particleperf}, we present the particle-level performance metrics of both models as a function of the CLD downstream dataset size.
We observe an improved performance from the \finetuned model over the \fromscratch model across all particle-level metrics.
For nearly all metrics, the \finetuned model trained on 100,000 events matches the performance of the \fromscratch model trained on 4 million events ($40\times$ less data).
All evaluation performance plots are produced with 20,000 events from the CLD downstream dataset that is reserved for testing.

\begin{figure*}[htbp]
    \centering
    \includegraphics[width=0.49\textwidth]{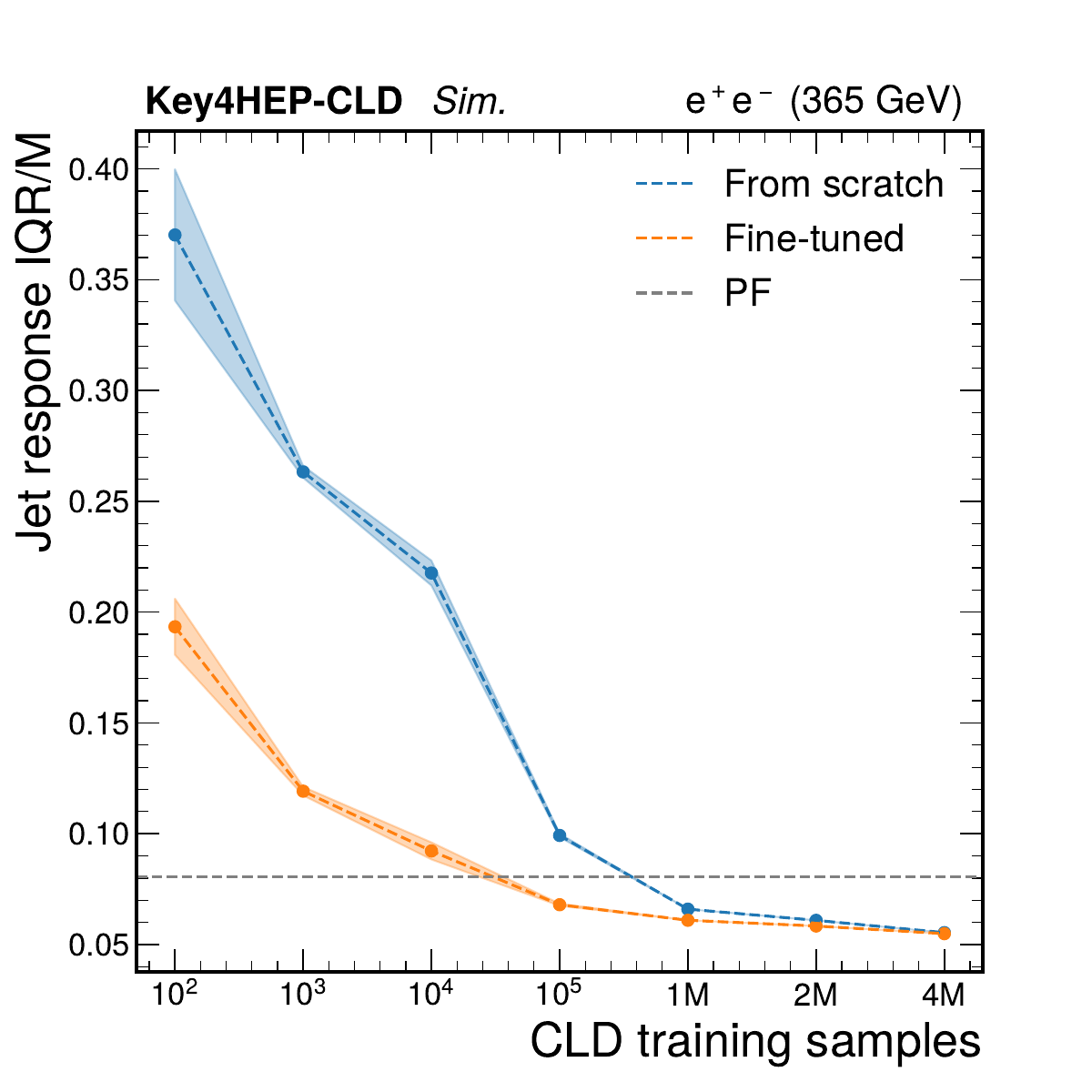}
    \includegraphics[width=0.49\textwidth]{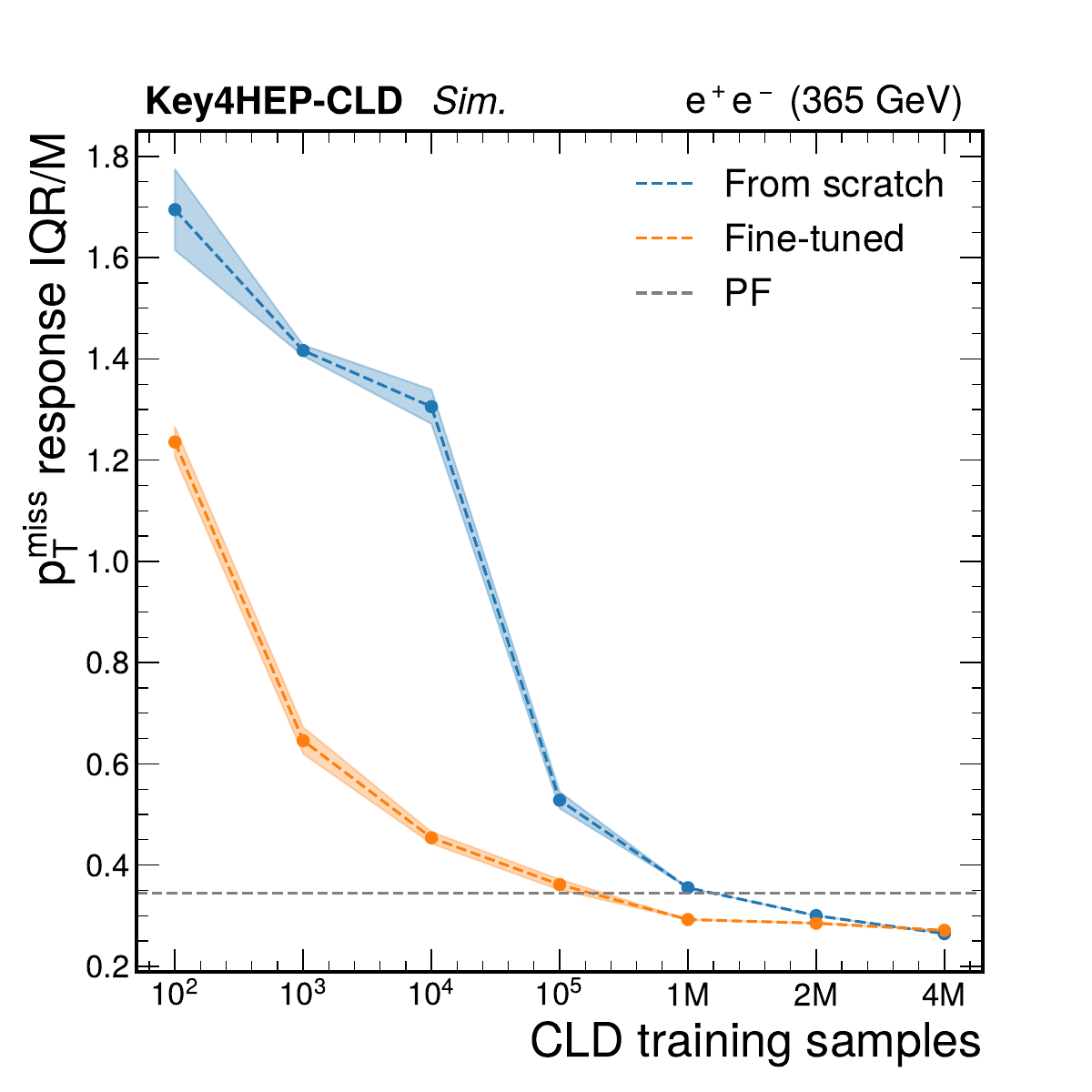}    
    \caption{The jet and \ptmiss performance as a function of the CLD dataset size, for the \finetuned model (orange), \fromscratch (blue), and the traditional PF algorithm (grey).
    Experiments on $<1 \mathrm{M}$ CLD training samples, are repeated three times with different random seeds, and the shaded uncertainty band covers the RMS uncertainty of the three runs, while the dotted line represents the mean performance.    
    }
    \label{fig:fccee_scaling_perf_physics}
\end{figure*}

We evaluate both models on event-level metrics, including the reconstruction accuracy of jets and missing transverse momentum, in Fig.~\ref{fig:fccee_scaling_perf_physics}.
The jet (\ptmiss) reconstruction accuracy is measured in terms of the jet (\ptmiss) resolution which is quantified by the inter-quartile range (IQR) and median of the jet (\ptmiss) response.
The reference jets, and \ptmiss, used to calculate the response are built using truth particles.
Since the event-level metrics are computed directly with respect to the truth (i.e. without any reference to the particle-level target definition that is arbitrary), we include the traditional rule-based PF algorithm in the comparison; referred to as ``PF'' in the figures, representing the current available implementation of the Pandora PF algorithm~\cite{Marshall:2012hh,Marshall:2012ry,Marshall:2015rfa}.
We observe that both ML models outperform PF as the training dataset increases.
Additionally, we observe the the \finetuned model is able to outperform PF with a much smaller dataset size compared to the \fromscratch model.
Both the \finetuned model and the \fromscratch model converge to the same performance at 4 million events.

\subsection{Ultimate MLPF performance on CLD dataset}
\label{sec:results_CLD_big}

\begin{figure*}[htbp]
    \centering
    \includegraphics[width=0.3\textwidth]{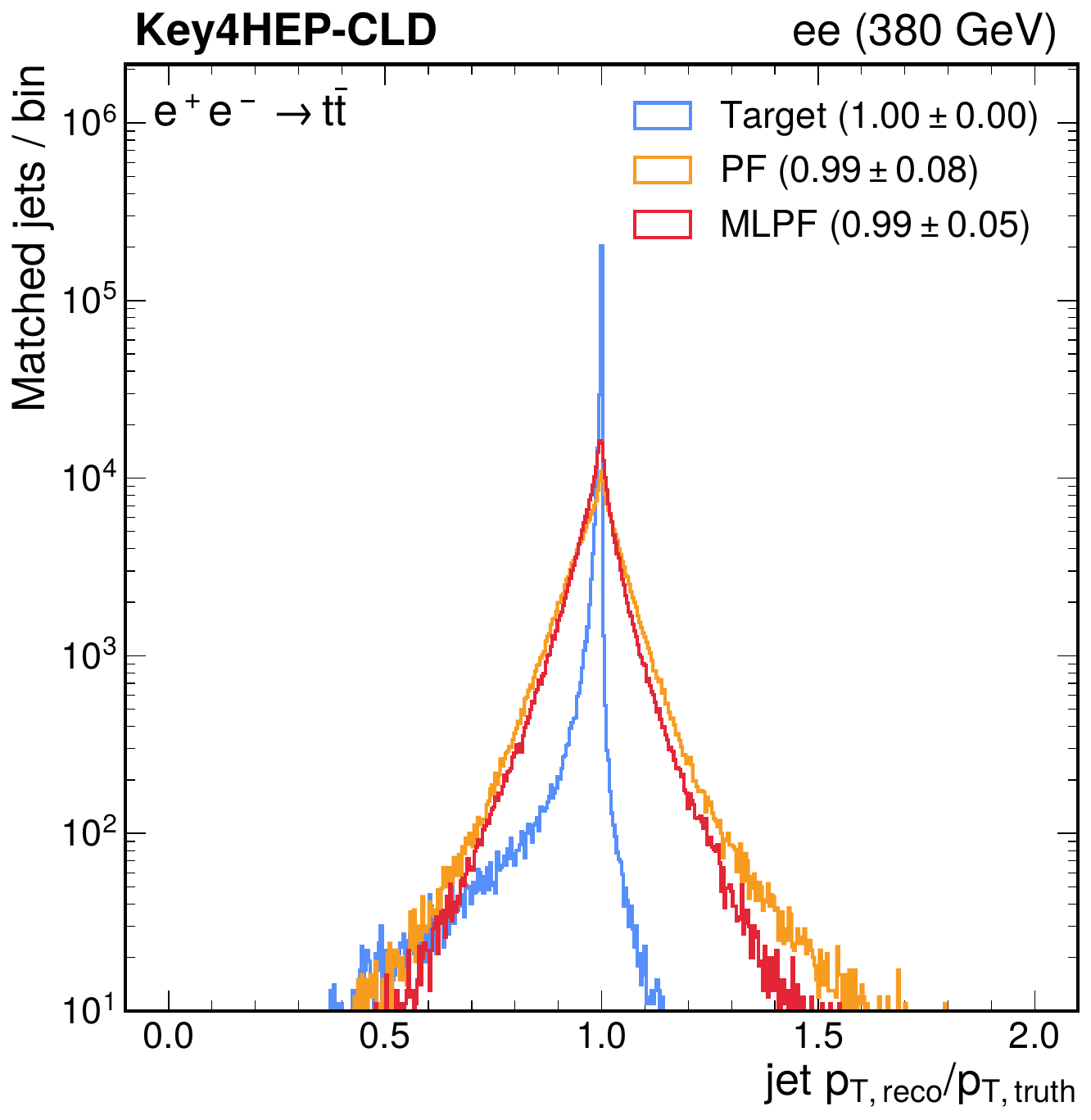}
    \includegraphics[width=0.3\textwidth]{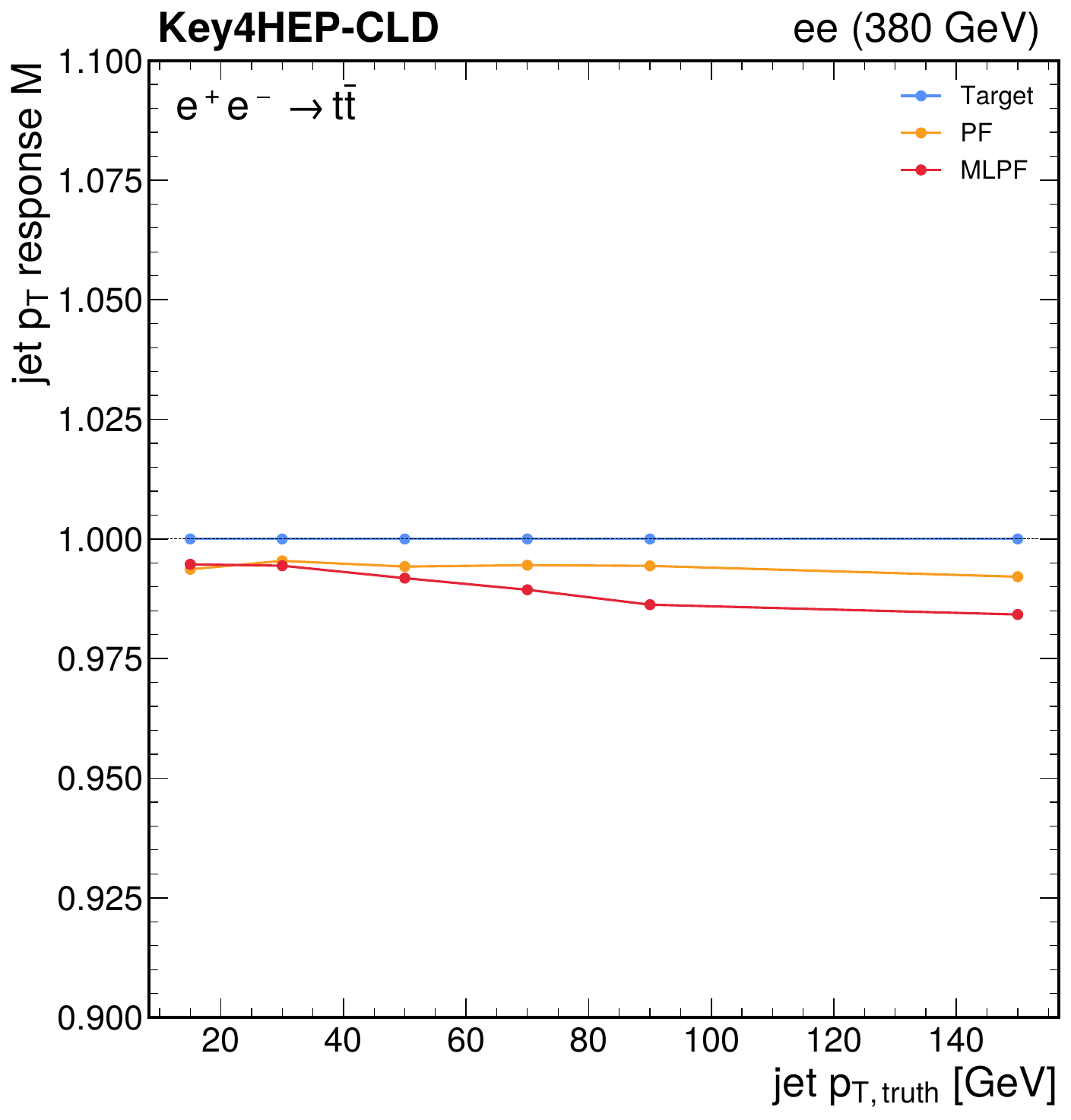}
    \includegraphics[width=0.3\textwidth]{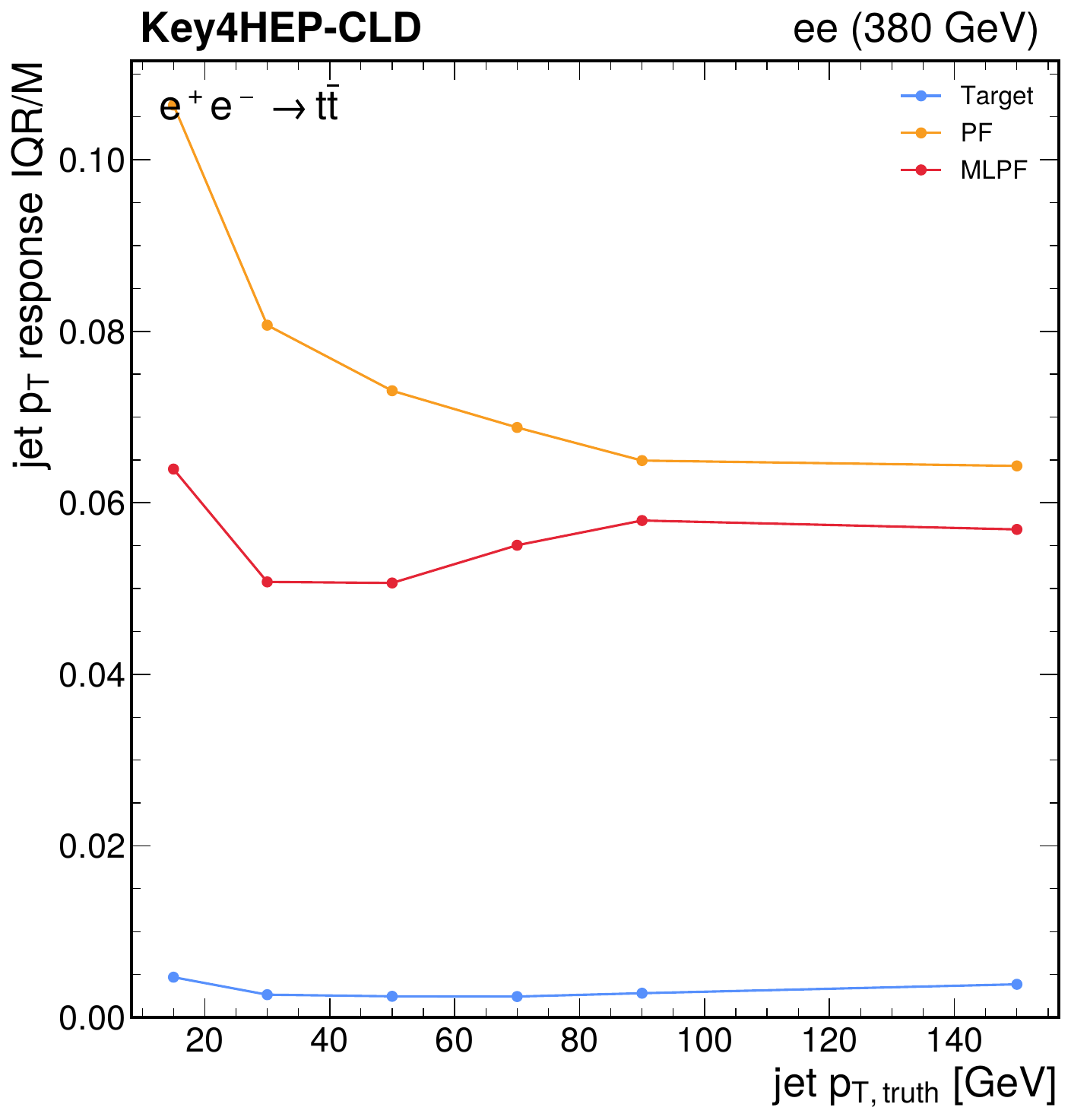}
    \caption{Jet response for target jets (blue), PF (orange), and MLPF (red); measured with respect to stable \pythia truth particles. The resolution of the target particle distribution, quantified by the interquartile range over the median, is significantly better than the resolution of jets from PF particles, as would be expected. The MLPF model outperforms PF when measured against the truth particles, both in the bulk and in the tails of the jet \ptmomentum spectrum.}
    \label{fig:cld_perf}
\end{figure*}

We present the jet performance of the MLPF model trained on 4 million CLD events in Fig.~\ref{fig:cld_perf}.
We compare MLPF, PF jets, and target jets.
Similar to Fig.~\ref{fig:target_matching}, we present the the inclusive jet response, the median of the jet response distribution as a function of \ptmomentum, and the jet resolution as a function of \ptmomentum quantified by the interquartile range over the median.
We observe that the resolution of the target jets is significantly better than the resolution of jets from PF particles and the resolution of the MLPF reconstructed jets is better than the baseline PF algorithm.

\subsection{The effect of the initial training dataset size}
\label{sec:results_backbone}

\begin{figure}[htbp]
    \centering
    \includegraphics[width=0.23\textwidth]{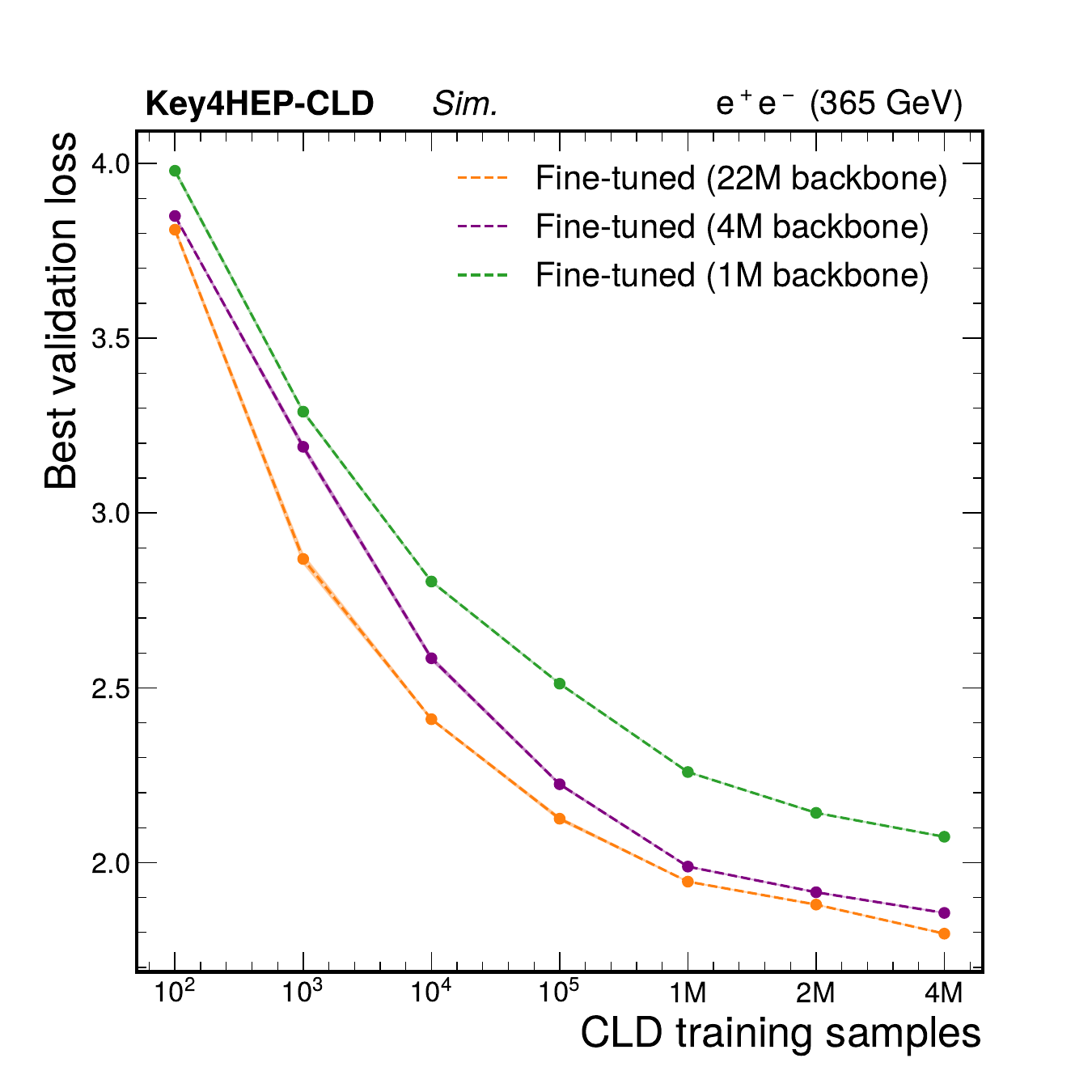} \\
    \includegraphics[width=0.23\textwidth]{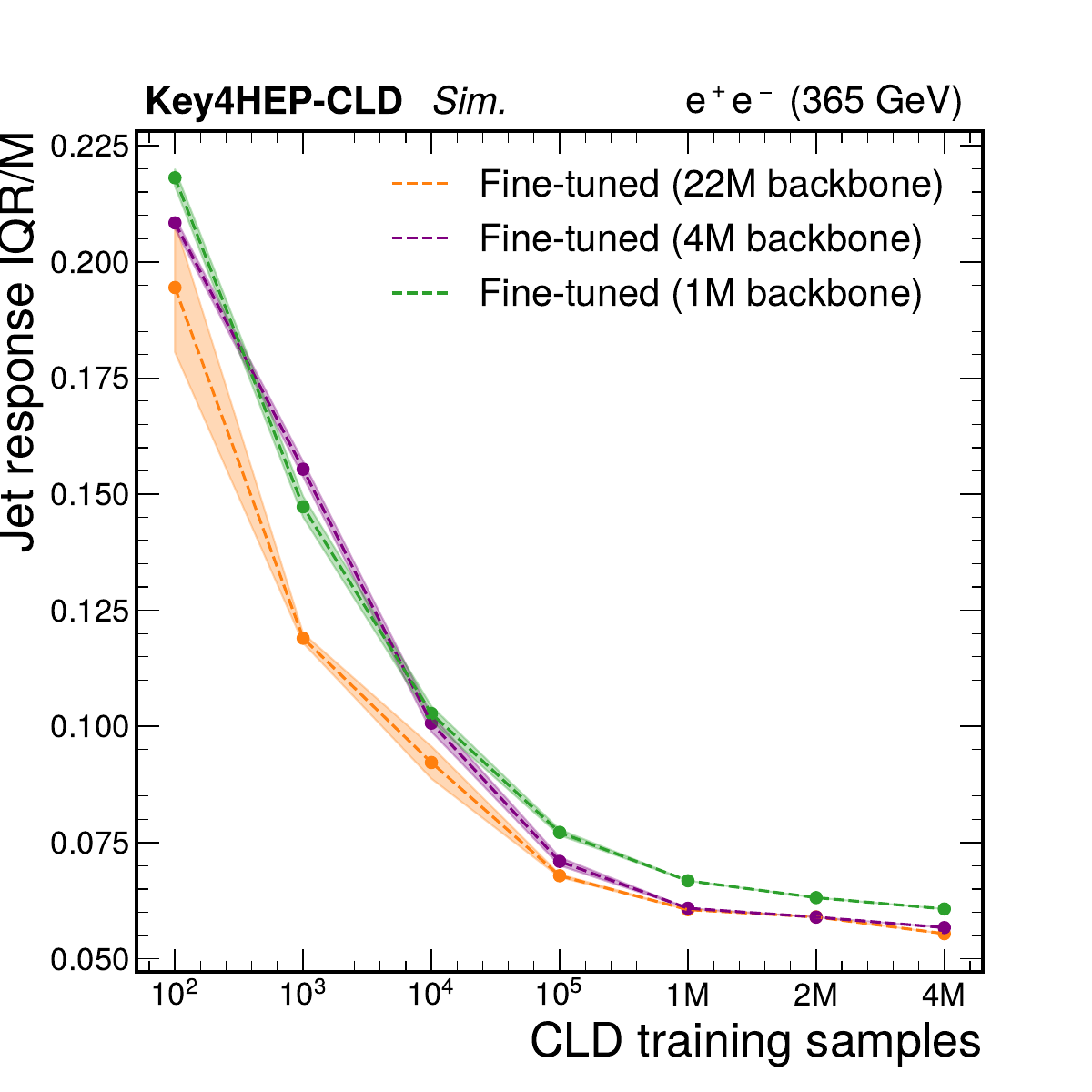}
    \includegraphics[width=0.23\textwidth]{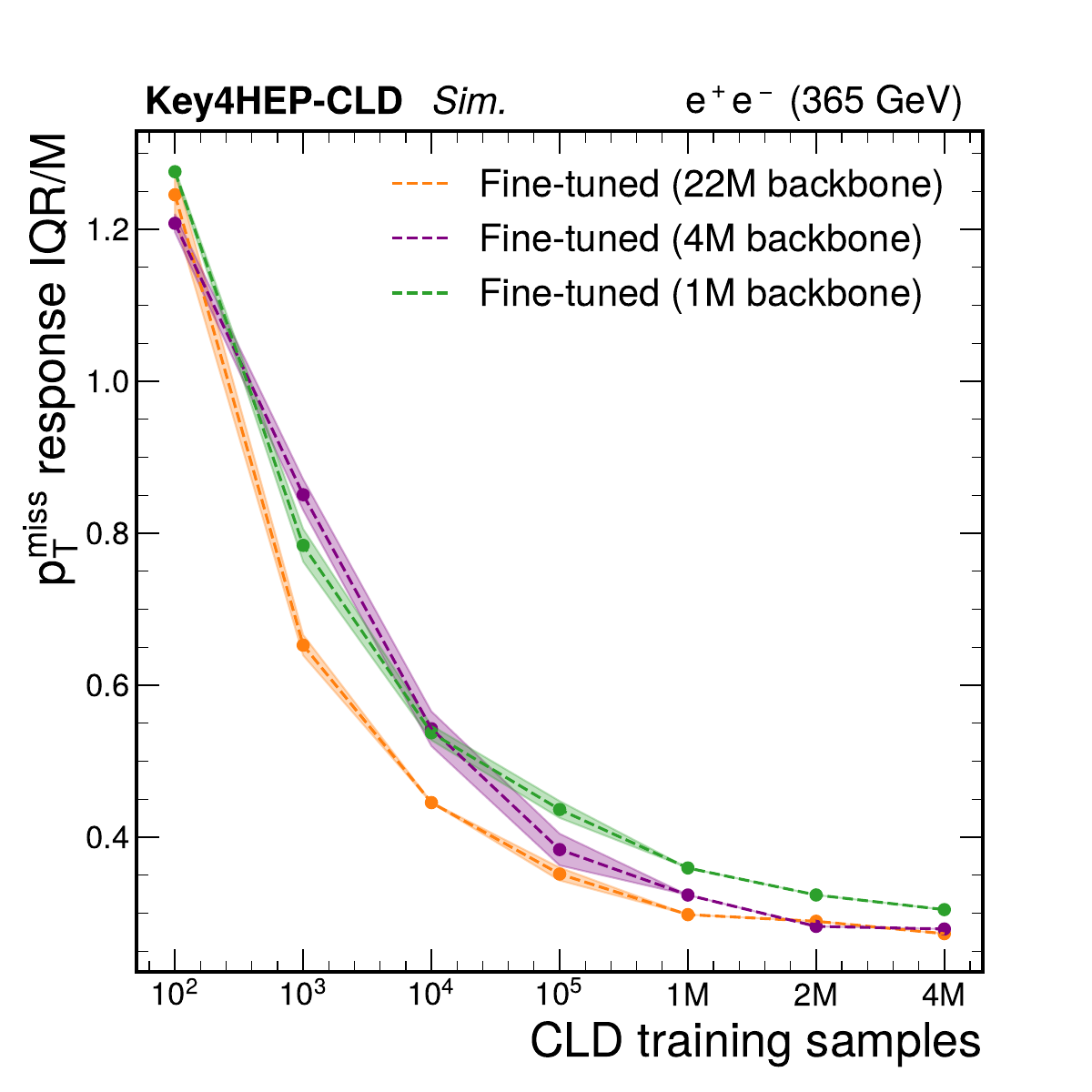}
    \caption{The best validation loss on 5,000 \ttbar validation samples (top), jet performance (bottom left) and \ptmiss performance (bottom right), as a function of the CLD dataset size; for the \finetuned model using the 22M backbone model (orange), and the \finetuned model using the 1M backbone model (green).
    Experiments on ${<}1$M CLD training samples, are repeated three times with different random seeds, and the shaded uncertainty band covers the RMS uncertainty of the three runs, while the dotted line represents the mean performance.      
    }
    \label{fig:fccee_scaling_backbone_study}
\end{figure}

Finally, we study the effect of the pre-training dataset size on the downstream performance by pre-training two additional transformer models on a smaller CLICdet dataset.
The first model is pre-trained on 1 million \ttbar CLIC events, and we refer to it as ``1M backbone'', and the second model is pre-trained on 4 million \ttbar CLIC events, and we refer to it as ``4M backbone''.
The pre-training follows the same strategy previously discussed.
In Fig.~\ref{fig:fccee_scaling_backbone_study}, we present a comparison between the original backbone model (labeled ``22M backbone'') and the two additional candidate backbone models trained on fewer events (``1M backbone'' and ``4M backbone''). 
We observe that at 100 CLD events, the three models have similar performance, but starting from 1,000 events, we observe a significant improvement in the performance of the 22M backbone model over the other two backbone models in terms of validation loss, jet, and \ptmiss performance.
This illustrates the possible improvements in ML model performance from larger datasets.

\section{Conclusion}

We have evaluated the transfer learning capabilities of the machine-learned particle-flow (MLPF) algorithm by pre-training the model on the dataset of electron-positron ($\Pe^+\Pe^-$) collisions at $\sqrt{s}=380\GeV$ in the Compact Linear Collider detector (CLICdet), and subsequently fine-tuning the model on a different dataset of $\Pe^+\Pe^-$ collisions at $\sqrt{s}=365\GeV$ in the CLIC-like detector (CLD) proposed for the Future Circular Collider (FCC-ee).
We observe that the fine-tuned model significantly outperforms a model trained from scratch on small subsets of the CLD dataset (${<}2$M events), with a $30\%$ improvement in jet resolution at $100,000$ events, as well as outperforming the current baseline implementation of the particle-flow algorithm.
The results indicate that pre-training large machine learning (ML) models on massive datasets can learn the underlying physics of particle reconstruction, and may ultimately improve the sensitivity of high energy physics analyses.
This work opens the door to rapid detector design and optimization using ML, where models can be designed and adapted to learn to reconstruct events at detectors with different geometries efficiently, enabling rapid comparisons between different detector designs on the basis of physics performance.

\section*{Acknowledgments}

F.M. is supported by an Hal{\i}c{\i}o\u{g}lu Data Science Institute (HDSI) fellowship. 
F.M. and J.D. are supported by the Research Corporation for Science Advancement (RCSA) under grant No. CS-CSA-2023-109, Alfred P. Sloan Foundation under grant No. FG-2023-20452, U.S. Department of Energy (DOE), Office of Science, Office of High Energy Physics Early Career Research program under Grant No. DE-SC0021187, and the U.S. National Science Foundation (NSF) Harnessing the Data Revolution (HDR) Institute for Accelerating AI Algorithms for Data Driven Discovery (A3D3) under Cooperative Agreement No. PHY-2117997. 
M.K. is supported by the US Department of Energy (DOE) under Grant No. DE-AC02-76SF00515.
J.P. is supported by the Estonian Research Council Grants No. PSG864, No. RVTT3-KBFI and by the European Regional Development Fund through the CoE Program Grant No. TK202.
D.G. is supported by the Future Circular Collider Innovation Study (FCCIS) project, which has received funding from the European Union's Horizon 2020 research and innovation programme under Grant No. 951754.
This work was also supported in part by NSF Awards No. CNS-1730158, No. ACI-1540112, No. ACI-1541349, No. OAC-1826967, No. OAC-2112167, No. CNS-2100237, No. CNS-2120019.
E.W. is supported by the SPECTRUM project which has received funding from the European Union's Horizon Europe programme (European Union Grant Agreement No. 101131550).

The authors contributed according to the contributor roles taxonomy (CRediT) categories as follows. 
F.M.: conceptualization, investigation, methodology, software, validation, data curation, writing (original draft).
J.P.: investigation, software, data curation, writing (original draft), resources, funding acquisition, supervision, writing (review and editing).
D.G.: conceptualization, writing (review and editing).
E.W.: investigation, software, writing (review and editing), resources.
M.Z.: software, validation, investigation.
M.K.: conceptualization, methodology, supervision, writing (review and editing).
J.D.: conceptualization, software, resources, funding acquisition, supervision, writing (review and editing).

\section*{Data availability}
The data that support the findings of this article are openly available~\cite{Pata2025, ZenodoCLD2025raw,ZenodoCLD2025TFDS}.

\appendix
\section{Visualization of the decay and simulation tree}
\begin{figure*}[htbp]
    \centering
    \includegraphics[width=0.85\textwidth]{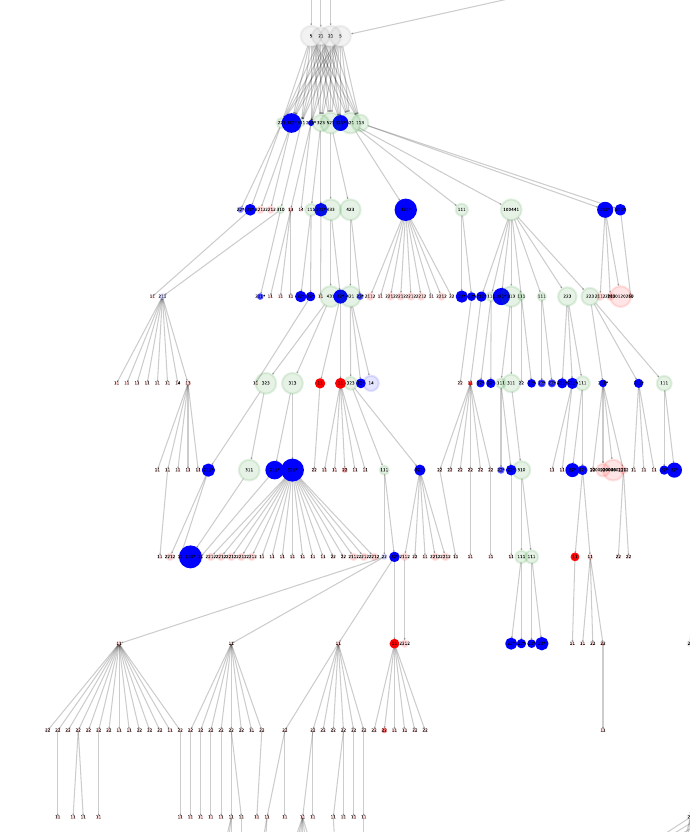}
    \caption{A snapshot of the \pythia decay and \GEANTfour simulation tree from the generator and simulator output for a full-simulation \ttbar event, showing the daughters of the two \PQb quarks. Each node is a particle from simulator, labeled with the PDG particle ID code~\cite{ParticleDataGroup:2024cfk}. The edges are the parent-daughter relationships between the particles. The color of the node signifies the generator status code (red: status 0, blue: status 1, green: status 2, gray: other). Particles that did not leave hits either directly or through their descendants are transparent. The node size is proportional to the energy of the particle. Nodes that have a * after the particle ID are included in the target particle set.}
    \label{fig:decay_tree}
\end{figure*}

Figure~\ref{fig:decay_tree} shows a snapshot of the \pythia decay and \GEANTfour simulation tree for a simulated \ttbar event from the CLICdet dataset.

\section{Input features}
\label{app:input_feat}

\begin{figure*}[htbp]
    \centering
    \includegraphics[width=0.4\textwidth]{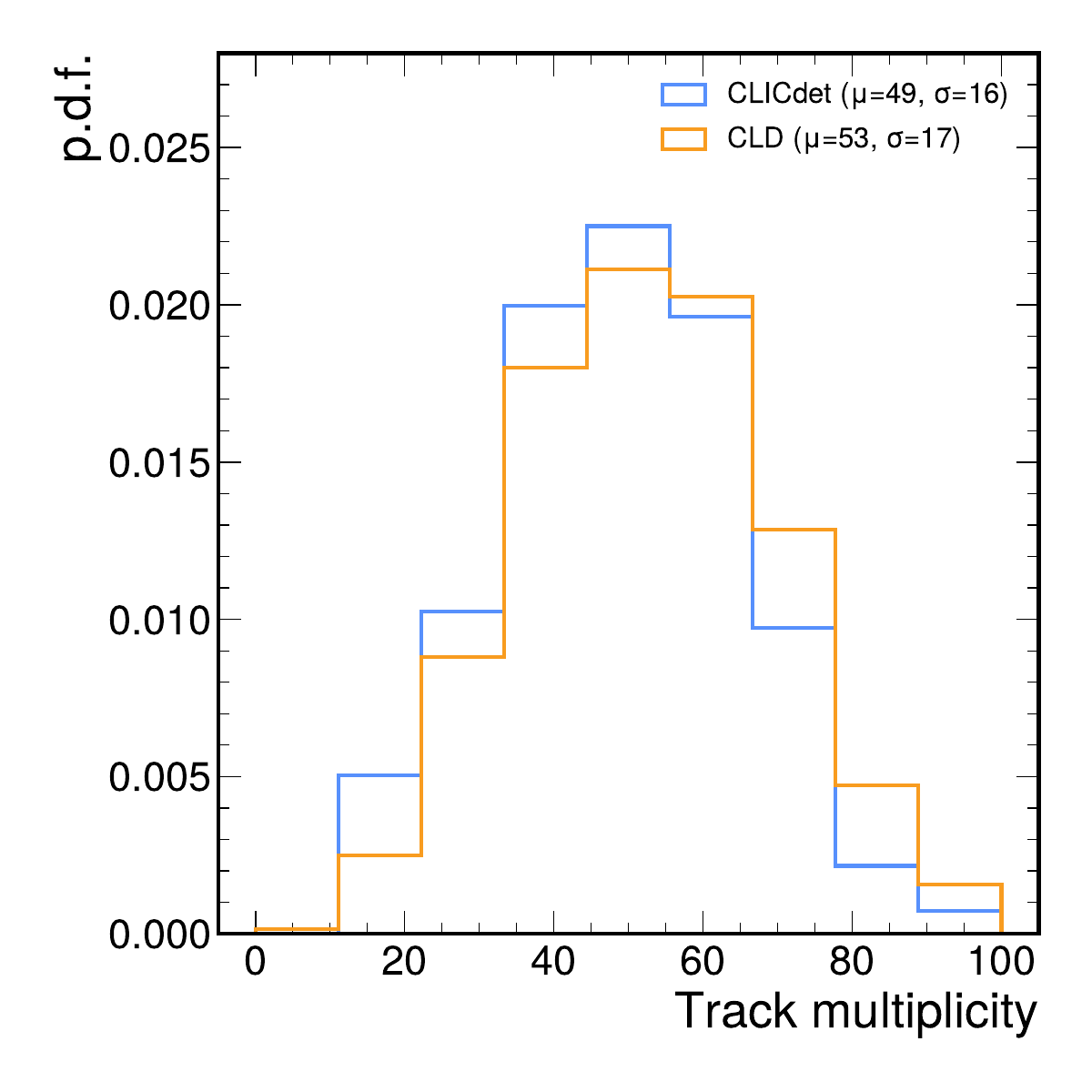}
    \includegraphics[width=0.4\textwidth]{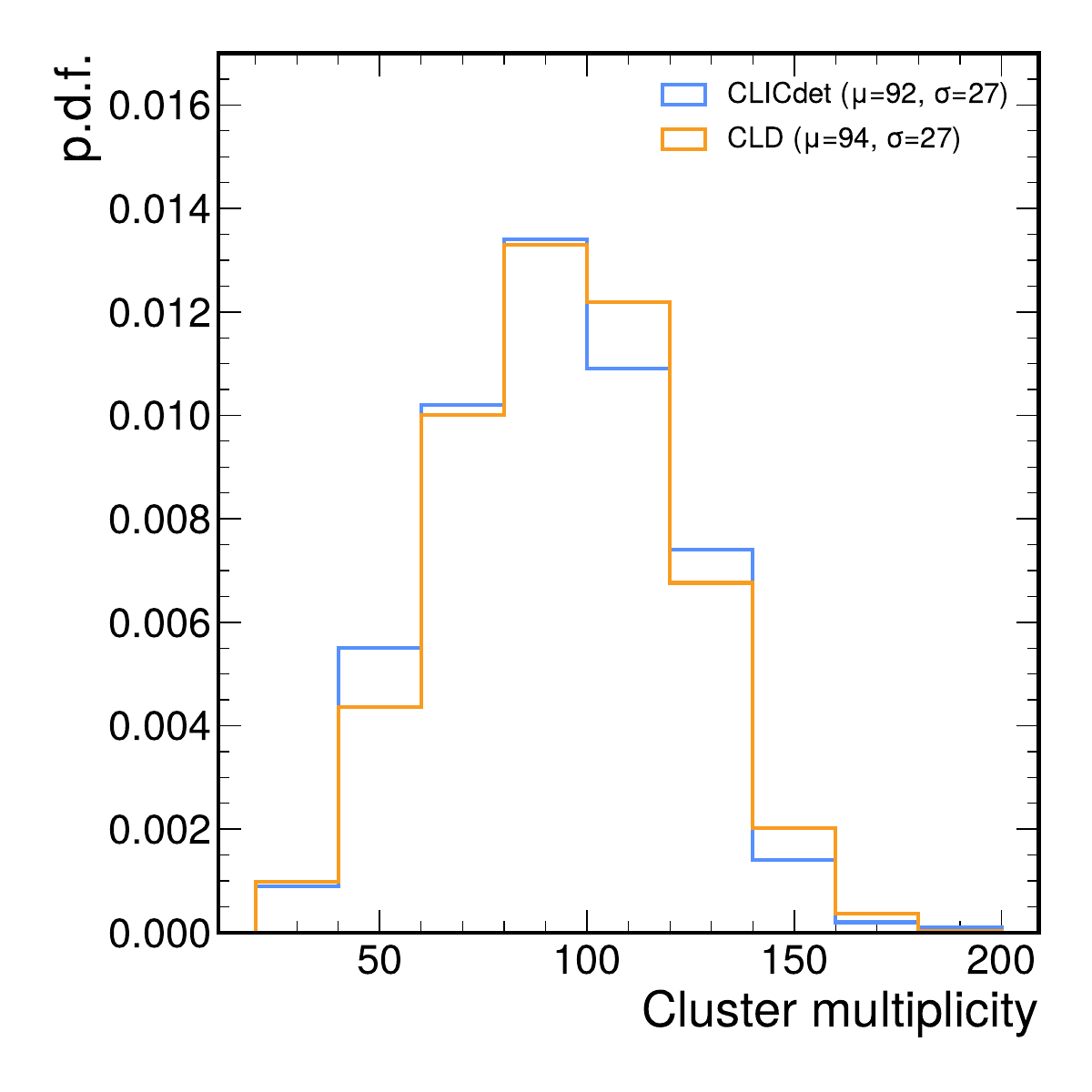} \\    
    \caption{The input multiplicity for reconstructed tracks (left), and clusters (right) per event; for the CLICdet dataset (blue) and the CLD dataset (orange).
    }
    \label{fig:input_mult}
\end{figure*}

On average, each event consists of around 50 tracks and 92 calorimeter clusters, which form our input set to the ML model.
The input track and cluster multiplicity for 5,000 events, comparing the CLICdet and CLD is shown in Fig.~\ref{fig:input_mult}.
The track and cluster input features are
\begin{eqnarray}
x_{\mathrm{track}}&=&[\ptmomentum, \eta, \sin \phi, \cos \phi,
p,\chi^2, n_\mathrm{dof},r_\mathrm{innermost},\nonumber\\
&&\tan \lambda, d_0, \Omega, z_0]\\
x_{\mathrm{cluster}}&=&[E_\mathrm{T}, \eta, \sin \phi, \cos \phi, E, x, y, z,
\theta, E_{\mathrm{ECAL}}, E_{\mathrm{HCAL}},\nonumber\\
&&E_{\mathrm{other}}, N_\mathrm{hits}, \sigma(x_{\mathrm{hits}}), \sigma(y_{\mathrm{hits}}), \sigma(z_{\mathrm{hits}})]
\end{eqnarray}

\begin{figure*}[htbp]
    \centering
    \includegraphics[width=0.24\textwidth]{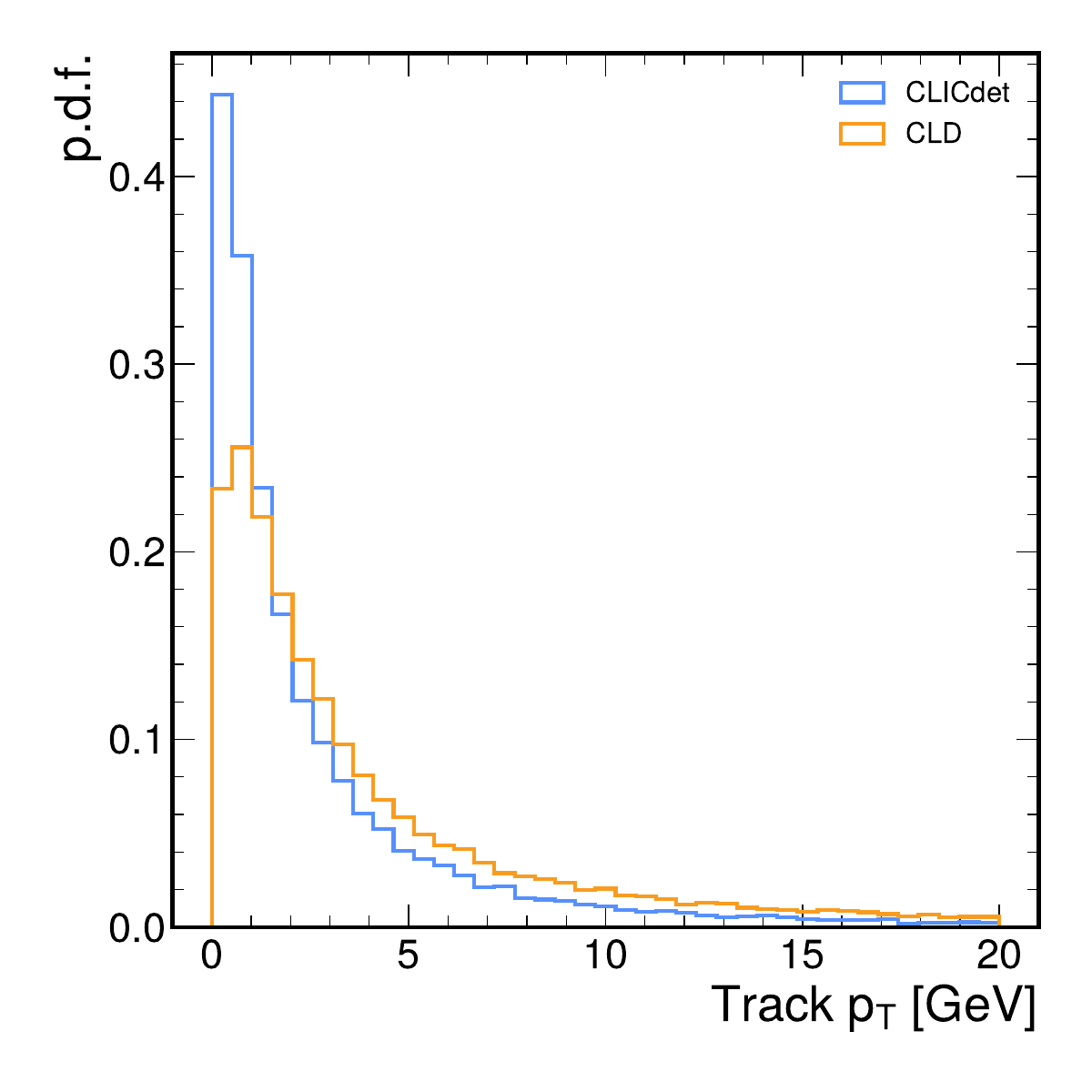}
    \includegraphics[width=0.24\textwidth]{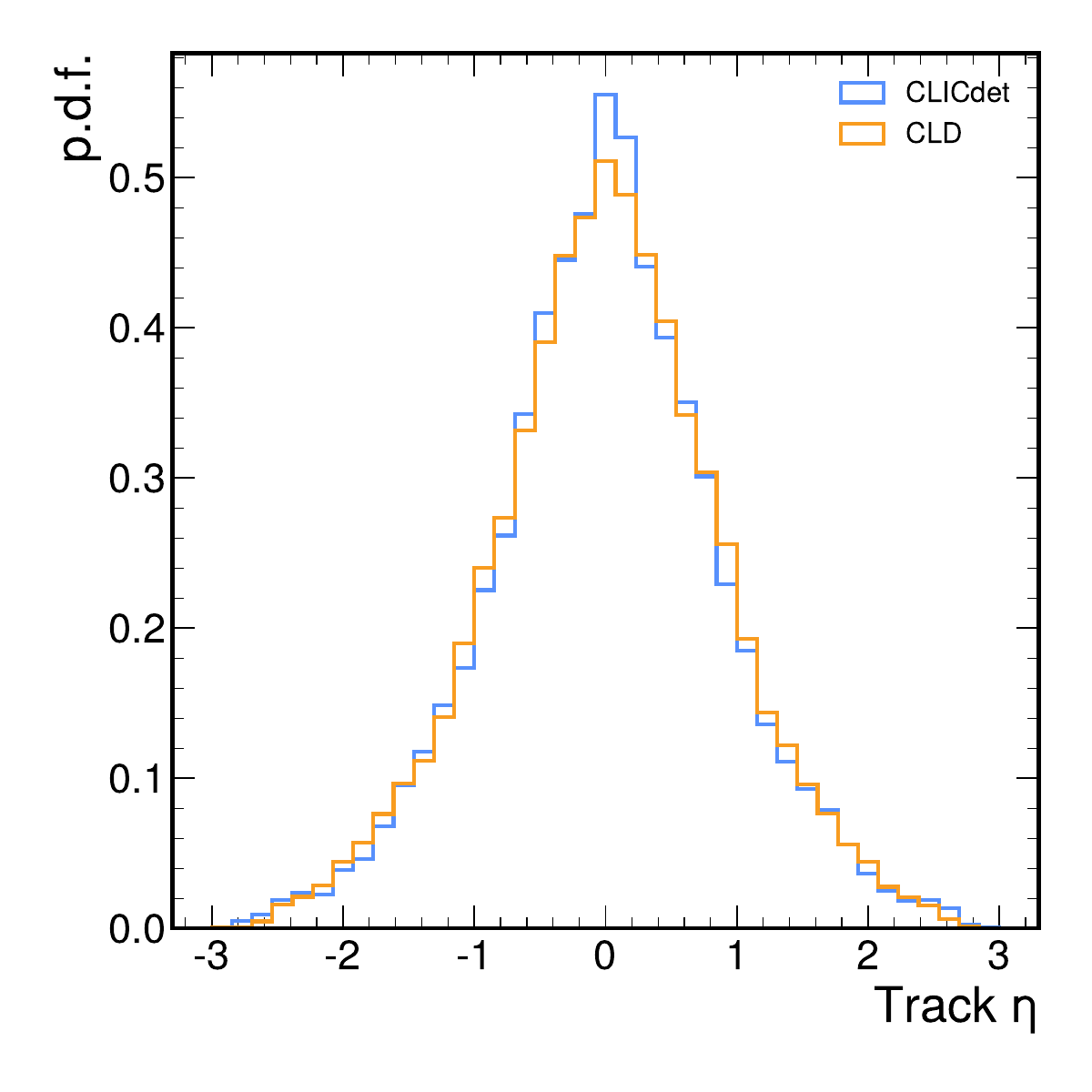}
    \includegraphics[width=0.24\textwidth]{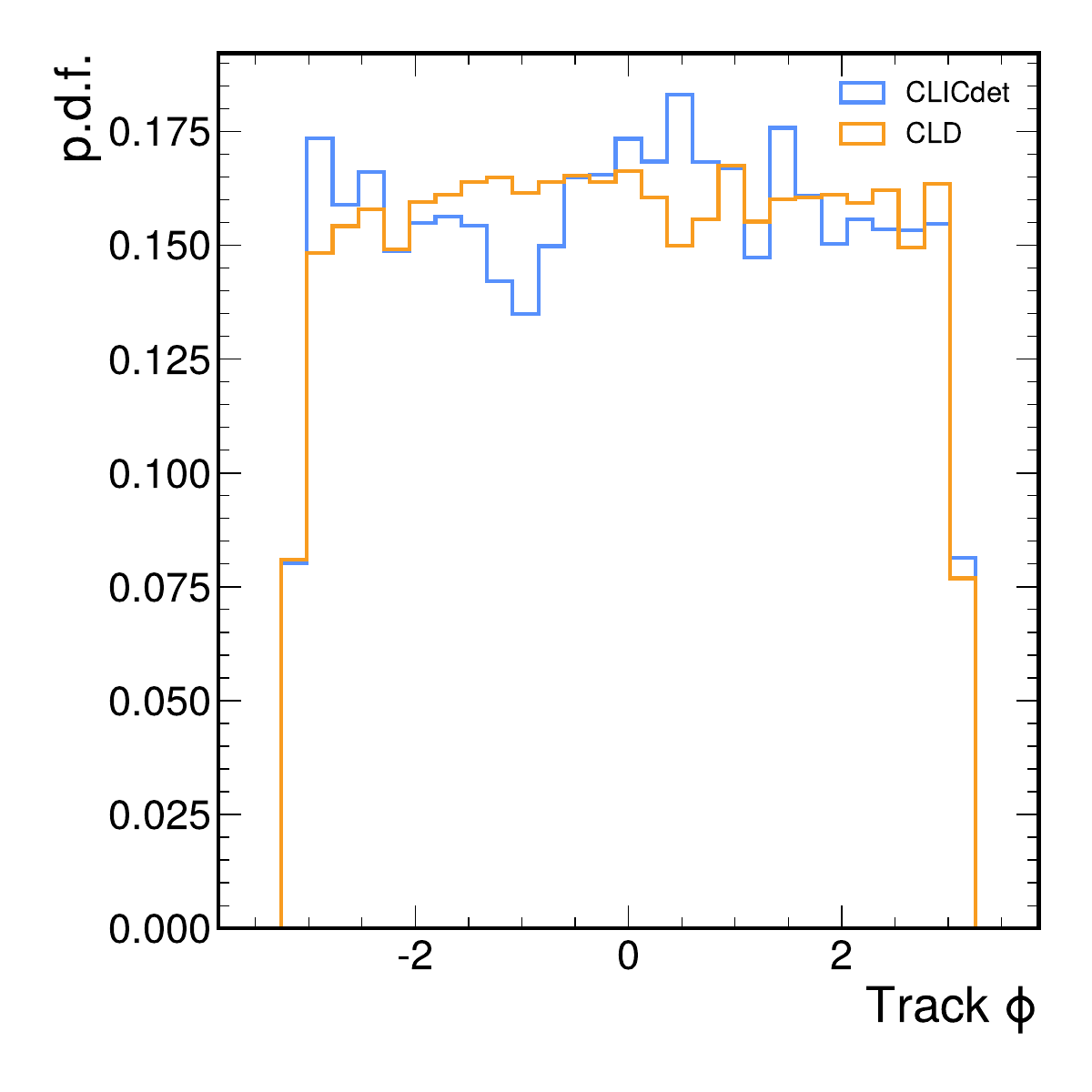}
    \includegraphics[width=0.24\textwidth]{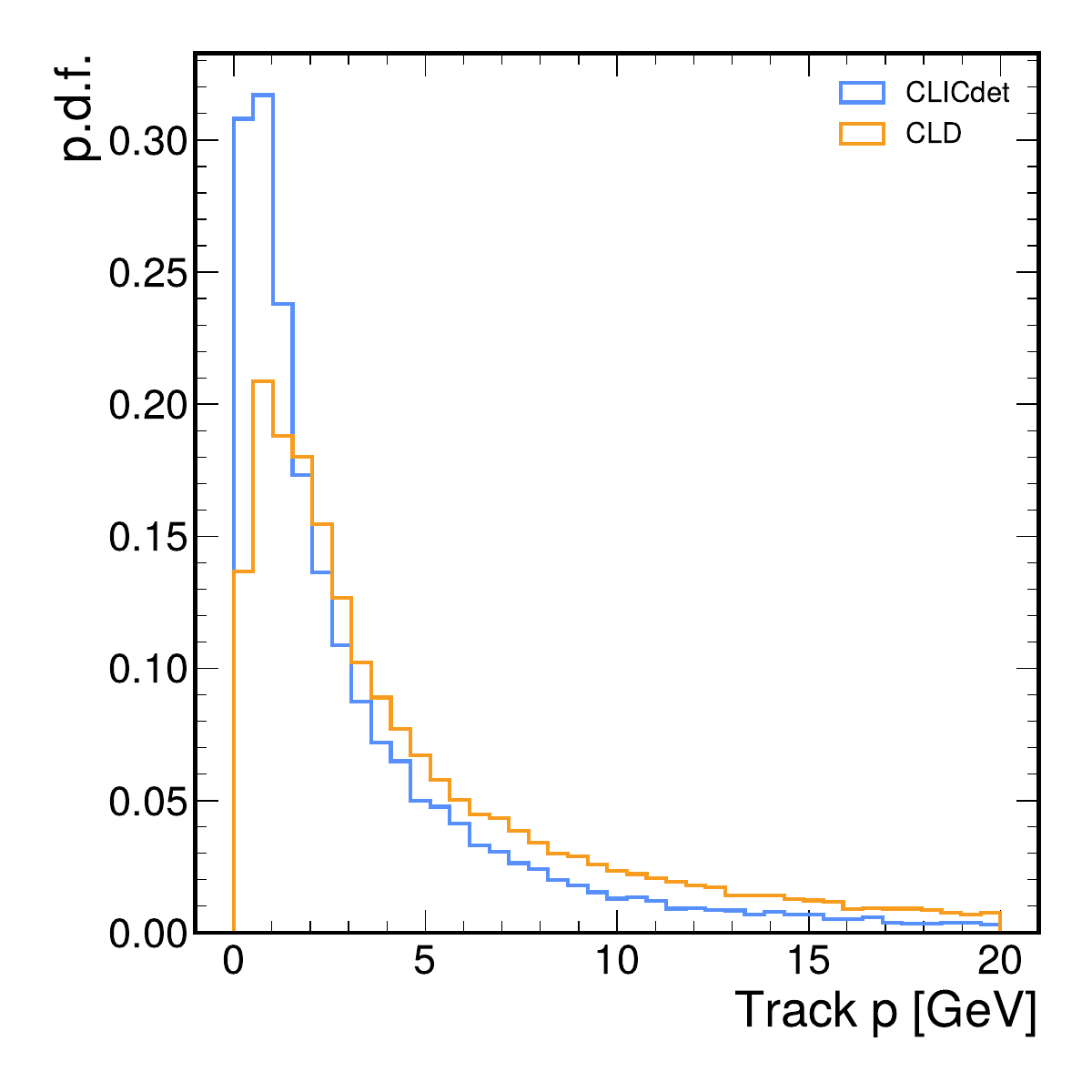} \\
    \includegraphics[width=0.24\textwidth]{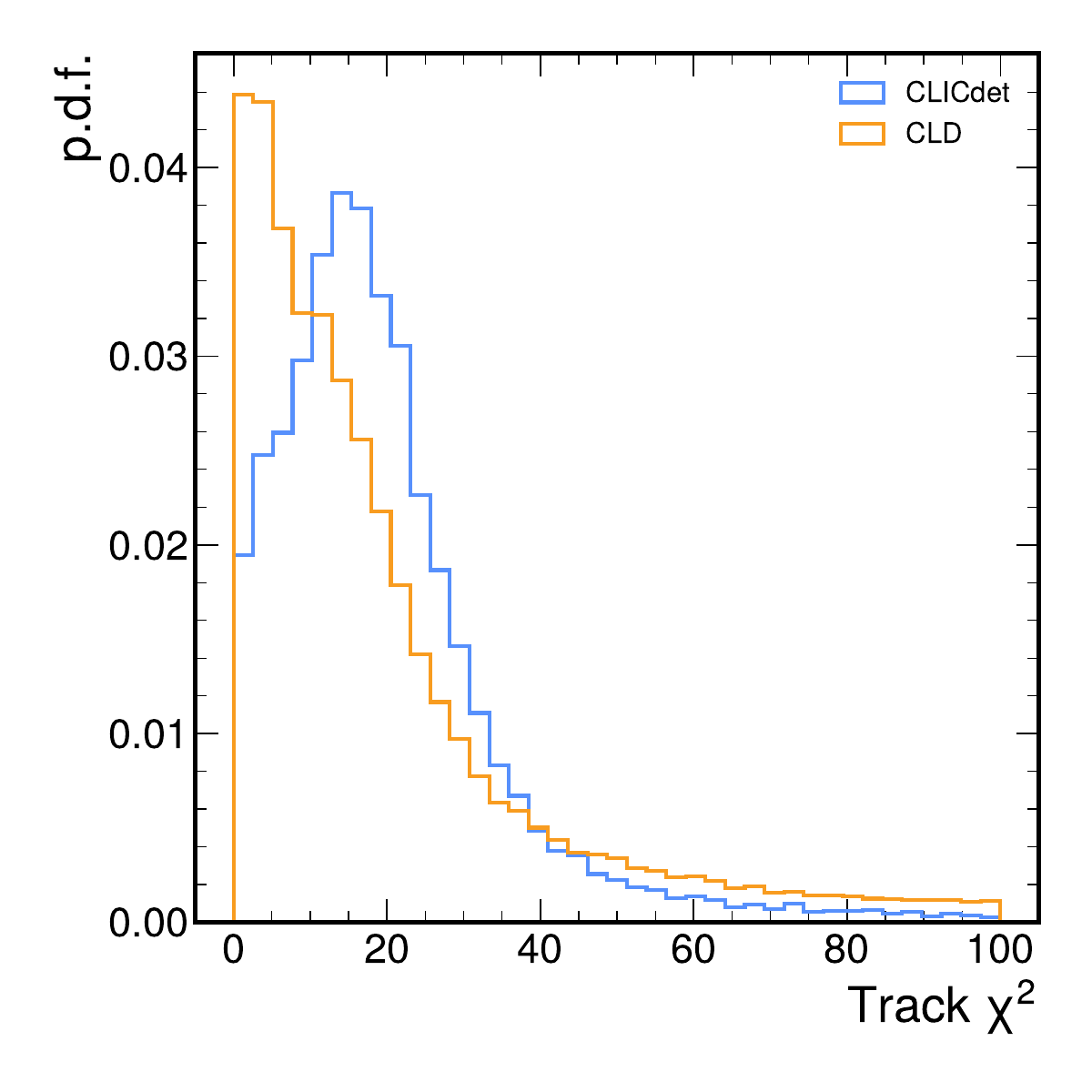}
    \includegraphics[width=0.24\textwidth]{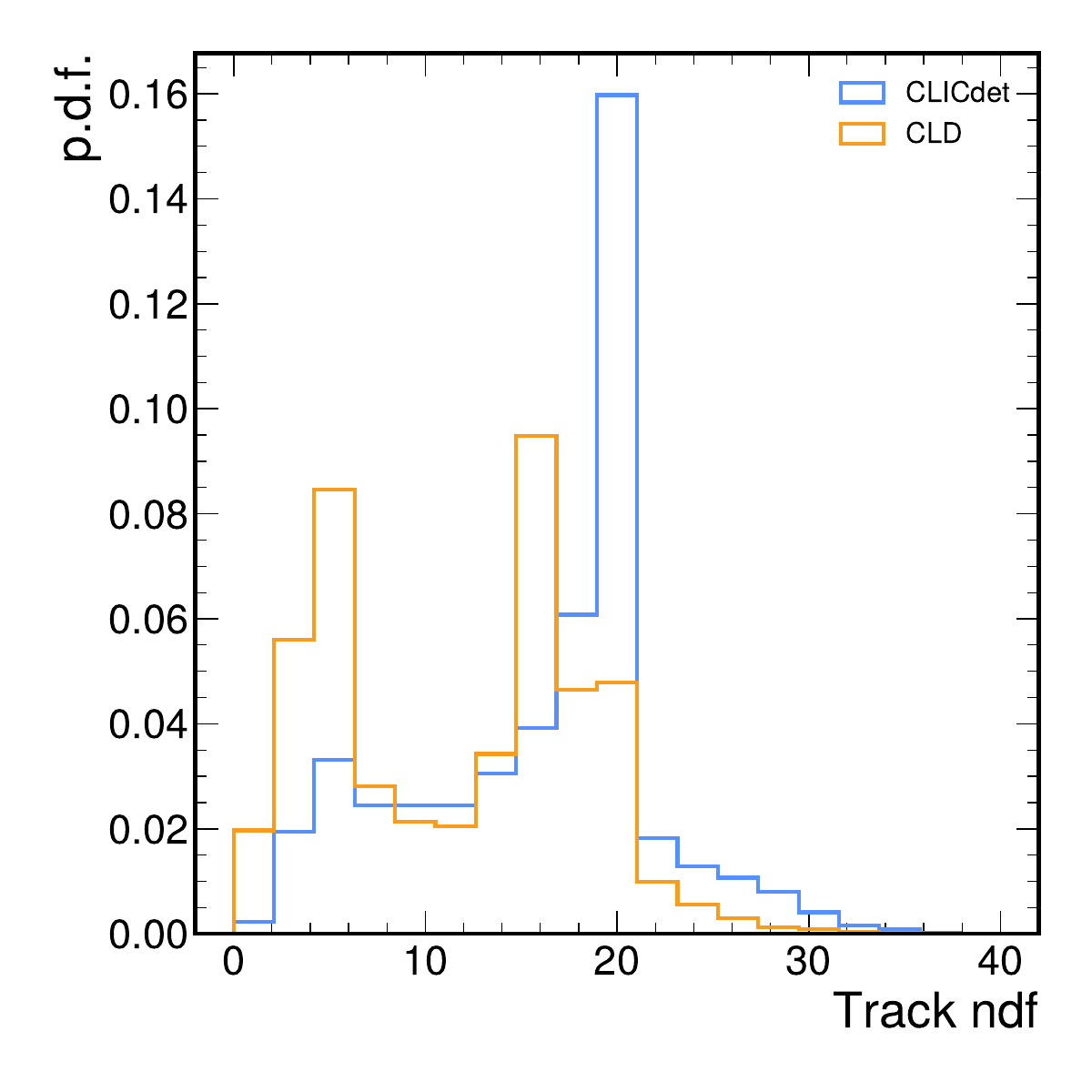} 
    \includegraphics[width=0.24\textwidth]{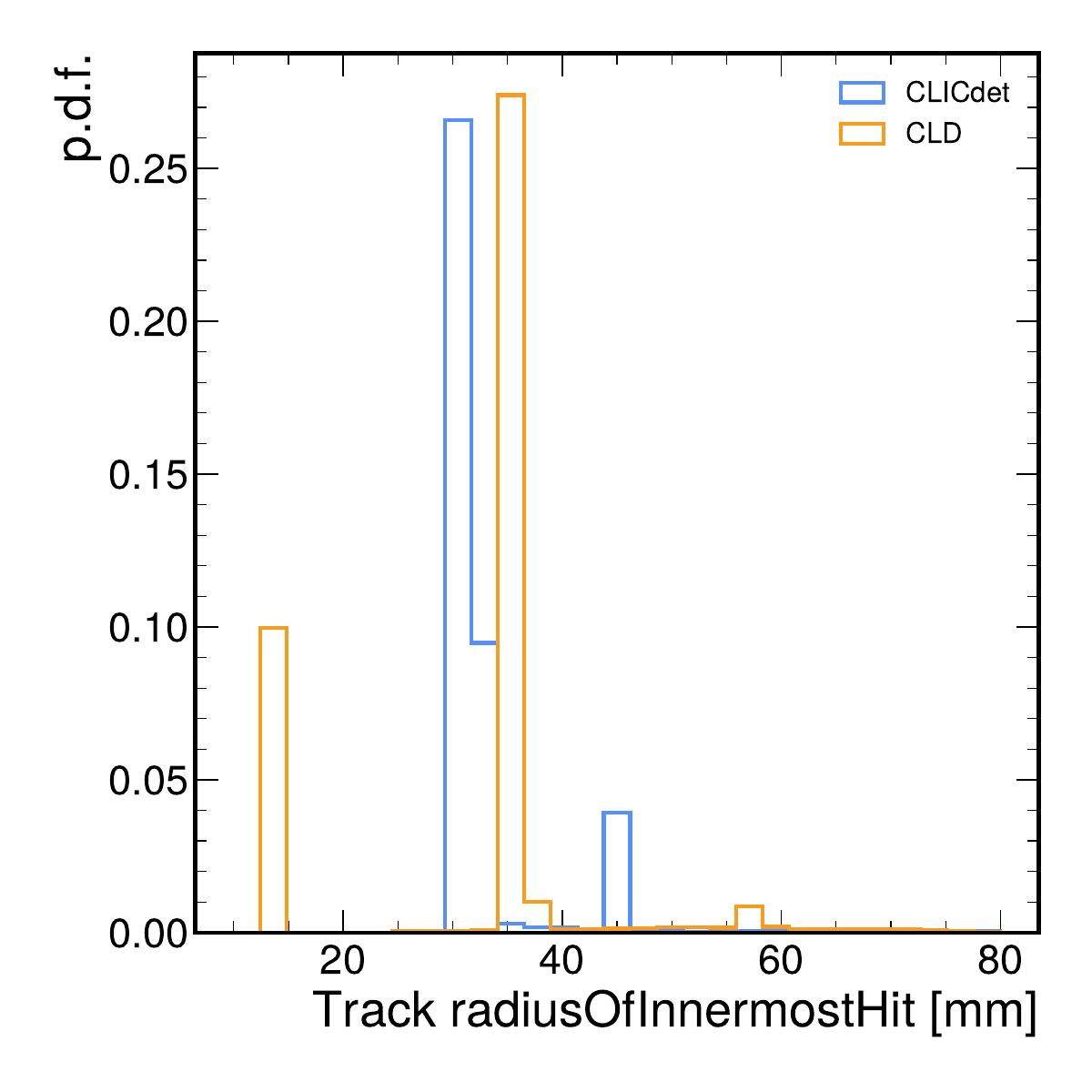}
    \includegraphics[width=0.24\textwidth]{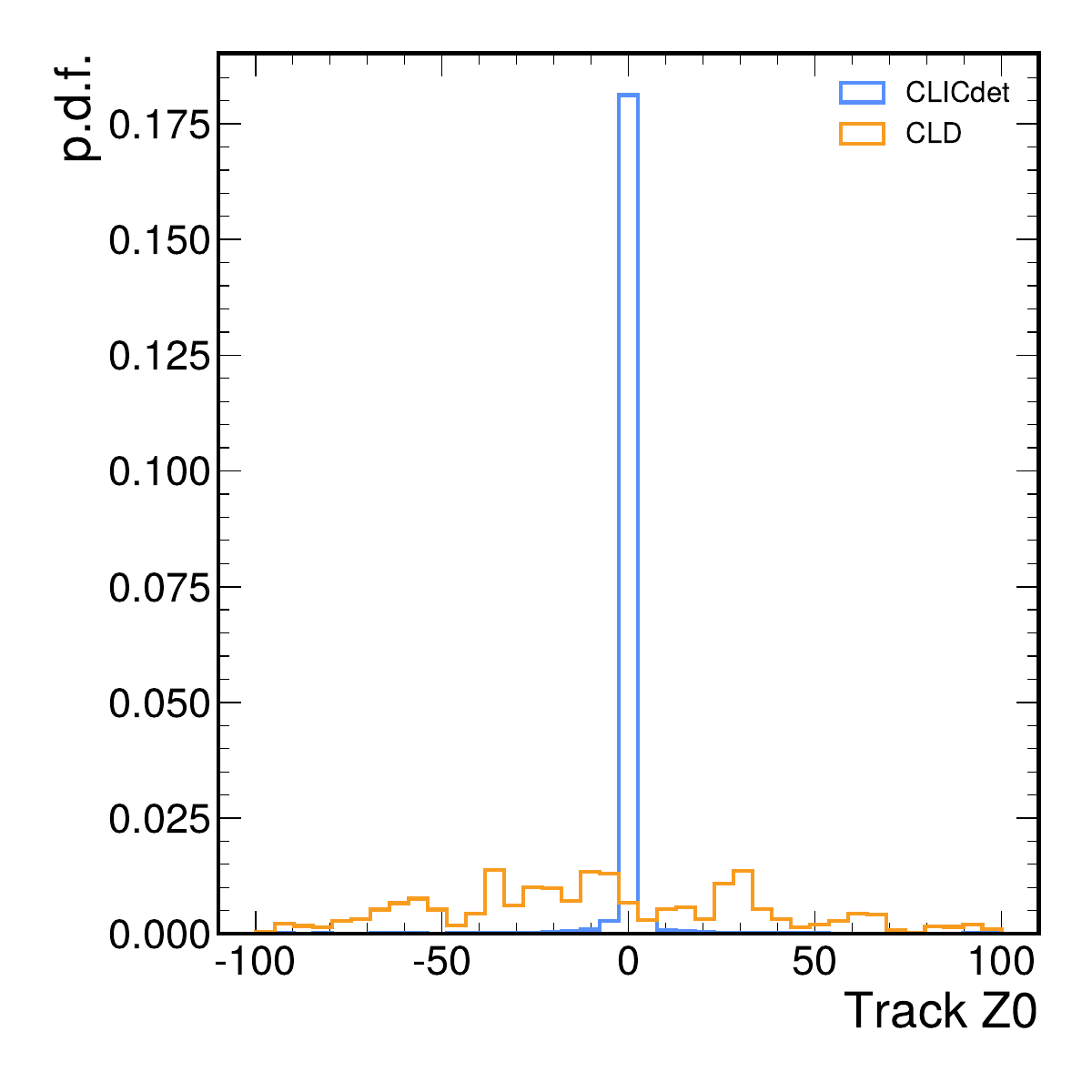} \\
    \includegraphics[width=0.24\textwidth]{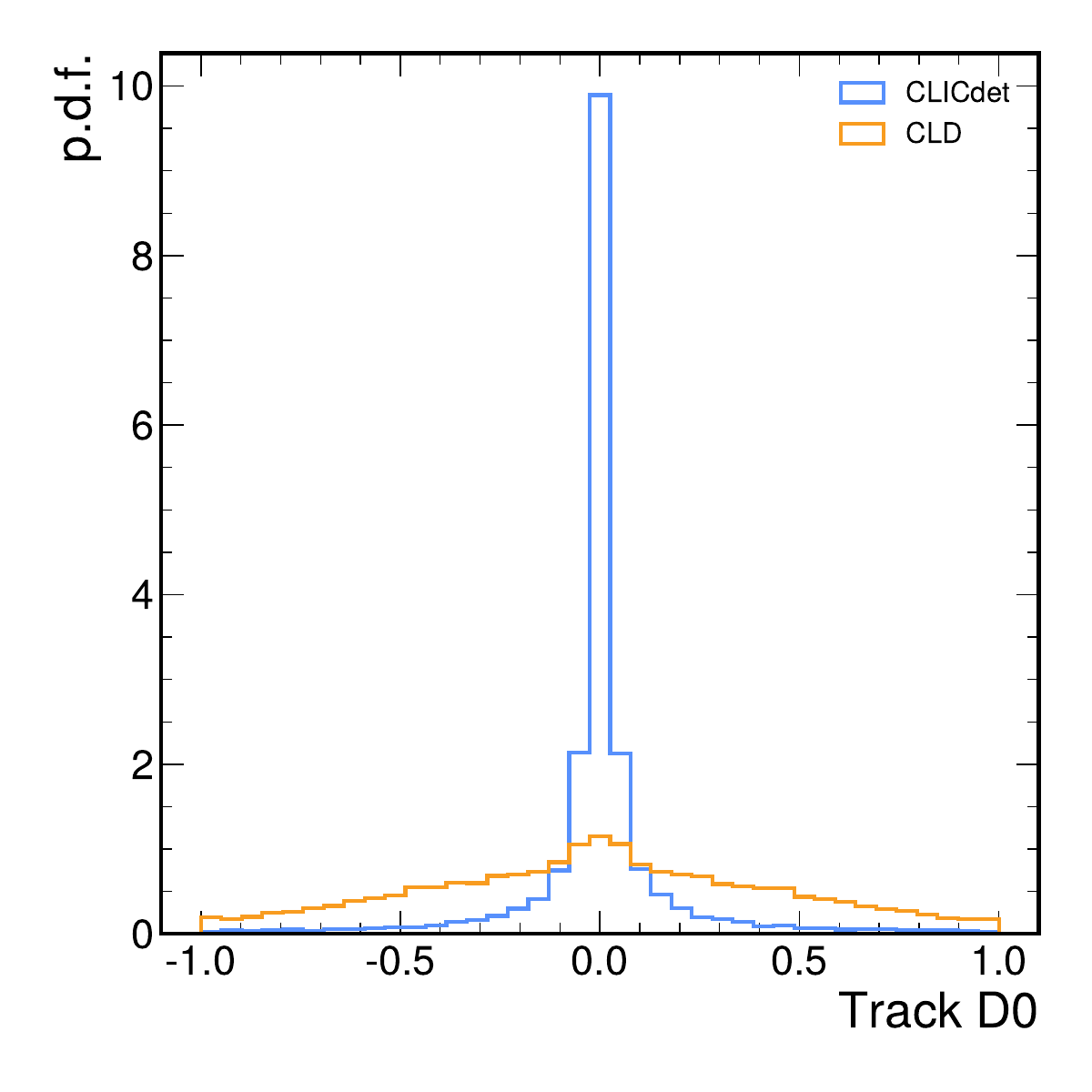}
    \includegraphics[width=0.24\textwidth]{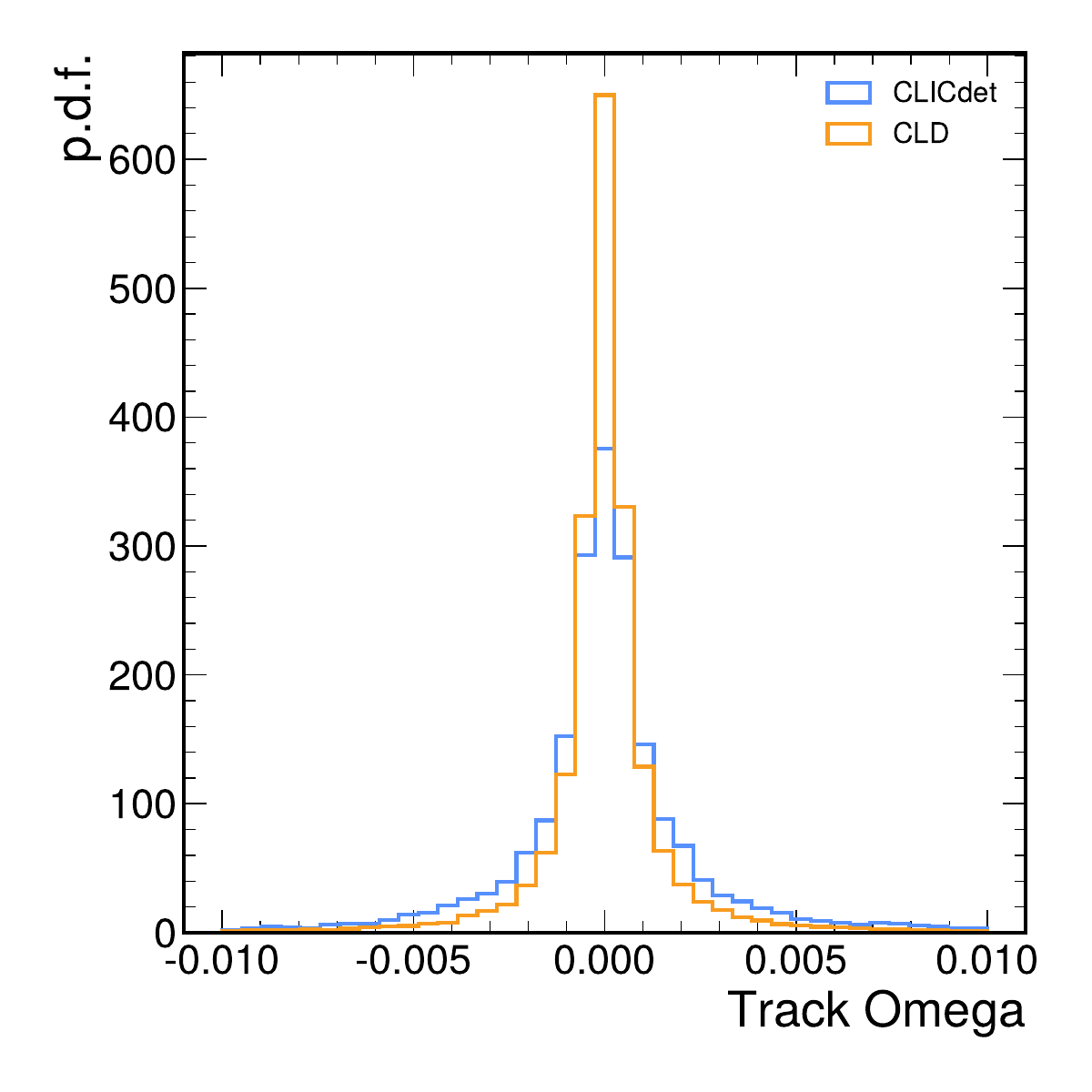}
    \includegraphics[width=0.24\textwidth]{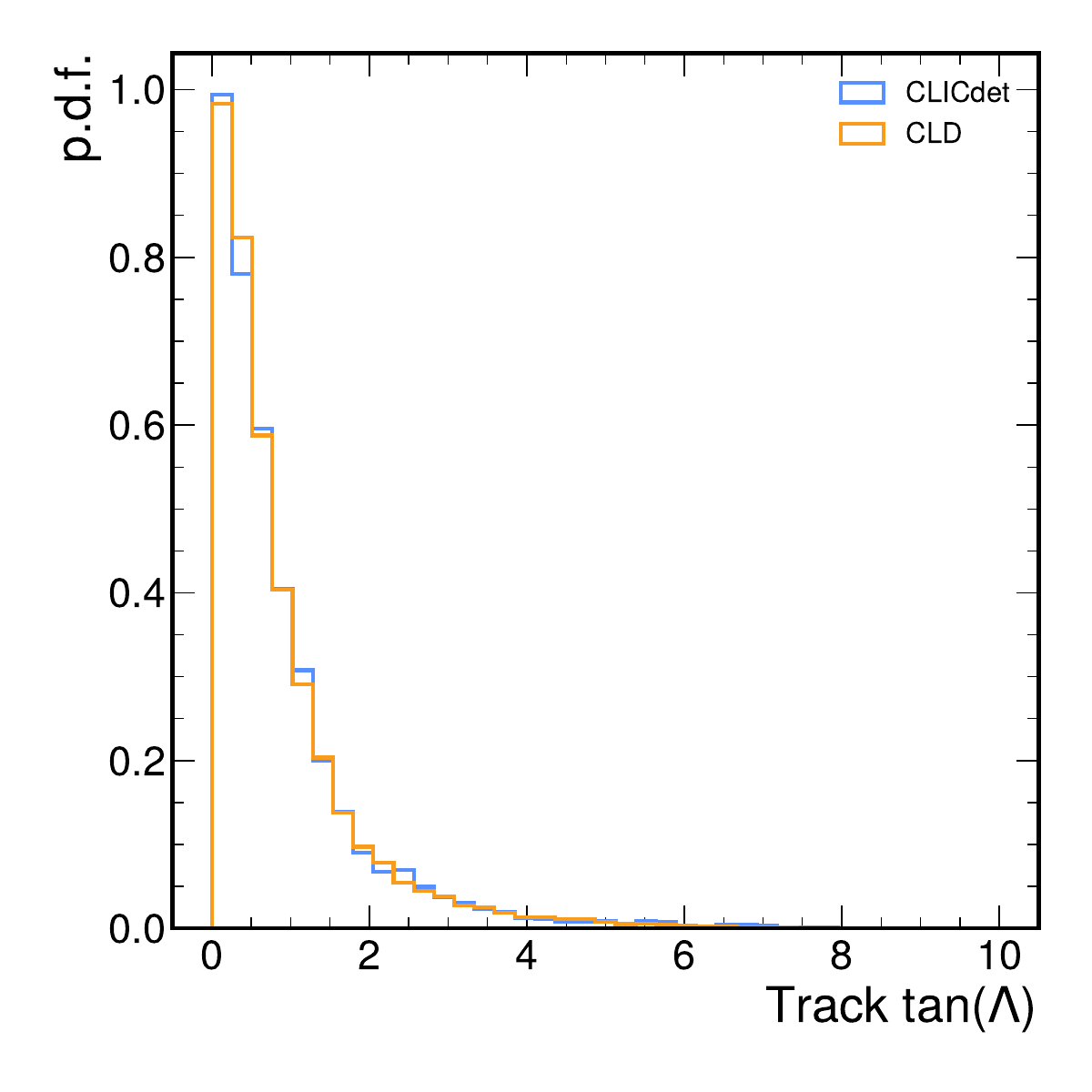} \\     
    \caption{The input feature distribution for reconstructed tracks in CLICdet and CLD.}
    \label{fig:input_dist_tracks}
\end{figure*}

\begin{figure*}[htbp]
    \centering
    \includegraphics[width=0.24\textwidth]{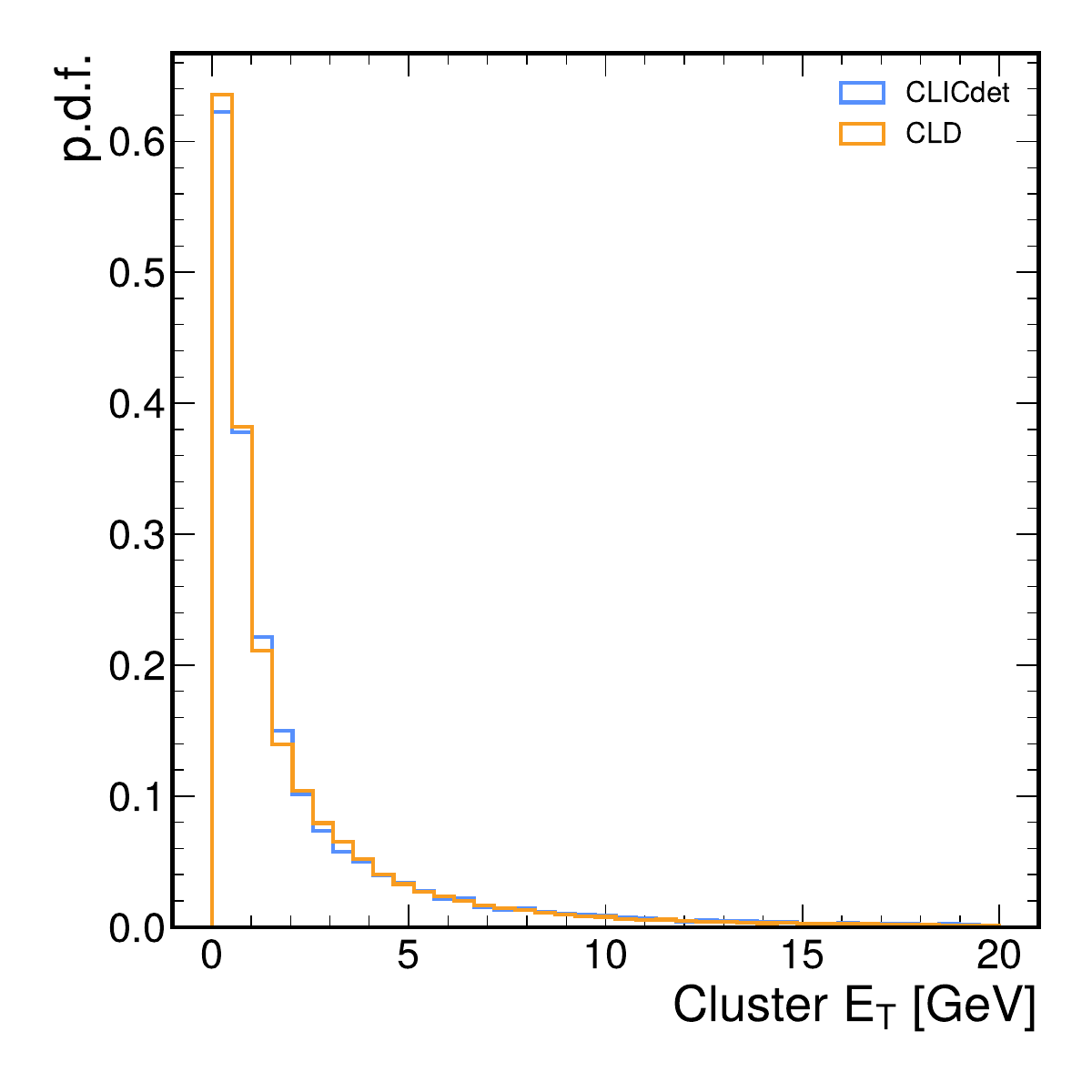}
    \includegraphics[width=0.24\textwidth]{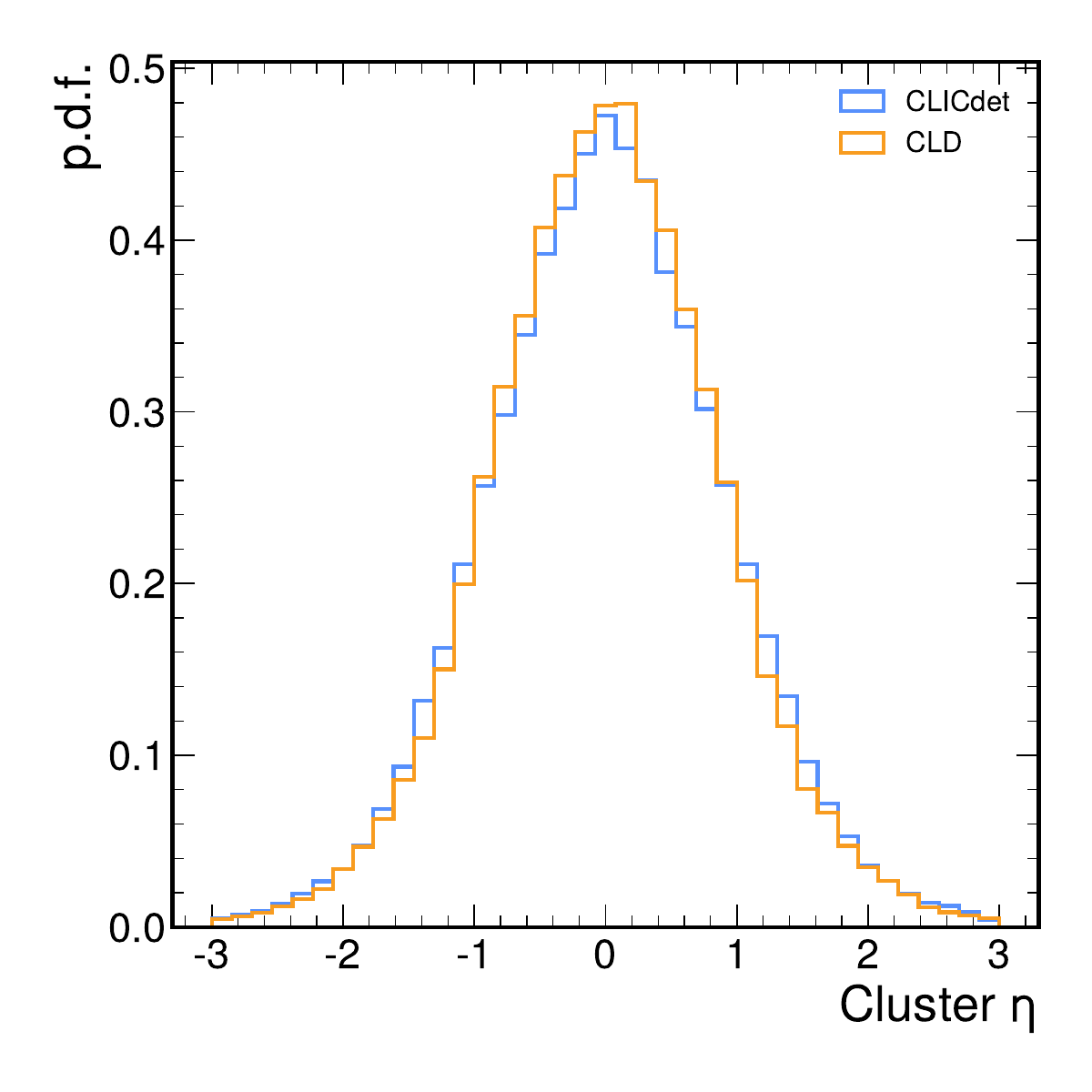}
    \includegraphics[width=0.24\textwidth]{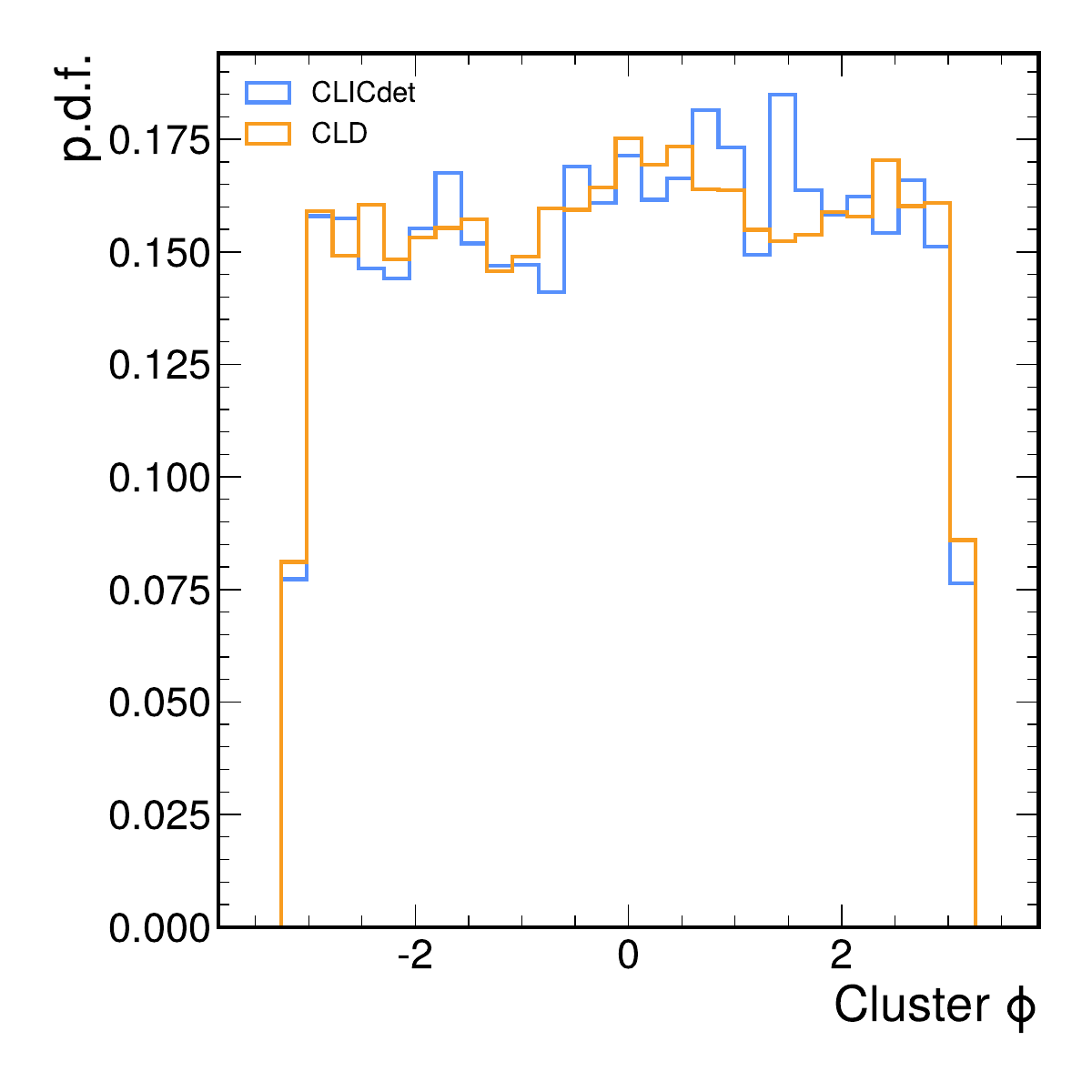} 
    \includegraphics[width=0.24\textwidth]{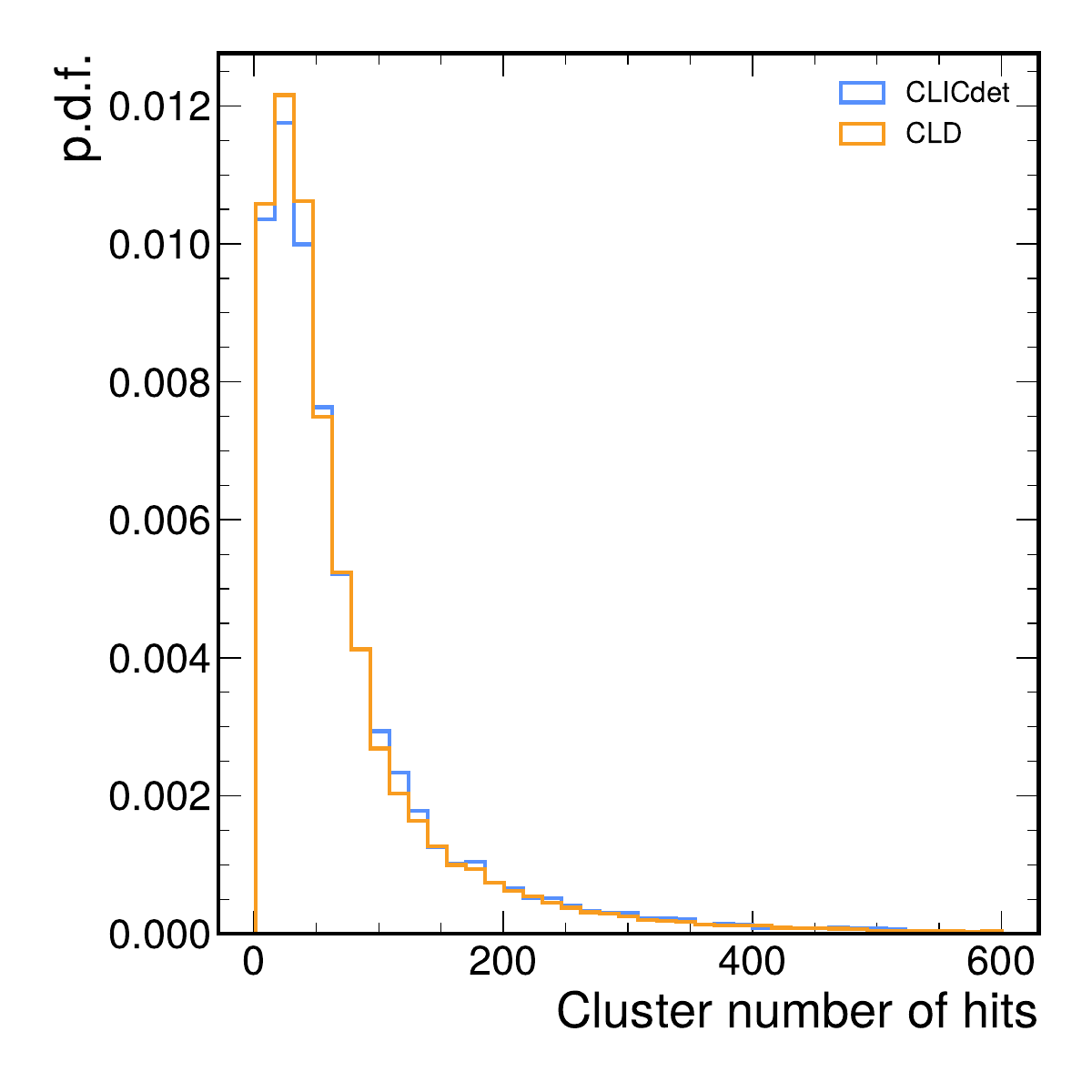} \\
    \includegraphics[width=0.24\textwidth]{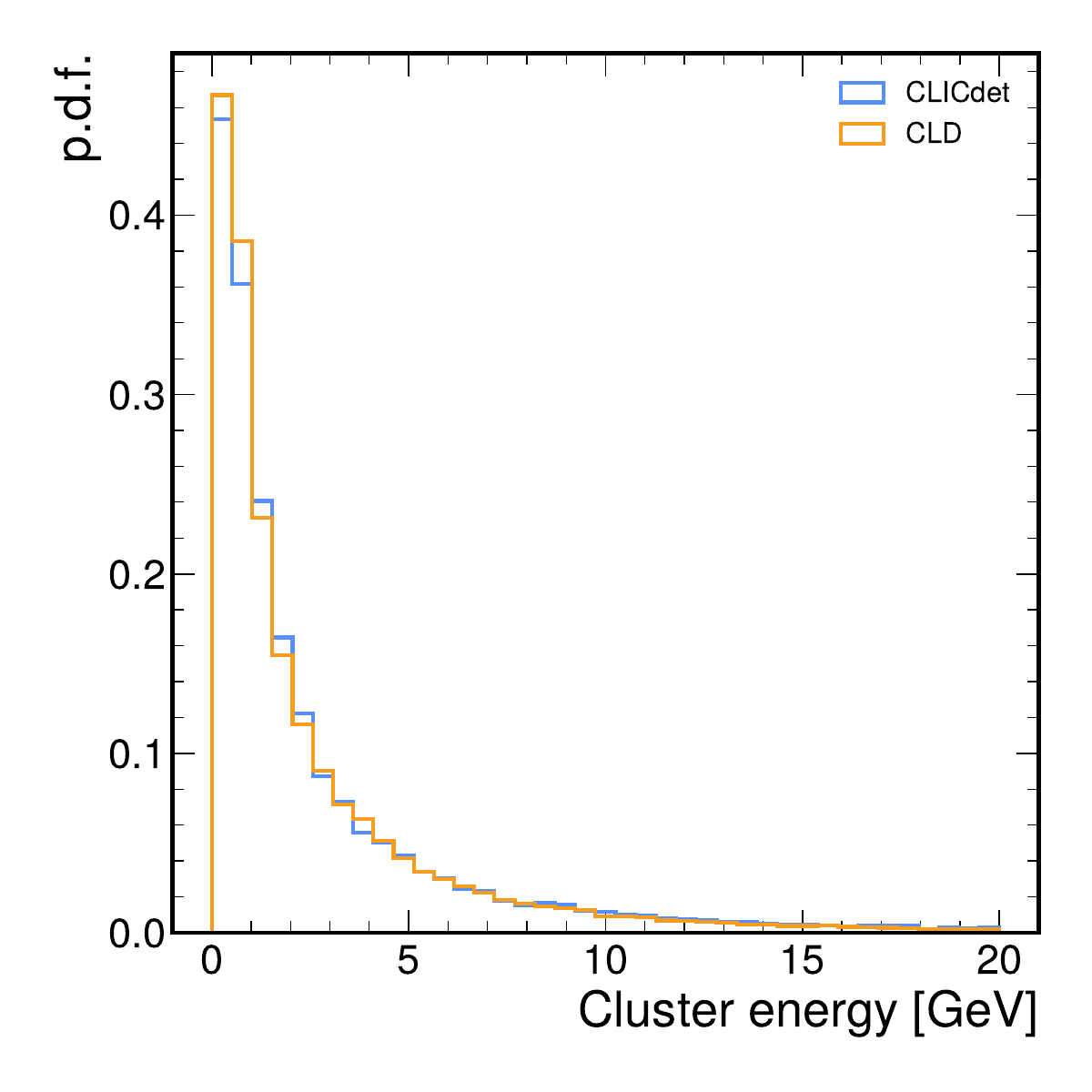}
    \includegraphics[width=0.24\textwidth]{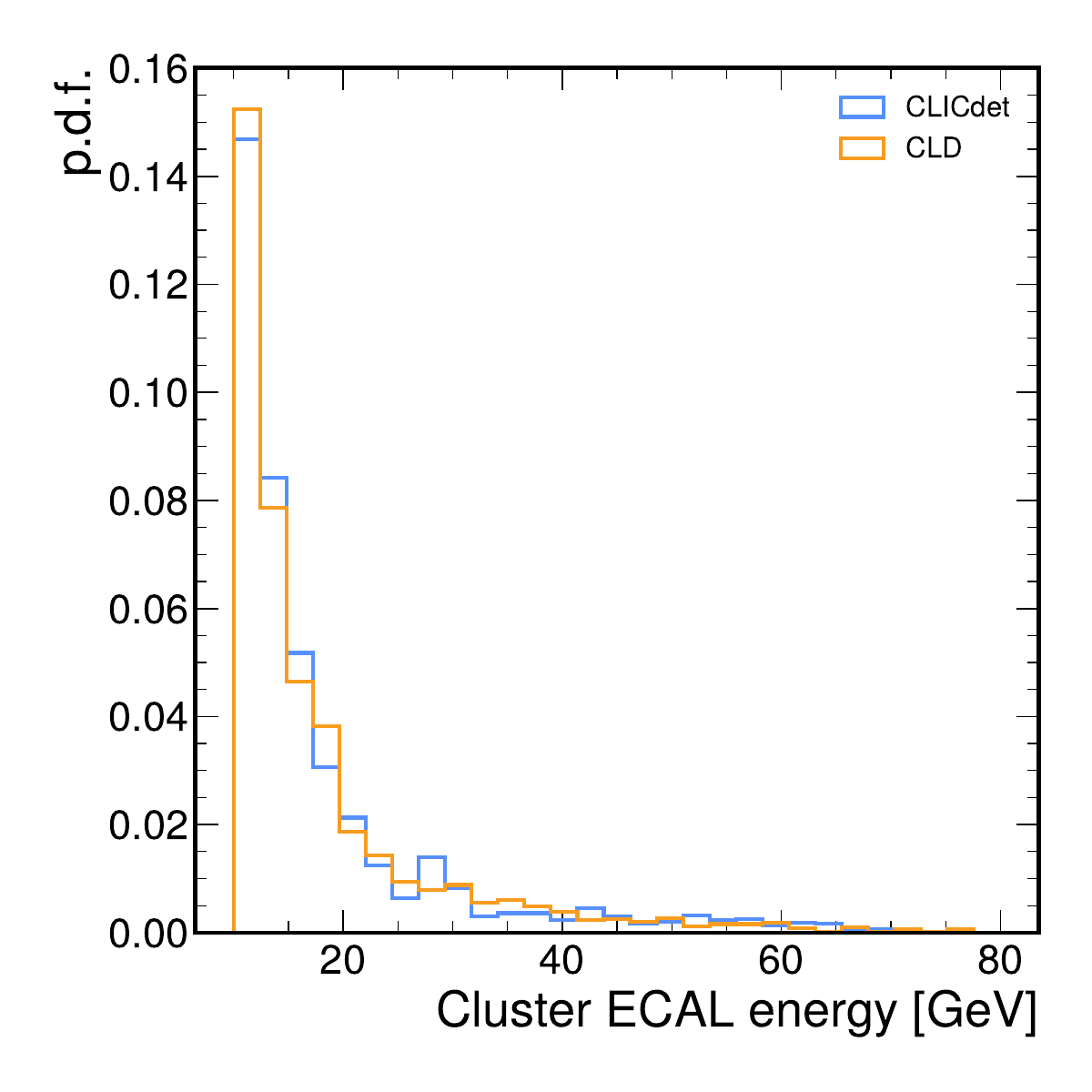}
    \includegraphics[width=0.24\textwidth]{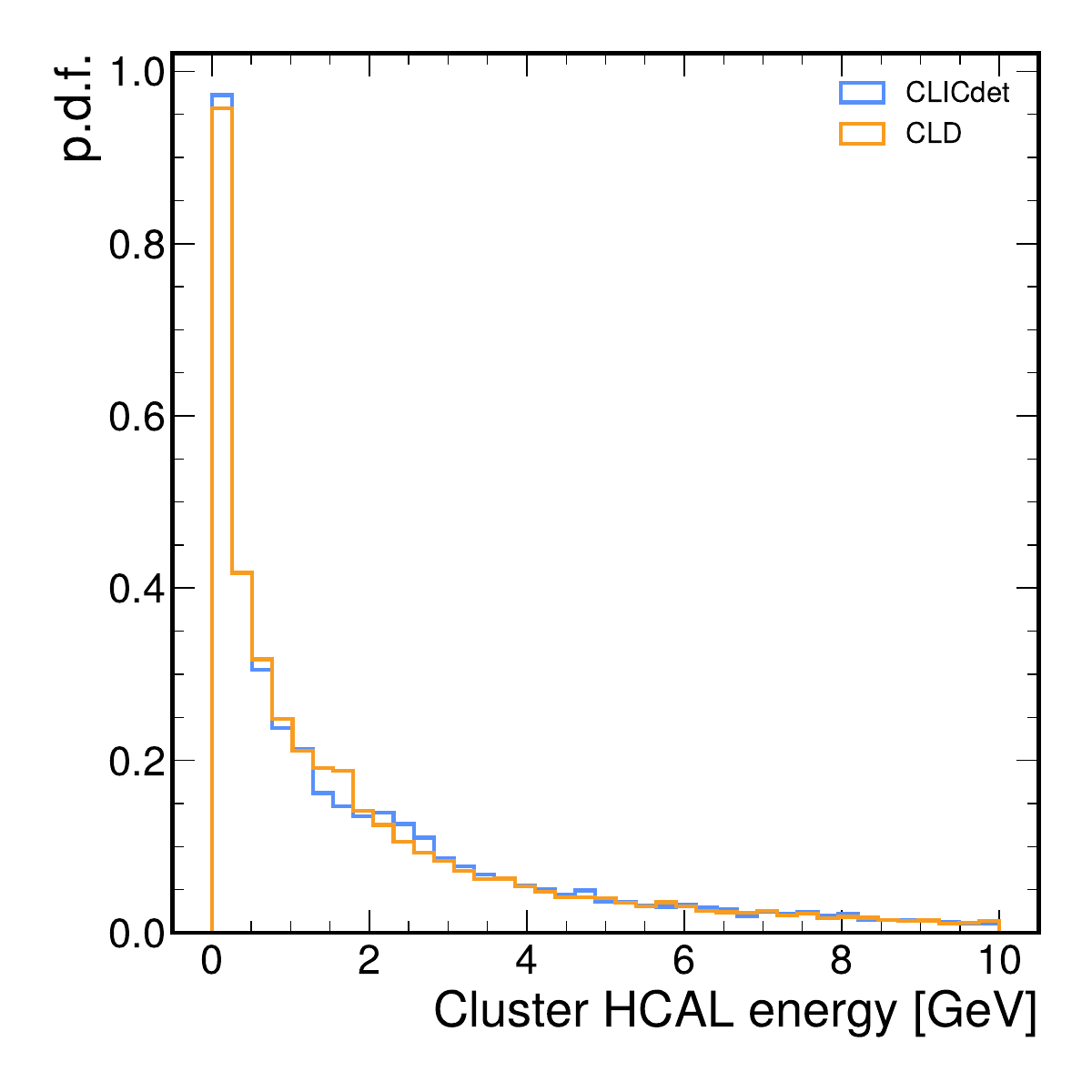}
    \includegraphics[width=0.24\textwidth]{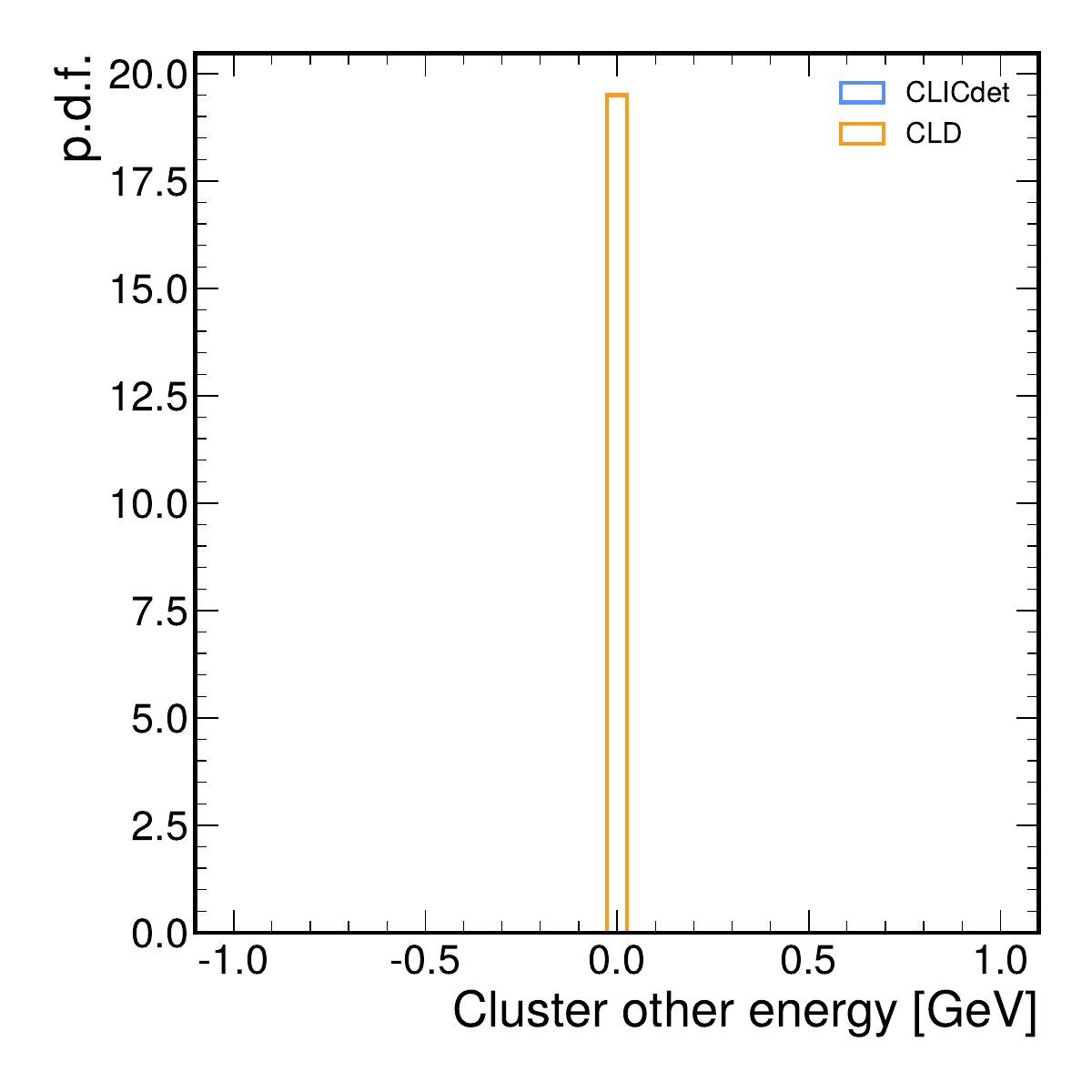} \\
    \includegraphics[width=0.24\textwidth]{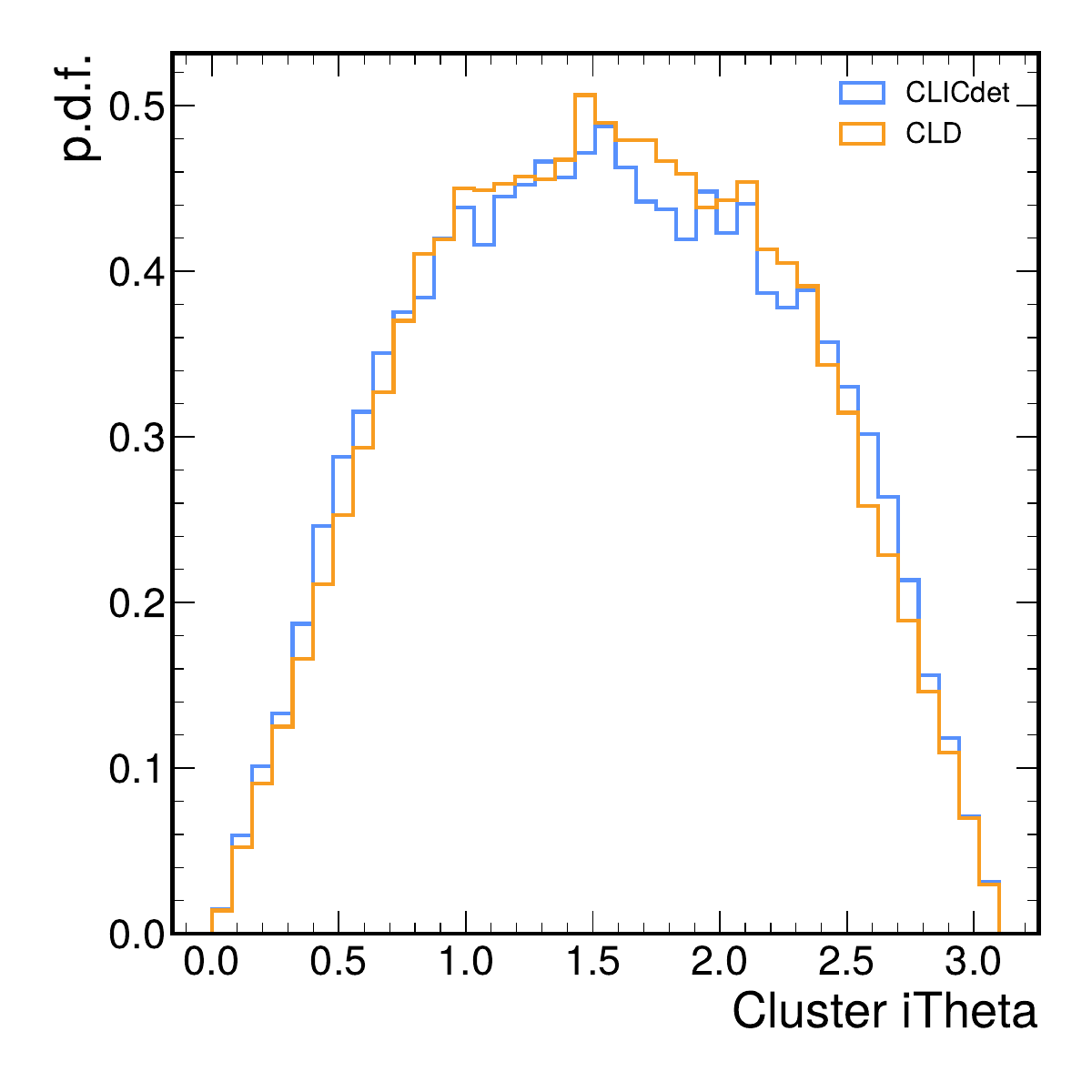}
    \includegraphics[width=0.24\textwidth]{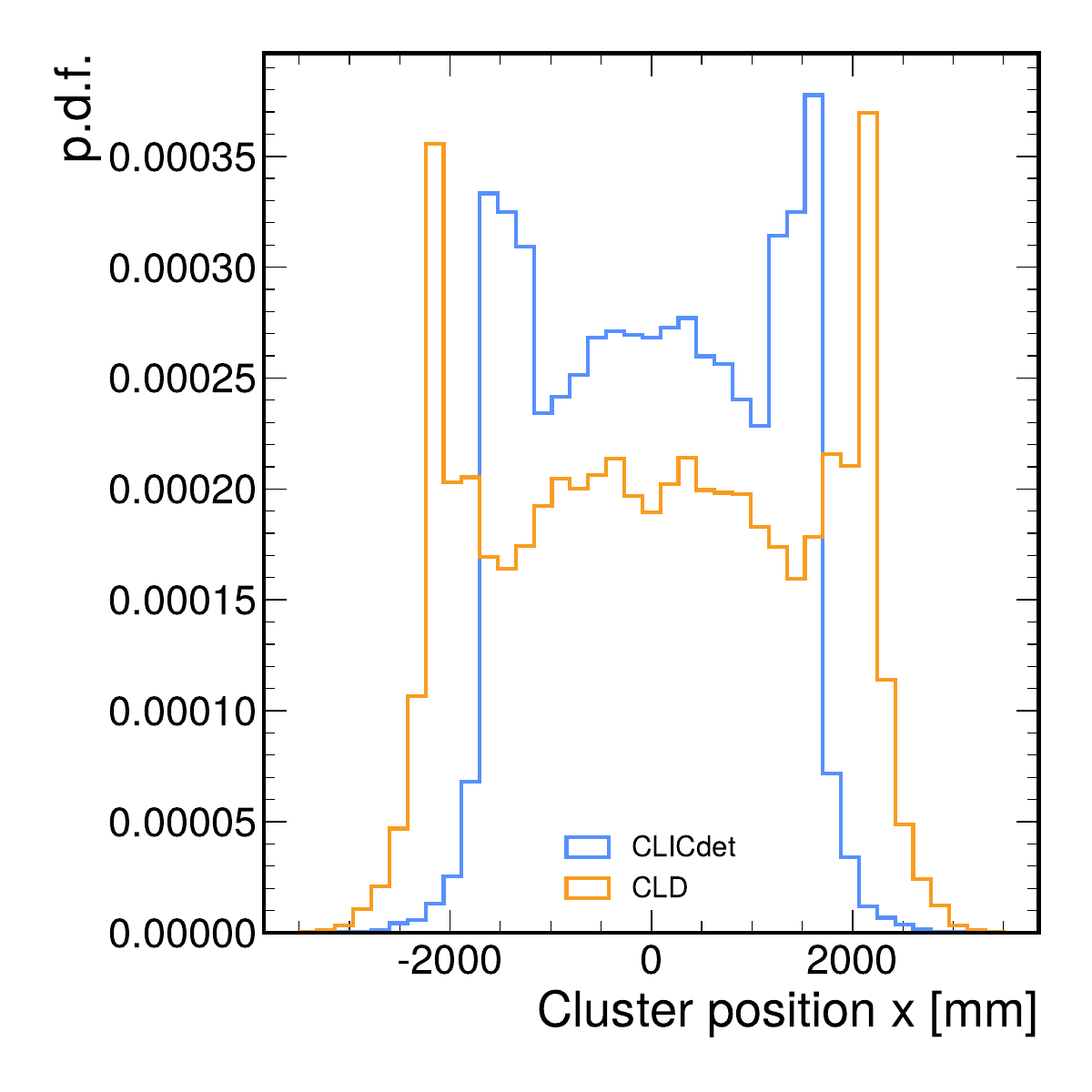}
    \includegraphics[width=0.24\textwidth]{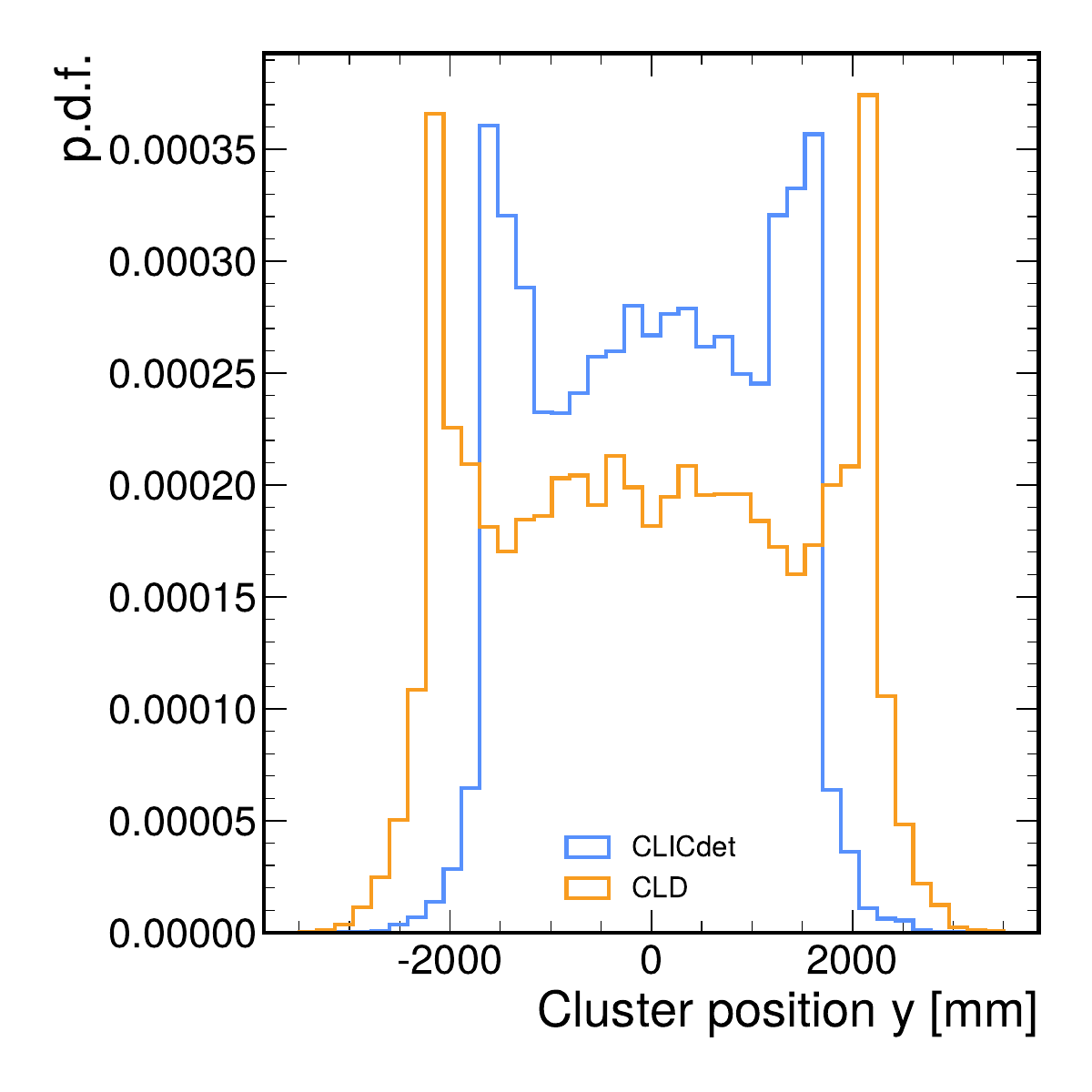}
    \includegraphics[width=0.24\textwidth]{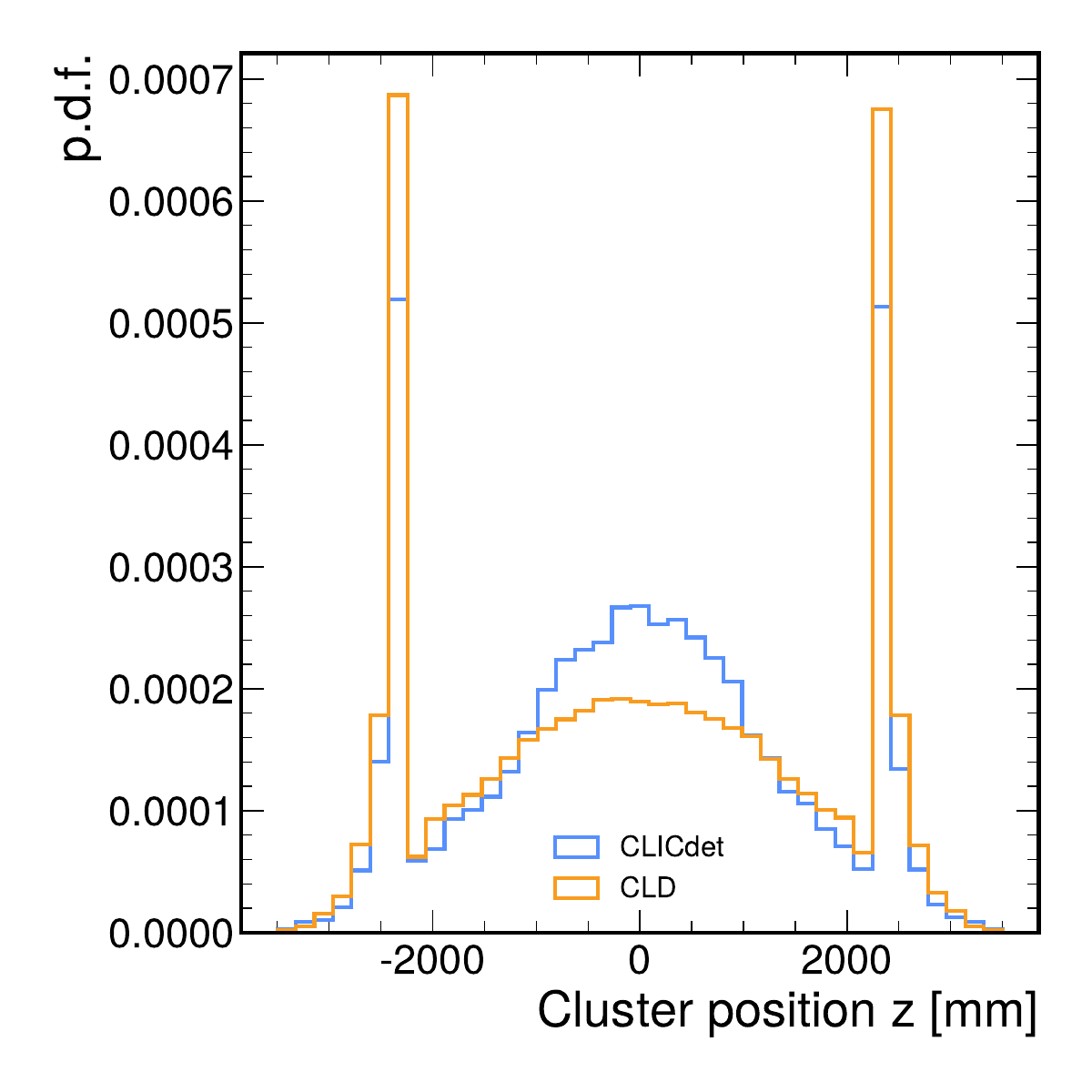} \\
    \includegraphics[width=0.24\textwidth]{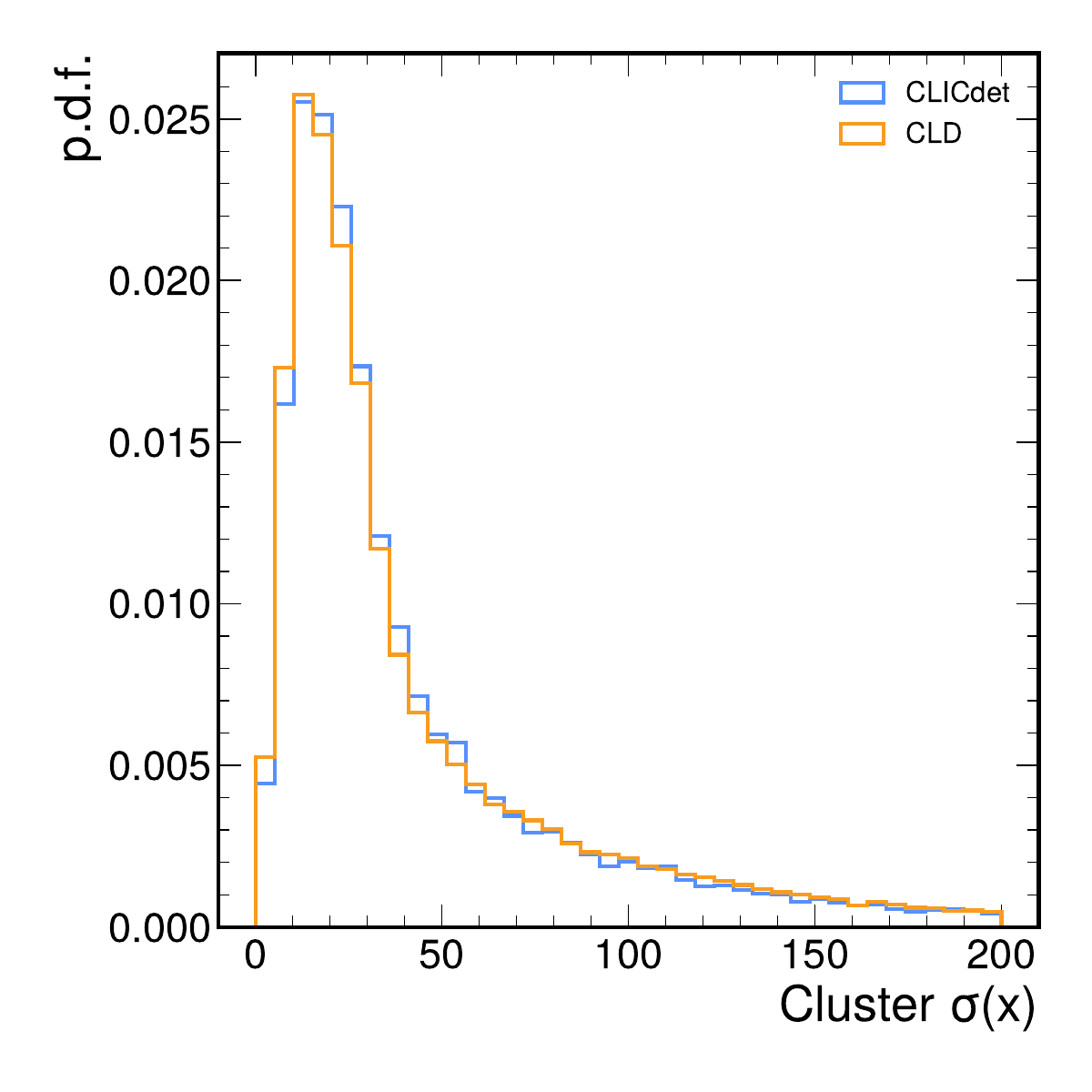}
    \includegraphics[width=0.24\textwidth]{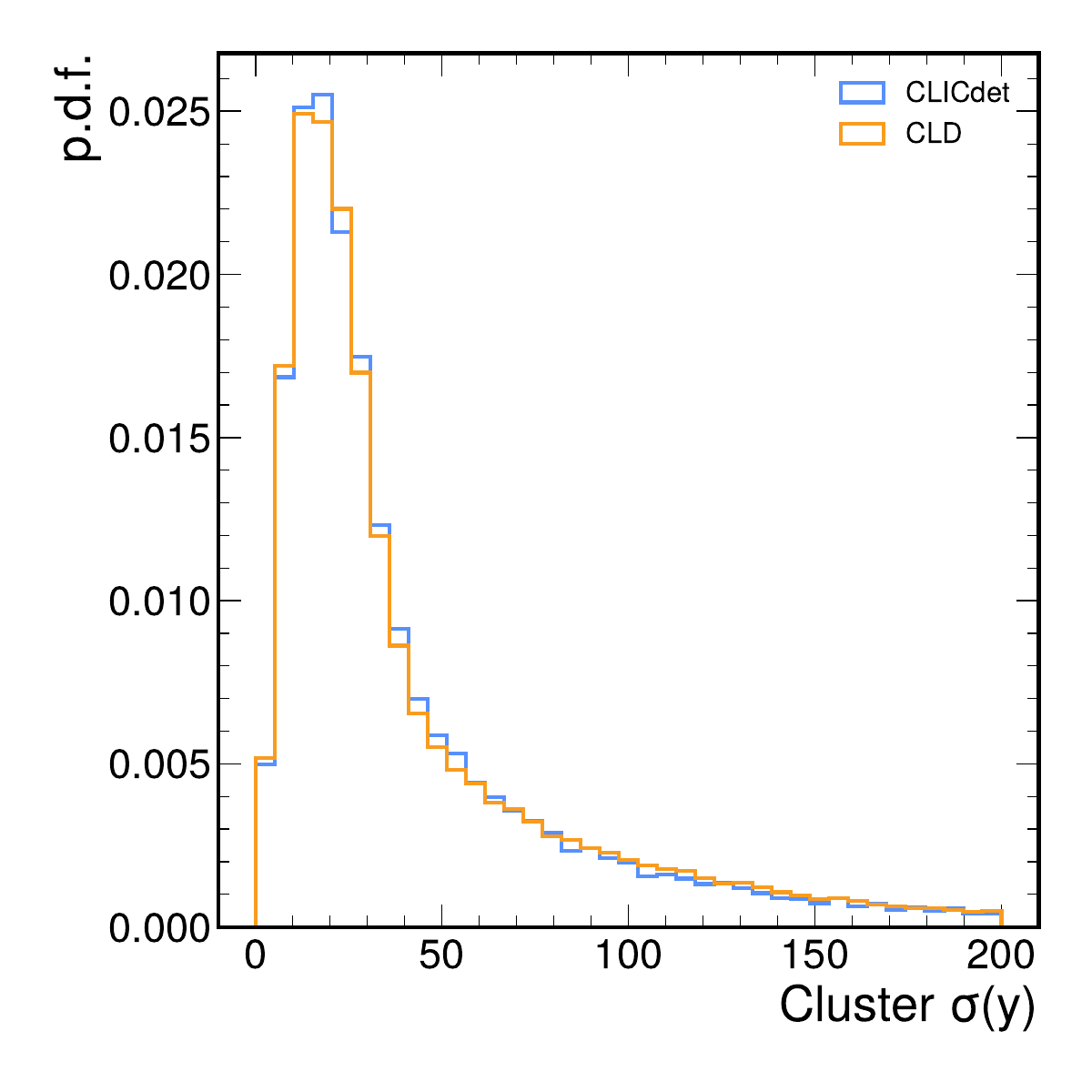}
    \includegraphics[width=0.24\textwidth]{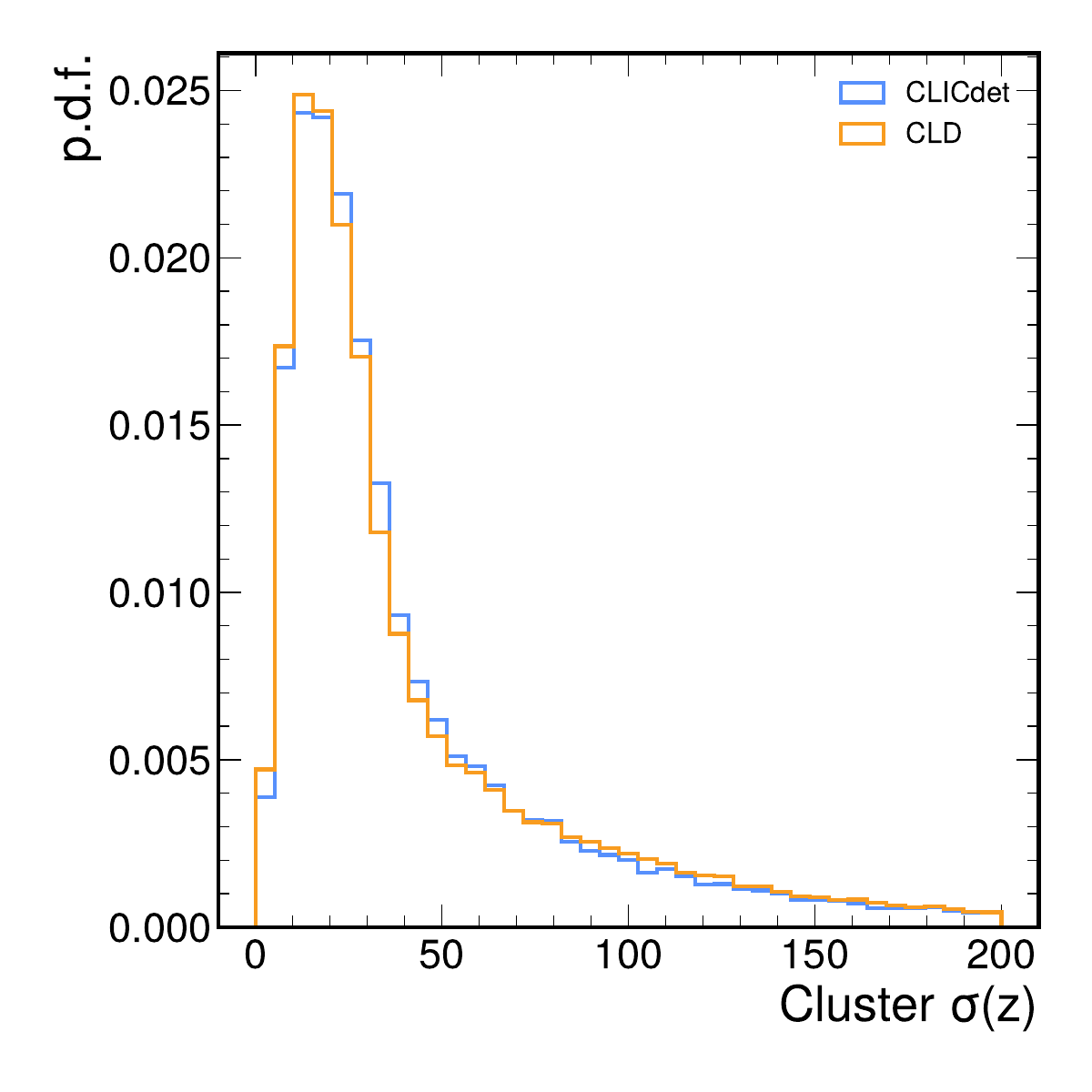} \\
    \caption{The input feature distribution for reconstructed clusters in CLICdet and CLD.}
    \label{fig:input_dist_clusters}
\end{figure*}

For input tracks, $p$ is the magnitude of the total 3D momentum, $\chi^2$ is the chi-squared test statistic of the track fit, $n_\mathrm{dof}$ is the number of degrees of freedom of the track fit, $r_\mathrm{innermost}$ is the radius of the innermost hit that has been used in the track fit, $\lambda$ is the dip angle of the track in $r$--$z$, $d_0$ is the transverse impact parameter, $\Omega$ is the signed curvature of the track, and $z_0$ is the longitudinal impact parameter.
For input clusters, $x,y,z$ are the Cartesian coordinates for the position of the cluster in mm, $\theta$ is the polar angle of the intrinsic direction of the cluster,  $E_{\mathrm{ECAL}}$ is the energy deposited in the electromagnetic calorimeters, $E_{\mathrm{HCAL}}$ is the energy deposited in the hadron calorimeters, $E_{\mathrm{other}}$ is the energy deposited in the muon chambers, $N_\mathrm{hits}$ is the number of hits in the calorimeters associated with the cluster, and $\sigma(x_{\mathrm{hits}}),\sigma(y_{\mathrm{hits}}), \sigma(z_{\mathrm{hits}})$ are the standard deviation of the energy weighted positions of the hits associated with the cluster.
The input feature distributions for CLICdet and CLD are shown for reconstructed tracks in Fig.~\ref{fig:input_dist_tracks}, and for clusters in Fig.~\ref{fig:input_dist_clusters}.
The main differences between the track distributions are the track fit parameters ($\chi^2$, $n_\mathrm{dof}$), and track impact parameters ($z_0$, $z_0$), due to the difference in magnetic field and vertex radius between the two detector configurations.
We suspect that the difference in track \ptmomentum distribution is due to the difference in center-of-mass energy between the samples generated for CLICdet (at $\sqrt{s}=380\GeV$) and CLD (at $\sqrt{s}=365\GeV$).
For clusters, the main difference between the two detector distributions is the cluster positions due to the different calorimeter positions.

\clearpage

\bibliographystyle{cms_unsrt}
\bibliography{references}

\end{document}